\documentclass[final,3p,times, onecolumn]{elsarticle}

\usepackage{latexsym}
\usepackage{bm}

\usepackage{amsmath}
\usepackage{graphicx}   
\usepackage{enumerate}   
\usepackage{amssymb}
\begin{document}

\begin{frontmatter}



\title{Bivectors in Newman-Penrose formalism in General Relativity - from electromagnetism to Weyl curvature tensor}

\author{Wytler Cordeiro dos Santos}
\ead{wytler@fis.unb.br}
\address{Universidade de
Bras\'\i lia, CEP 70910-900, DF, Brasil}

\begin{abstract}
The use of the bivectors in the General Relativity with Newman-Penrose formalism is important to the description of the exact solutions of the Einstein's field equations. 
This review is devoted to introduce the basic ideas with calculation details of the bivectors in Newman-Penrose formalism through conversion of the complex self-dual electromagnetic field from orthornormal Minkowski basis to bivectors in the complex null tetrad basis.
Furthermore, in this context, it is obtained the complex coefficients of the Weyl tensor in Newman-Penrose formalism.
\end{abstract}


\end{frontmatter}

% \linenumbers

%\label{}

\section{Introduction}

The Newman-Penrose null tetrad or pseudo-orthornormal basis is a mathematical formalism applied to General Relativity that is very useful to construct and analyze the exact solutions of Einstein's field equations. This mathematical formalism was first proposed by Newman and Penrose  in the year of 1962 \cite{Newman} and since then several researchers have used this mathematical formalism in the description of exact solutions of General Relativity \cite{Kramer, Griffiths}. In General Relativity there are mathematical objects skew-symmetric in one or two pairs of indices, for instance, the field-strenght tensor of electromagnetism and the Riemann curvature  tensor. These antisymmetric tensors of second order can be associated to a bivector, 2-form, determined by  exterior product,
\begin{equation}
\label{bivector}
 {\bm X} = X_{\mu\nu}~dx^{\mu}\wedge dx^{\nu}.
\end{equation}
There are several difficulties to understand the Newman-Penrose null tetrad applied to skew-symmetric bivectors, such as the electromagnetic tensor and the Weyl tensor, that one observe that since the pioneering work of Newman and Penrose to the classical reference of H. Stephani et al. \cite{Kramer} they did not present some important details. For the use of the bivectors in the General Relativity with Newman-Penrose formalism, it is necessary to introduce a basis to bivector space. It is obtained with the aid of the electromagnetic field tensor and from this on, the basis of bivector space can be extended to the Weyl curvature tensor. 
When converting the complex self-dual electromagnetic field to a bivector in Newman-Penrose formalism, it is possible to write the Maxwell equations in terms of the spin coefficients of Newman-Penrose, the known equations of Newman-Penrose equations of electromagnetism. Writing beyond, one can obtain the complex coefficients of the Weyl tensor in Newman-Penrose formalism. These details are matters to be discussed in this review.

\section{The Newman-Penrose coordinate system}

Our focus is on Riemannian manifold with Lorentzian signature of spacetime metric: $(-+++)$, where $\{{\bm E}_{\mu}\} = \left\{\dfrac{\partial}{\partial t},\dfrac{\partial}{\partial x}, \dfrac{\partial}{\partial y},\dfrac{\partial}{\partial z} \right\}$ is an orthornormal tetrad of the Lorentz frame, with Greek letter span $\mu = 0,1,2,3$ and with its dual basis $dx^{0}=dt$, $dx^{1}=dx$, $dx^{2}=dy$ and $dx^{3}=dz$.
From orthornormal tetrad of the Lorentz frame, we can express a non-coordinate basis of vectors in a system of locally inertial coordinate by linear combinations $\hat{\bm e}_{\alpha} = {e_{\alpha}}^{\mu} {{\bm E}_{\mu}} $, where $\alpha$ is an index of the pseudo-orthornormal basis or null tetrad formalism due to Newman and Penrose \cite{Newman, Kramer, Griffiths, Wytler1} and $\mu$ is an index of the coordinate basis.
Thus, we introduce a pseudo-orthornormal basis or null tetrad formalism due to Newman and Penrose \cite{Newman, Kramer, Griffiths, Wytler1}, where $\{\hat{\bm e}_{\alpha} \} = \{ \bm{k}, \bm{l}, \bm{m}, \bar{\bm m} \}$ is defined by,
\begin{equation}
\label{vector_basis}
 \begin{cases}
  \hat{\bm e}_{0} = \bm{k} = {e_{0}}^{\mu} {{\bm E}_{\mu}} = k^{\mu} \dfrac{\partial}{\partial x^{\mu}},\\[8pt]
  \hat{\bm e}_{1} = \bm{l} = {e_{1}}^{\mu} {{\bm E}_{\mu}} =l^{\mu} \dfrac{\partial}{\partial x^{\mu}},\\[8pt]
  \hat{\bm e}_{2} = \bm{m} = {e_{2}}^{\mu} {{\bm E}_{\mu}} = m^{\mu} \dfrac{\partial}{\partial x^{\mu}},\\[8pt]
  \hat{\bm e}_{3} = \bar{\bm m} = {e_{3}}^{\mu} {{\bm E}_{\mu}} = \bar{m}^{\mu} \dfrac{\partial}{\partial x^{\mu}}.
 \end{cases}
\end{equation}
The coordinate changes between orthornormal tetrad of the Lorentz frame and pseudo-orthornormal Newman-Penrose frame are
\begin{equation}
 u = \frac{1}{\sqrt{2}}(t-x)\hspace*{1cm} \mbox{and} \hspace*{1cm} v = \frac{1}{\sqrt{2}}(t+x), \nonumber
\end{equation}
where the variables $u$ and $v$ are  retarded and advanced  null coordinates.
In similar way, the $yz$-plane is parameterized in terms of the complex coordinate,
\begin{equation}
 \zeta = \frac{1}{\sqrt{2}}(y+ iz)\hspace*{1cm} \mbox{and} \hspace*{1cm} \bar{\zeta} = \frac{1}{\sqrt{2}}(y-iz). \nonumber
\end{equation}
With these, in a coordinate basis \cite{Wytler1}, we can write the tetrad field $k^{\mu}, l^{\mu}, m^{\mu}$ and $\bar{m}^{\mu}$  as
\begin{equation}
\label{coordinate_basis_vector}
 (k^{\mu}) = \frac{1}{\sqrt{2}} \begin{pmatrix} 1 \cr 1 \cr 0 \cr 0             
           \end{pmatrix}, \hspace*{1cm}
 (l^{\mu}) = \frac{1}{\sqrt{2}} \begin{pmatrix} 1 \cr -1 \cr 0 \cr 0             
           \end{pmatrix}, \hspace*{1cm}
 (m^{\mu}) = \frac{1}{\sqrt{2}} \begin{pmatrix} 0 \cr 0 \cr 1 \cr -i             
           \end{pmatrix}, \hspace*{1cm}  \mbox{and} \hspace*{1cm} 
  (\bar{m}^{\mu}) = \frac{1}{\sqrt{2}} \begin{pmatrix} 0 \cr 0 \cr 1 \cr i             
           \end{pmatrix}.         
\end{equation}
The elements ${e_{\alpha}}^{\mu}$ can be put in a matrix, where  $\alpha$ is the index of matrix line and $\mu$ is the index of matrix column, thus we have that,
\begin{equation}
 \label{e_matrix}
 ({e_{\alpha}}^{\mu}) = \frac{1}{\sqrt{2}} \begin{pmatrix}
                         1 & 1 & 0 & 0 \cr
                         1 & -1 & 0 & 0 \cr
                         0 & 0 & 1 & -i \cr
                          0 & 0 & 1 & i
                        \end{pmatrix}.
\end{equation}
With this matrix we obtain relationship between pseudo-orthornormal Newman-Penrose frame and orthornormal Minkowski frame by $\dfrac{\partial}{\partial x^{\alpha}} = ({e_{\alpha}}^{\mu})\dfrac{\partial}{\partial x^{\mu}}$, resulting in
\begin{equation}
 \label{coordinate_basis_vector_2}
 \frac{\partial}{\partial v} = \frac{1}{\sqrt{2}} \left(\frac{\partial}{\partial t} + \frac{\partial}{\partial x}\right), \hspace*{0.5cm} \frac{\partial}{\partial u} = \frac{1}{\sqrt{2}} \left(\frac{\partial}{\partial t} - \frac{\partial}{\partial x}\right), \hspace*{0.5cm}\frac{\partial}{\partial \zeta} = \frac{1}{\sqrt{2}} \left(\frac{\partial}{\partial y} -i \frac{\partial}{\partial z}\right) \hspace*{0.5cm} \mbox{and} \hspace*{0.5cm} \frac{\partial}{\partial \bar{\zeta}} = \frac{1}{\sqrt{2}} \left(\frac{\partial}{\partial y} +i \frac{\partial}{\partial z}\right).
\end{equation}
With these we obtain the rule, $\dfrac{\partial v}{\partial v} = \dfrac{\partial u}{\partial u} = \dfrac{\partial\zeta}{\partial \zeta} = \dfrac{\partial \bar{\zeta}}{\partial \bar{\zeta}} = 1$.

%%%%

The dual basis, the 1-form are given by $\tilde{\bm\theta}^{\alpha} = {\omega^{\alpha}}_{\mu} dx^{\mu}$,
\begin{equation}
\label{NP_basis_dual_vector_00}
\tilde{\bm \theta}^{0} = dv = dt+dx, \hspace*{0.5cm}\tilde{\bm \theta}^{1} = du = dt - dx, \hspace*{0.5cm}\tilde{\bm \theta}^{2} = d\zeta = dy + i dz, \hspace*{0.5cm} \mbox{and} \hspace*{0.5cm}\tilde{\bm \theta}^{3} = d\bar{\zeta} = dy - idz.
\end{equation}
We can represent the matrix $({\omega^{\alpha}}_{\mu})$ as,
\begin{equation}
 \label{omega_matrix}
 ({\omega^{\alpha}}_{\mu}) = \frac{1}{\sqrt{2}} \begin{pmatrix}
                         1 & 1 & 0 & 0 \cr
                         1 & -1 & 0 & 0 \cr
                         0 & 0 & 1 & i \cr
                          0 & 0 & 1 & -i
                        \end{pmatrix},
\end{equation}
and again we have that $\alpha$ is the index of matrix line and $\mu$ is the index of matrix column. Observe that we put the transposed matrix of $({e_{\alpha}}^{\mu})$ as $({e^{\mu}}_{\alpha})$ we obtain that the matrix product  $({e^{\mu}}_{\alpha}) ({\omega^{\alpha}}_{\nu})$ results in the identity matrix $({\delta^{\mu}}_{\nu})$.

It is important to note that
\begin{equation}
\label{complex_conjugate_1-form}
 \overline{\tilde{\bm\theta}^{3}} =\tilde{\bm \theta}^{2}.
\end{equation}
When one apply the complex conjugate operation on a tensor that contains indices 2 and 3, it will result in an exchange of index 2 by 3 and index 3 by 2, for instance, $\overline{\bm \Gamma}_{02} = {\bm \Gamma}_{03}$.

The complex null tetrad dual basis is related to an orthornormal basis by
\begin{equation}
\label{dual_basis}
 \begin{cases}
 \tilde{\bm \theta}^{0} = {\omega^{0}}_{\mu} dx^{\mu} = -l_{\mu} dx^{\mu} \cr
 \tilde{\bm \theta}^{1} =  {\omega^{1}}_{\mu} dx^{\mu} = -k_{\mu} dx^{\mu} \cr
 \tilde{\bm \theta}^{2} =  {\omega^{2}}_{\mu} dx^{\mu} = \bar{m}_{\mu} dx^{\mu} \cr
 \tilde{\bm \theta}^{2} =  {\omega^{3}}_{\mu} dx^{\mu} = m_{\mu} dx^{\mu}.
 \end{cases}
\end{equation}
In the Minkowski frame we have that $ k_{\mu} = \eta_{\mu\nu} k^{\mu}$, $ l_{\mu} = \eta_{\mu\nu} l^{\mu}$, $ m_{\mu} = \eta_{\mu\nu} m^{\mu}$ and $ \bar{m}_{\mu} = \eta_{\mu\nu} \bar{m}^{\mu}$ such taht,
\begin{equation}
\label{coordinate_basis_dual_vector}
 (l_{\mu}) = \frac{1}{\sqrt{2}} \begin{pmatrix} -1 \cr -1 \cr 0 \cr 0             
           \end{pmatrix}, \hspace*{1cm}
 (k_{\mu}) = \frac{1}{\sqrt{2}} \begin{pmatrix} -1 \cr 1 \cr 0 \cr 0             
           \end{pmatrix}, \hspace*{1cm}
 (m_{\mu}) = \frac{1}{\sqrt{2}} \begin{pmatrix} 0 \cr 0 \cr 1 \cr -i             
           \end{pmatrix}, \hspace*{1cm}  \mbox{and} \hspace*{1cm} 
  (\bar{m}_{\mu}) = \frac{1}{\sqrt{2}} \begin{pmatrix} 0 \cr 0 \cr 1 \cr i             
           \end{pmatrix}.         
\end{equation}
The only nonzero contractions between (\ref{coordinate_basis_vector}) and (\ref{coordinate_basis_dual_vector}) are $k^{\mu}l_{\mu} = -1$ and $m^{\mu}\bar{m}_{\mu} = 1$.

In the pseudo-orthornormal Newman-Penrose non-coordinate basis, the tetrad field terms $k^{\alpha}$, $l^{\alpha}$, $m^{\alpha}$ and $\bar{m}^{\alpha}$ in matrix form with aid of equation (\ref{omega_matrix})  are given by $(k^{\alpha}) = ({\omega^{\alpha}}_{\mu})(k^{\mu})$, $(l^{\alpha}) = ({\omega^{\alpha}}_{\mu})(l^{\mu})$, $(m^{\alpha}) = ({\omega^{\alpha}}_{\mu})(m^{\mu})$ and $(\bar{m}^{\alpha}) = ({\omega^{\alpha}}_{\mu})(\bar{m}^{\mu})$ where we obtain the below column matrices, 
\begin{equation}
\label{NP_basis_vector}
 (k^{\alpha})= \begin{pmatrix}
              1 \cr 0 \cr 0 \cr 0
             \end{pmatrix},
\hspace*{1cm}
 (l^{\alpha})= \begin{pmatrix}
              0 \cr 1 \cr 0 \cr 0
             \end{pmatrix},
\hspace*{1cm}
 (m^{\alpha})= \begin{pmatrix}
              0 \cr 0 \cr 1 \cr 0
             \end{pmatrix}  
\hspace*{1cm} \mbox{and} \hspace*{1cm} 
 (\bar{m}^{\alpha})= \begin{pmatrix}
              0 \cr 0 \cr 0 \cr 1
             \end{pmatrix}.
\end{equation}
In this approach we use $\dfrac{\partial}{\partial x^{\alpha}} = \left(\dfrac{\partial}{\partial v},\dfrac{\partial}{\partial u}, \dfrac{\partial}{\partial \zeta},\dfrac{\partial}{\partial \bar{\zeta}}    \right)$.

The covariant terms  $k_{\alpha}$, $l_{\alpha}$, $m_{\alpha}$ and $\bar{m}_{\alpha}$ in the pseudo-orthornormal Newman-Penrose non-coordinate basis are obtained with aid of equation (\ref{e_matrix}) where  in matrix form are given by $(k_{\alpha}) = ({e_{\alpha}}^{\mu})(k_{\mu})$, $(l^{\alpha}) = ({e_{\alpha}}^{\mu})(l_{\mu})$, $(m^{\alpha}) = ({e_{\alpha}}^{\mu})(m_{\mu})$ and $(\bar{m}^{\alpha}) = ({e_{\alpha}}^{\mu})(\bar{m}_{\mu})$ ,
\begin{equation}
\label{NP_basis_dual_vector}
 (k_{\alpha})= \begin{pmatrix}
              0 \cr -1 \cr 0 \cr 0
             \end{pmatrix},
\hspace*{1cm}
 (l_{\alpha})= \begin{pmatrix}
              -1 \cr 0 \cr 0 \cr 0
             \end{pmatrix},
\hspace*{1cm}
 (m_{\alpha})= \begin{pmatrix}
              0 \cr 0 \cr 0 \cr 1
             \end{pmatrix}  
\hspace*{1cm} \mbox{and} \hspace*{1cm} 
 (\bar{m}_{\alpha})= \begin{pmatrix}
              0 \cr 0 \cr 1 \cr 0
             \end{pmatrix},
\end{equation}
where in this approach we use $dx^{\alpha} = (dv, du, d\zeta, d\bar{\zeta})$.
The orthogonality identity below is valid for the Newman-Penrose coordinate,
\begin{equation}
 \langle \hat{\bm e}_{\alpha} ,\tilde{\bm \theta}^{\beta} \rangle ={\delta_{\alpha}}^{\beta},
\end{equation}
with condition ${e_{\alpha}}^{\mu} {\omega^{\beta}}_{\mu} ={\delta_{\alpha}}^{\beta}$.
The only nonzero contractions between (\ref{NP_basis_vector}) and (\ref{NP_basis_dual_vector}) are
\begin{equation}
\label{contraction_tetrad}
 k^{\alpha}l_{\alpha} = l^{\alpha}k_{\alpha} = -1 \hspace*{1cm}\mbox{and}
\hspace*{1cm} m^{\alpha}\bar{m}_{\alpha} = \bar{m}^{\alpha}m_{\alpha} = 1.
\end{equation}

The metric tensor is defined by,
\begin{equation}
\label{metric_N-P_01}
 {\bm \gamma} = -2\tilde{\bm \theta}^{0} \otimes\tilde{\bm \theta}^{1} + 2\tilde{\bm \theta}^{2} \otimes\tilde{\bm \theta}^{3}.
\end{equation}
The components of the metric tensor in Newman-Penrose basis are displayed,
\begin{equation}
    \label{components_g_0}
 (\gamma_{\alpha\beta}) = \begin{pmatrix}
                         0 & -1 & 0 & 0 \cr
                         -1 & 0 & 0 & 0 \cr
                         0 & 0 & 0 & 1 \cr
                         0 & 0 & 1 & 0
                        \end{pmatrix}
\hspace*{0.5cm}
\mbox{and the inverse matrix}
\hspace*{0.5cm}
(\gamma^{\alpha\beta}) = \begin{pmatrix}
                         0 & -1 & 0 & 0 \cr
                         -1 & 0 & 0 & 0 \cr
                         0 & 0 & 0 & 1 \cr
                         0 & 0 & 1 & 0
                        \end{pmatrix},
\end{equation}
satisfying $\gamma_{\alpha\beta}\gamma^{\beta\epsilon} = {\delta_{\alpha}}^{\epsilon}$.

The relationship between a metric in a coordinate basis and the rigid frame of pseudo-orthornormal basis of Newman-Penrose formalism are given with aid of tetrad field matrices ${e_{\alpha}}^{\mu}$ and ${\omega^{\alpha}}_{\mu}$ satisfying
\begin{equation}
 {e_{\alpha}}^{\mu}{e_{\beta}}^{\nu} g_{\mu\nu} = \gamma_{\alpha\beta} \hspace*{1cm} \mbox{and} \hspace*{1cm} {\omega^{\alpha}}_{\mu}{\omega^{\beta}}_{\nu}\gamma_{\alpha\beta} = g_{\mu\nu}.
\end{equation}
The components metric tensor in the pseudo-orthornormal Newman-Penrose non-coordinate basis are,
\begin{equation}
 \label{components_g_1}
 \gamma_{\alpha\beta} = -\left(l_{\alpha}k_{\beta} + l_{\beta}k_{\alpha} \right)+
 \left(m_{\alpha}\bar{m}_{\beta} + m_{\beta}\bar{m}_{\alpha} \right).
\end{equation}

%%%%%%%%%%%%%%%%%%%%%%%%%%%%%%%%%%%%%%%%%%%%%%%%%%%%%%%%
%%%%%%%%%%%%%%%%%%%%%%%%%%%%%%%%%%%%%%%%%%%%%%%%%%%%%%%%
%%%%%%%%%%%%%%%%%%%%%%%%%%%%%%%%%%%%%%%%%%%%%%%%%%%%%%%%
%%%%%%%%%%%%%%%%%%%%%%%%%%%%%%%%%%%%%%%%%%%%%%%%%%%%%%%%
%%%%%%%%%%%%%%%%%%%%%%%%%%%%%%%%%%%%%%%%%%%%%%%%%%%%%%%%

In order to obtain the dual bivector in Newman-Penrose formalism, it is necessary to define the Levi-Civita 4-form. We choose an oriented basis in an orthornormal Minkowski tetrad $\{dx^{\mu}\}$ where the Levi-Civita 4-form is a volume element measure in Minkowski spacetime defined by \cite{Kramer, Hawking, De Felice},
\begin{equation}
 \label{Levi_Civita_4-form_Minkowski}
 {\bm \epsilon} = 4!~dx^{0} \wedge dx^{1} \wedge dx^{2} \wedge dx^{3}.  
\end{equation}
This 4-form has components,
\begin{equation}
 \epsilon_{\mu\nu\rho\sigma} = 4!~{\delta^{0}}_{[\mu}{\delta^{1}}_{\nu}{\delta^{2}}_{\rho}{\delta^{3}}_{\sigma]} =  \left({\delta^{0}}_{\mu}{\delta^{1}}_{\nu}{\delta^{2}}_{\rho}{\delta^{3}}_{\sigma} - {\delta^{0}}_{\mu}{\delta^{1}}_{\nu}{\delta^{3}}_{\rho}{\delta^{2}}_{\sigma} + \cdots + {\delta^{4}}_{\mu}{\delta^{3}}_{\nu}{\delta^{2}}_{\rho}{\delta^{1}}_{\sigma} - {\delta^{4}}_{\mu}{\delta^{3}}_{\nu}{\delta^{1}}_{\rho}{\delta^{2}}_{\sigma}\right), \nonumber
\end{equation}
and we have that,
\begin{equation}
 \epsilon_{0123} = +1. \nonumber
\end{equation}
We can relate the existence of the metric to define in a coordinate basis the 4-form of a volume element,
\begin{equation}
 \label{Levi_Civita_4-form_NP_01}
 {\bm \eta} =  \sqrt{-g}~ {\bm\epsilon},  
\end{equation}
where $\sqrt{-g} = \sqrt{-\det(g_{\mu\nu})}$. Thus, the the components of the 4-form $\bm{\eta}$ can be write as,
\begin{equation}
 \eta_{\mu\nu\rho\sigma} = \sqrt{-g}~\epsilon_{\mu\nu\rho\sigma}. \nonumber
\end{equation}
Transforming this in Newman-Penrose coordinates by use of tetrads ${e_{\alpha}}^{\mu}$ we have,
\begin{equation}
 \eta_{\alpha\beta\gamma\delta} = \eta_{\mu\nu\rho\sigma}{e_{\alpha}}^{\mu}{e_{\beta}}^{\nu}{e_{\gamma}}^{\rho}{e_{\delta}}^{\sigma} = \sqrt{-g}~\epsilon_{\mu\nu\rho\sigma}{e_{\alpha}}^{\mu}{e_{\beta}}^{\nu}{e_{\gamma}}^{\rho}{e_{\delta}}^{\sigma} \nonumber.
\end{equation}
And then we can compute $\eta_{0123}$ such as,
\begin{equation}
 \eta_{0123} = \sqrt{-g}~\epsilon_{\mu\nu\rho\sigma}{e_{0}}^{\mu}{e_{1}}^{\nu}{e_{2}}^{\rho}{e_{3}}^{\sigma} = \sqrt{-g}~\epsilon_{\mu\nu\rho\sigma}~k^{\mu}l^{\nu}m^{\rho}\bar{m}^{\sigma} \nonumber,
\end{equation}
where we use the fact that ${e_{0}}^{\mu} = k^{\mu}$, ${e_{1}}^{\mu} = l^{\mu}$, ${e_{2}}^{\mu} = m^{\mu}$  and ${e_{3}}^{\mu} = \bar{m}^{\mu}$ and we have that,
\begin{equation}
 \epsilon_{\mu\nu\rho\sigma}~k^{\mu}l^{\nu}m^{\rho}\bar{m}^{\sigma} = \epsilon_{0123}k^{0}l^{1}m^{2}\bar{m}^{3} + \cdots + \epsilon_{1032}k^{1}l^{0}m^{3}\bar{m}^{2} = -i. \nonumber
\end{equation}
We can relate the pseudo-orthornormal Newman-Penrose basis with orthornormal Minkowski  basis with $g_{\mu\nu} = \mbox{diag}(-1,1,1,1)$ where $\sqrt{-g} = 1$ and we have that $\eta_{0123} = -i$. Thus, the components of Levi-Civita 4-form are given by
\begin{equation}
 \label{Levi_Civita_4-form_NP_02}
 \eta_{\alpha\beta\gamma\delta} = \begin{cases}
        -i,   \hspace*{1cm} \mbox{for even permutations of 0, 1, 2, 3} \cr
        +i, \hspace*{1cm} \mbox{for odd permutations of 0, 1, 2, 3.}
                                  \end{cases}
 \end{equation}

%%%%%%%%%%%%%%%%%%%%%%%%%%%%%%%%%%%%%%%%%%%%%%%%%%%%%%%%
%%%%%%%%%%%%%%%%%%%%%%%%%%%%%%%%%%%%%%%%%%%%%%%%%%%%%%%%
%%%%%%%%%%%%%%%%%%%%%%%%%%%%%%%%%%%%%%%%%%%%%%%%%%%%%%%%
%%%%%%%%%%%%%%%%%%%%%%%%%%%%%%%%%%%%%%%%%%%%%%%%%%%%%%%%
%%%%%%%%%%%%%%%%%%%%%%%%%%%%%%%%%%%%%%%%%%%%%%%%%%%%%%%%

Newman and Penrose classify 12 independent complex linear combination of the Ricci rotation coefficients and called {\it spin coefficients}
as follows \cite{Wytler1},
\begin{equation}
 \label{spin_coefficient_01}
 -\kappa = \Gamma_{200} = \bar{\Gamma}_{300} = m^{\mu} k^{\nu}\nabla_{\nu} k_{\mu},
\end{equation}
\begin{equation}
     \label{spin_coefficient_02}
 -\rho =  \Gamma_{320} =  \bar{\Gamma}_{230} = m^{\mu} \bar{m}^{\nu}\nabla_{\nu} k_{\mu},
\end{equation}
\begin{equation}
    \label{spin_coefficient_03}
 -\sigma = \Gamma_{220} =  \bar{\Gamma}_{330} =  m^{\mu} m^{\nu}\nabla_{\nu} k_{\mu},  
\end{equation}
\begin{equation}
    \label{spin_coefficient_04}
 -\tau = \Gamma_{210} = \bar{\Gamma}_{310} =  m^{\mu} l^{\nu}\nabla_{\nu} k_{\mu},  
\end{equation}
\begin{equation}
    \label{spin_coefficient_05}
 \nu = \Gamma_{311} = \bar{\Gamma}_{211}= \bar{m}^{\mu} l^{\nu}\nabla_{\nu} l_{\mu},
\end{equation}
\begin{equation}
    \label{spin_coefficient_06}
 \mu = \Gamma_{321} = \bar{\Gamma}_{231} =\bar{m}^{\mu} m^{\nu}\nabla_{\nu} l_{\mu}, 
\end{equation}
\begin{equation}
    \label{spin_coefficient_07}
 \lambda =  \Gamma_{331} = \bar{\Gamma}_{211} = \bar{m}^{\mu} \bar{m}^{\nu}\nabla_{\nu} l_{\mu}, 
\end{equation}
\begin{equation}
    \label{spin_coefficient_08}
 \pi = \Gamma_{301} = \bar{\Gamma}_{201} = \bar{m}^{\mu} k^{\nu}\nabla_{\nu} l_{\mu}, 
\end{equation}
\begin{equation}
    \label{spin_coefficient_09}
 -\epsilon =\frac{1}{2}\left(\Gamma_{100} - \Gamma_{302} \right) = 
	    \frac{1}{2}\left ( l^{\mu} k^{\nu}\nabla_{\nu} k_{\mu} - \bar{m}^{\mu} k^{\nu}\nabla_{\nu} m_{\mu} \right), 
\end{equation}
\begin{equation}
	    \label{spin_coefficient_10}
  -\beta =\frac{1}{2}\left(\Gamma_{120} - \Gamma_{322} \right) = 
	    \frac{1}{2}\left ( l^{\mu} m^{\nu}\nabla_{\nu} k_{\mu} - \bar{m}^{\mu} m^{\nu}\nabla_{\nu} m_{\mu} \right), 
\end{equation}
\begin{equation}
	    \label{spin_coefficient_11}
  \gamma =\frac{1}{2}\left(\Gamma_{011} - \Gamma_{213} \right) = 
	    \frac{1}{2}\left ( k^{\mu} l^{\nu}\nabla_{\nu} l_{\mu} - m^{\mu} l^{\nu}\nabla_{\nu} \bar{m}_{\mu} \right), 
\end{equation}
\begin{equation}
	    \label{spin_coefficient_12}
  \alpha = \frac{1}{2}\left(\Gamma_{031} - \Gamma_{233} \right) = 
	    \frac{1}{2}\left ( k^{\mu} \bar{m}^{\nu}\nabla_{\nu} l_{\mu} - m^{\mu} \bar{m}^{\nu}\nabla_{\nu} \bar{m}_{\mu} \right).	    
\end{equation}

In the Newman-Penrose formalism it is efficient to use Cartan's method for the calculation of curvature. First, we must calculate the connection 1-forms by use of first Cartan's structure equation,
\begin{equation}
\label{1a_equacao_de_Cartan}
 \begin{cases}
  d\tilde{\bm\theta}^0 =  {\bm \Gamma}_{01}\wedge \tilde{\bm\theta}^0 + {\bm \Gamma}_{12}\wedge\tilde{\bm\theta}^2
  + \overline{\bm \Gamma}_{12} \wedge\tilde{\bm\theta}^3, \cr
d\tilde{\bm\theta}^1  = {\bm \Gamma}_{01}\wedge \tilde{\bm\theta}^1 + {\bm \Gamma}_{02}\wedge \tilde{\bm\theta}^2
+ \overline{\bm \Gamma}_{02}\wedge \tilde{\bm\theta}^3, \cr
d\tilde{\bm\theta}^2  = \overline{\bm \Gamma}_{02}\wedge \tilde{\bm\theta}^0 + \overline{\bm \Gamma}_{12}\wedge \tilde{\bm\theta}^1
+ {\bm \Gamma}_{23}\wedge \tilde{\bm\theta}^2.
 \end{cases}
\end{equation}
Then we must use the second Cartan's structure equation,
\begin{equation}
\label{2a_equacao_de_Cartan}
 \begin{cases}
  \bm \Theta_{03} = d\bm\Gamma_{03} + {\bm \Gamma}_{03} \wedge ( \bm\Gamma_{01}+ \bm \Gamma_{23}),\cr
   \bm \Theta_{12} = d\bm\Gamma_{12} - \bm\Gamma_{12} \wedge (\bm\Gamma_{01} +  \bm \Gamma_{23} ),\cr
   \bm \Theta_{01} +  \bm \Theta_{23} = d( \bm\Gamma_{01}+ \bm\Gamma_{23}) - 2\bm \Gamma_{03}\wedge \bm\Gamma_{12},
 \end{cases}
\end{equation}
that yelds the curvature 2-forms $\bm{\Theta}_{\alpha\beta} = \dfrac{1}{2} R_{\alpha\beta\gamma\delta}~\tilde{\bm\theta}^{\gamma} \wedge \tilde{\bm\theta}^{\delta}$.
The above system of curvature 2-forms has proved very useful in Newman-Penrose formalism, where exact solutions of Einstein equations are obtained \cite{Kramer, Griffiths}.

\section{Electromagnetism in Newman-Penrose formalism}

The electromagnetic field is described by a vector potential $A_{\mu} = (\phi,A_x,A_y,A_z)$ where the field-strenght tensor of electromagnetic field is given by $F_{\mu\nu} = \partial_{\mu} A_{\nu} - \partial_{\nu} A_{\mu}$, where it is antisymmetric tensor, $F_{\mu\nu} = -F_{\nu\mu}$. In a Minkowski frame the electromagnetic field-strenght tensor can be displayed in matrix form as \cite{Wytler2},
\begin{equation}
\label{electromagnetic_field-strenght}
 (F_{\mu\nu}) = \begin{pmatrix}
                0 & -E_x & - E_y & -E_z \cr
                E_x & 0 & B_z & - B_y \cr
                E_y & -B_z & 0 & B_x \cr
                E_z & B_y & -B_x & 0
               \end{pmatrix}.
\end{equation}
It is stand out that in the field-strenght tensor of electromagnetic we have the components of 3-vector electric $\bm{E} = (E_x, E_y, E_z)$ and 3-vector magnectic $\bm{B} = (B_x, B_y, B_z)$.

\subsection{Bivectors and duality}

In accordance with definition (\ref{bivector}), a bivector is a antisymmetric tensor, and the dual bivector, also antisymmetric tensor, is a pseudotensor . In an orthornormal Minkowski basis the components of a pseudotensor is defined by,
\begin{equation}
 \widetilde{F}_{\mu\nu} = \frac{1}{2} \epsilon_{\mu\nu\rho\sigma}F^{\rho\sigma}.
\end{equation}
With $\epsilon_{0123} = 1$ we have the dual electromagnetic bivector as 
\begin{equation}
 \widetilde{F}_{\mu\nu} =
 \begin{pmatrix}
                0 & B_x &  B_y & B_z \cr
                -B_x & 0 & E_z & - E_y \cr
                -B_y & -E_z & 0 & E_x \cr
                -B_z & E_y & -E_x & 0
               \end{pmatrix}.
\end{equation}
A complex bivector is defined by,
\begin{equation}
 {\cal F}_{\mu\nu} = F_{\mu\nu} + i \widetilde{F}_{\mu\nu}, 
\end{equation}
such that it is self-dual \cite{Wytler2},
\begin{equation}
\label{sel-dual_property_1}
 \widetilde{\cal F}_{\mu\nu} = - i {\cal F}_{\mu\nu}.
\end{equation}
In an inertial frame in which the observer is at rest, the four-velocity is $u^{\mu} = (1,0,0,0)$. If we project the electromagnetic complex bivector ${\cal F}_{\mu\nu} $ in the direction of the timelike unit vector $u^{\mu}$, we have,
\begin{equation}
 {\cal F}_{\mu\nu}u^{\nu} = F_{\mu} =  \begin{pmatrix}
    0 \cr F_x  \cr F_y \cr F_z
                            \end{pmatrix}
 =\begin{pmatrix}
    0 \cr E_x - i B_x  \cr E_y - i B_y \cr E_z - i B_z
                            \end{pmatrix} .
\end{equation}

It is very useful to define,
\begin{equation}
\label{identidade_I}
 I_{\mu\nu\rho\sigma} = \frac{1}{4} \left(g_{\mu\nu\rho\sigma}+ i \epsilon_{\mu\nu\rho\sigma} \right)
\end{equation}
where $g_{\mu\nu\rho\sigma} = g_{\mu\rho} g_{\nu\sigma} - g_{\mu\sigma} g_{\nu\rho} $. With this tensor we obtain
\begin{equation}
\label{identidade_I_F_1}
 I_{\mu\nu\rho\sigma}F^{\rho\sigma} =
 \frac{1}{4} \left(g_{\mu\nu\rho\sigma}+ i \epsilon_{\mu\nu\rho\sigma} \right)F^{\rho\sigma} = 
 \frac{1}{4} \left(2F_{\mu\nu} + i2\widetilde{F}_{\mu\nu} \right)=
 \frac{1}{2} {\cal F}_{\mu\nu}.
\end{equation}
 We obtain too,
\begin{equation}
 I_{\mu\nu\rho\sigma}{\cal F}^{\rho\sigma} =
 \frac{1}{4} \left(g_{\mu\nu\rho\sigma}+ i \epsilon_{\mu\nu\rho\sigma} \right){\cal F}^{\rho\sigma} = 
 \frac{1}{4} \left(2{\cal F}_{\mu\nu} + i2\widetilde{\cal F}_{\mu\nu} \right), \nonumber
\end{equation} 
with the self-dual property from equation(\ref{sel-dual_property_1}), we have that
\begin{equation}
\label{identidade_I_F_2}
 I_{\mu\nu\rho\sigma}{\cal F}^{\rho\sigma} = {\cal F}_{\mu\nu}.
\end{equation}

It is useful to calculate the contraction,
\begin{eqnarray}
 I_{\mu\nu\rho\sigma}I^{\mu\nu\rho\sigma} &=& \frac{1}{4} \left(g_{\mu\rho} g_{\nu\sigma} - g_{\mu\sigma} g_{\nu\rho} + i \epsilon_{\mu\nu\rho\sigma} \right)\cdot \frac{1}{4} \left(g^{\mu\rho} g^{\nu\sigma} - g^{\mu\sigma} g^{\nu\rho} + i \epsilon^{\mu\nu\rho\sigma} \right)\cr
 &=& \frac{1}{16} \bigg( g_{\mu\rho} g_{\nu\sigma}g^{\mu\rho} g^{\nu\sigma} - g_{\mu\rho} g_{\nu\sigma}g^{\mu\sigma} g^{\nu\rho} - g_{\mu\sigma} g_{\nu\rho}g^{\mu\rho} g^{\nu\sigma} + g_{\mu\sigma} g_{\nu\rho} g^{\mu\sigma} g^{\nu\rho} \cr
 & &+i g_{\mu\rho} g_{\nu\sigma}\epsilon^{\mu\nu\rho\sigma} -i g_{\mu\sigma} g_{\nu\rho}\epsilon^{\mu\nu\rho\sigma} + i g^{\mu\rho} g^{\nu\sigma} \epsilon_{\mu\nu\rho\sigma}-i g^{\mu\sigma} g^{\nu\rho}\epsilon_{\mu\nu\rho\sigma} - \epsilon_{\mu\nu\rho\sigma}\epsilon^{\mu\nu\rho\sigma} \bigg), \nonumber
\end{eqnarray} 
for every contraction $g_{\mu\rho} g_{\nu\sigma}\epsilon^{\mu\nu\rho\sigma} =0$ and for $\epsilon_{\mu\nu\rho\sigma}\epsilon^{\mu\nu\rho\sigma}=-24 $, it results in
\begin{equation}
\label{contracao II}
 I_{\mu\nu\rho\sigma}I^{\mu\nu\rho\sigma} = 3.
\end{equation}

In order to convert the field-strenght tensor of electromagnetic from Minkowski frame to the Newman-Penrose formalism we could take the definition $F_{\alpha\beta} = \partial_{\alpha} A_{\beta} - \partial_{\beta} A_{\alpha}$ and apply the non-coordinate basis, where $\partial_{\alpha} = {e_{\alpha}}^{\mu}\partial_{\mu}$ with components obtained in the equation (\ref{coordinate_basis_vector_2}) and $A_{\alpha} =  {e_{\alpha}}^{\mu}A_{\mu} = \frac{1}{\sqrt{2}}(\phi + A_x, \phi - A_x, A_y - iA_z, A_y + iA_z)$. For example,
\begin{equation}
 F_{02} = \left( \partial_{v} A_{\zeta} - \partial_{\zeta} A_{v} \right) =  \frac{1}{2}\left[( \partial_{t} + \partial_x)(A_y - i A_z) - (\partial_y - i\partial_z)(\phi + A_x) \right] = \frac{1}{2} \left[-(E_y-iB_y) + i(E_z-iB_z) \right]. \nonumber
\end{equation}
But instead of doing it this way, we can obtain the components of 
field-strenght tensor of electromagnetic by $F_{\alpha\beta} = F_{\mu\nu}{e_{\alpha}}^{\mu}{e_{\beta}}^{\nu}$ by multiplying the corresponding matrices (\ref{electromagnetic_field-strenght}) and (\ref{e_matrix}) such that $(F_{\alpha\beta}) = ({e_{\alpha}}^{\mu})(F_{\mu\nu})({e^{\nu}}_{\beta})$, where the matrix $({e^{\nu}}_{\beta})$ is a transposed matrix of $({e_{\beta}}^{\nu})$. Thus we have that,
\begin{equation}
 (F_{\alpha\beta}) = \frac{1}{2} \begin{pmatrix}
                         1 & 1 & 0 & 0 \cr
                         1 & -1 & 0 & 0 \cr
                         0 & 0 & 1 & -i \cr
                          0 & 0 & 1 & i
                        \end{pmatrix}\cdot \begin{pmatrix}
                0 & -E_x & - E_y & -E_z \cr
                E_x & 0 & B_z & - B_y \cr
                E_y & -B_z & 0 & B_x \cr
                E_z & B_y & -B_x & 0
               \end{pmatrix}\cdot\begin{pmatrix}
                         1 & 1 & 0 & 0 \cr
                         1 & -1 & 0 & 0 \cr
                         0 & 0 & 1 & 1 \cr
                          0 & 0 & -i & i
                        \end{pmatrix}, \nonumber
\end{equation}
then the field-strenght tensor of electromagnetic field in Newman-Penrose formalism results in
\begin{equation}
 \label{electromagnetic_field-strenght_NP}
 (F_{\alpha\beta}) = \frac{1}{2} \begin{pmatrix}
        0 & 2E_x & -F_y + iF_z & -\bar{F}_y - i \bar{F}_z \cr
        -2E_x & 0 & -\bar{F}_y + i \bar{F}_z & -F_y - iF_z \cr
        F_y - iF_z & \bar{F}_y - i \bar{F}_z & 0 & 2i B_x \cr
        \bar{F}_y + i \bar{F}_z & F_y + iF_z & -2iB_x & 0 
                                 \end{pmatrix},
\end{equation}
where $F_y = E_y -i B_y$ and $ F_z = E_z - i B_z$. We must note that $F_{03} = \bar{F}_{02}$ and $F_{13} = \bar{F}_{12}$. Based on this result, we can advance in discussions about complex self-dual electromagnetic field.

\subsection{The Lorentz invariants, complex self-dual electromagnetic field and definitions of bivectors in Newman-Penrose non-coordinate basis} 

The laws of Physics in the Minkowski spacetime are invariant under Lorents transformations, that are linear and performing transformations between the coordinates of a local inertial frame ${\cal O}$ and other local inertial frame ${\cal O}'$ expressed for electromagnetic field tensor by $F'^{\alpha\beta} = {\Lambda^{\alpha}}_{\gamma}{\Lambda^{\beta}}_{\delta}F^{\gamma\delta}$, where it results in two Lorentz invariants. Several details about this topic are covered in this reference \cite{Wytler2}. 
The first Lorentz invariant is given by,
\begin{equation}
\label{Invariant_1}
 I_1=\frac{1}{2} F_{\alpha\beta}F^{\alpha\beta}
\end{equation}
where we can start with the covariant tensor of eletromagnetic field
\begin{equation}
\label{electromagnetic_tensor}
 (F_{\alpha\beta}) = \begin{pmatrix}
                    0 & F_{01} & F_{02} & F_{03} \cr
                    -F_{01} & 0 & F_{12} & F_{13} \cr
                    -F_{02} & -F_{12} & 0 & F_{23} \cr
                    -F_{03} & -F_{13} & -F_{23} & 0
                   \end{pmatrix}.
\end{equation}
The first Lorentz invariant (\ref{Invariant_1}) in Newman-Penrose coordinate basis is obtained when we calculate the contravariant tensor of eletromagnetic field in Newman-Penrose system to contract with covariante (\ref{electromagnetic_tensor}). Therefore, the contravariant tensor of eletromagnetic field in Newman-Penrose system is given with aid of metric tensor in  Newman-Penrose coordinate basis (\ref{components_g_0}),
\begin{equation}
 F^{\alpha\beta} = \gamma^{\alpha\delta}F_{\delta\epsilon}\gamma^{\epsilon\beta}, \nonumber
\end{equation}
such as
\begin{equation}
 (F^{\alpha\beta}) = \begin{pmatrix}
                      0 & -1 & 0 & 0 \cr
                      -1 & 0 & 0 & 0 \cr
                      0 & 0 & 0 & 1\cr
                      0 & 0 & 1 & 0
                     \end{pmatrix}
 \begin{pmatrix}
                    0 & F_{01} & F_{02} & F_{03} \cr
                    -F_{01} & 0 & F_{12} & F_{13} \cr
                    -F_{02} & -F_{12} & 0 & F_{23} \cr
                    -F_{03} & -F_{13} & -F_{23} & 0
                   \end{pmatrix}
                   \begin{pmatrix}
                      0 & -1 & 0 & 0 \cr
                      -1 & 0 & 0 & 0 \cr
                      0 & 0 & 0 & 1\cr
                      0 & 0 & 1 & 0
                     \end{pmatrix} \nonumber
\end{equation}
thus, the contravariant tensor of eletromagnetic field is
\begin{equation}
 (F^{\alpha\beta}) =
\begin{pmatrix}
                    0 & -F_{01} & -F_{13} & -F_{12} \cr
                    F_{01} & 0 & -F_{03} & -F_{02} \cr
                    F_{13} & F_{03} & 0 & -F_{23} \cr
                    F_{12} & -F_{02} & F_{23} & 0
                   \end{pmatrix}.
\end{equation}
We have that the first Lorentz invariant (\ref{Invariant_1}) is therefore
\begin{equation}
 \label{Invariant_1A}
 I_1 = -(F_{01})^2 - (F_{23})^2 - 2F_{02}F_{13} - 2F_{03}F_{12}.
\end{equation}

The second  Lorentz invariant is given by
\begin{equation}
\label{Invariant_2}
 I_2=\frac{1}{2} \widetilde{F}_{\alpha\beta}F^{\alpha\beta}.
\end{equation}
Now the dual tensor $\widetilde{F}_{\alpha\beta}$ in Newman-Penrose non-coordinate basis is given with aid of Levi-Civita tensor (\ref{Levi_Civita_4-form_NP_02}) by,
\begin{equation}
 \widetilde{F}_{\alpha\beta} = \frac{1}{2}\eta_{\alpha\beta\gamma\delta}F^{\gamma\delta}.
\end{equation}
Thus, we have the values,
$$ \widetilde{F}_{01} =  \eta_{0123}F^{23} = iF_{23}$$
$$ \widetilde{F}_{02} =  \eta_{0213}F^{13} = -iF_{02}$$
$$ \widetilde{F}_{03} =  \eta_{0312}F^{12} = iF_{03}$$
$$ \widetilde{F}_{12} =  \eta_{1203}F^{03} = -iF_{12}$$
$$ \widetilde{F}_{13} =  \eta_{1302}F^{02} = -iF_{13}$$
$$ \widetilde{F}_{23} =  \eta_{2301}F^{01} = iF_{01}$$
Thus the dual of electromagnetic tensor is,
\begin{equation}
\label{dual_electromagnetic_NP}
 (\widetilde{F}_{\alpha\beta}) = i \begin{pmatrix}
                    0 & F_{23} & -F_{02} & F_{03} \cr
                    -F_{23} & 0 & F_{12} & -F_{13} \cr
                    F_{02} & -F_{12} & 0 & F_{01} \cr
                    -F_{03} & F_{13} & -F_{01} & 0
                   \end{pmatrix}.
\end{equation}
With this result we can write the second Lorentz invariant (\ref{Invariant_2}) sucha as
\begin{equation}
\label{Invariant_2A}
 I_2=-2i(F_{01}F_{23} - F_{02}F_{13}+F_{03}F_{12}).
\end{equation}
The complete Lorentz invariant given by,
\begin{equation}
 {\cal F}_{\alpha\beta}{\cal F}^{\alpha\beta} = 2F_{\alpha\beta}F^{\alpha\beta} + 2i \widetilde{F}_{\alpha\beta}F^{\alpha\beta} = 4I_1 + 4iI_2,
\end{equation}
results in
\begin{equation}
\label{Invariant_3}
 {\cal F}_{\alpha\beta}{\cal F}^{\alpha\beta} = -4[(F_{01} - F_{23})^2 + 4 F_{02}F_{13}].
\end{equation}

With aid of the equations (\ref{electromagnetic_tensor}) and (\ref{dual_electromagnetic_NP}), the matrix of tensor ${\cal F}_{\alpha\beta} = F_{\alpha\beta} + i \widetilde{F}_{\alpha\beta}$ is displayed as
\begin{equation}
\label{matrix_complex_electomagnetic_tensor}
 ({\cal F}_{\alpha\beta}) =  \begin{pmatrix}
                    0 & F_{01}-F_{23} & 2F_{02} & 0 \cr
                    -(F_{01} - F_{23}) & 0 & 0 & 2F_{13} \cr
                    -2F_{02} & 0 & 0 & -(F_{01}-F_{23}) \cr
                    0 & -2F_{13} & F_{01}-F_{23} & 0
                   \end{pmatrix}.
\end{equation}
The covariant tensor is calculated and displayed as
\begin{equation}
 ({\cal F}^{\alpha\beta}) =  \begin{pmatrix}
                    0 & -(F_{01}-F_{23}) & -2F_{13} & 0 \cr
                    F_{01} - F_{23} & 0 & 0 & -2F_{02} \cr
                    2F_{13} & 0 & 0 & F_{01}-F_{23} \cr
                    0 & 2F_{02} & -(F_{01}-F_{23}) & 0
                   \end{pmatrix}.
\end{equation}
We can calculate the dual tensor $\widetilde{\cal F}_{\alpha\beta}$ and displayed as,
\begin{equation}
 (\widetilde{\cal F}_{\alpha\beta}) = -i \begin{pmatrix}
                    0 & F_{01}-F_{23} & 2F_{02} & 0 \cr
                    -(F_{01} - F_{23}) & 0 & 0 & 2F_{13} \cr
                    -2F_{02} & 0 & 0 & -(F_{01}-F_{23}) \cr
                    0 & -2F_{13} & F_{01}-F_{23} & 0
                   \end{pmatrix},
\end{equation}
and we conclude that $\widetilde{\cal F}_{\alpha\beta} = -i {\cal F}_{\alpha\beta}$, i.e. the complex electromagnetic is self-dual.

%
%%%%%%%%%%%%%%%%%%%%%%%%%%%%%%%%%%%%%%%%%%%%%%%%%%
%

The electromagnetic bivector is defined as a 2-form, 
\begin{equation}
\label{electromagnetic_2-form}
 {\bm F} = \frac{1}{4} {\cal F}_{\alpha\beta}\, \tilde{\bm\theta}^{\alpha} \wedge \tilde{\bm\theta}^{\beta},
\end{equation}
we can calculate in Newman-Penrose coordinate system, where we indentify from matrix (\ref{matrix_complex_electomagnetic_tensor}), 
$${\cal F}_{01} = F_{01}-F_{23},$$ 
$${\cal F}_{02} = 2F_{02},$$ 
$${\cal F}_{13} = 2F_{13}$$
and 
$${\cal F}_{23} = -(F_{01}-F_{23})$$
such they produce
\begin{equation}
 {\bm F} = \frac{1}{2} \left[(F_{01} -F_{23})\, \tilde{\bm\theta}^{0} \wedge \tilde{\bm\theta}^{1} + 2F_{02}\, \tilde{\bm\theta}^{0} \wedge \tilde{\bm\theta}^{2} + 2F_{13}\, \tilde{\bm\theta}^{1} \wedge \tilde{\bm\theta}^{3} - (F_{01} -F_{23})\, \tilde{\bm\theta}^{2} \wedge \tilde{\bm\theta}^{3} \right],\nonumber
\end{equation}
with $\tilde{\bm\theta}^{1} \wedge \tilde{\bm\theta}^{3} = - \tilde{\bm\theta}^{3} \wedge \tilde{\bm\theta}^{1} $ we can write it as
\begin{equation}
\label{bivector_NP_coordinate}
 {\bm F} =  F_{02}\, \tilde{\bm\theta}^{0} \wedge \tilde{\bm\theta}^{2} - F_{13}\, \tilde{\bm\theta}^{3} \wedge \tilde{\bm\theta}^{1} +
\frac{1}{2} (F_{01} -F_{23})(\tilde{\bm\theta}^{0} \wedge \tilde{\bm\theta}^{1} -\tilde{\bm\theta}^{2} \wedge \tilde{\bm\theta}^{3})
\end{equation}
A basis for a bivectorial space associated with the coordinates of Newman-Penrose system, can be built from above equation,
\begin{equation}
\begin{cases}
 {\bm Z}^{1} = \tilde{\bm\theta}^{0} \wedge \tilde{\bm\theta}^{2}, \cr
 {\bm Z}^{2} = \tilde{\bm\theta}^{3} \wedge \tilde{\bm\theta}^{1}, \cr
 {\bm Z}^{3} =\tilde{\bm\theta}^{0} \wedge \tilde{\bm\theta}^{1} -\tilde{\bm\theta}^{2} \wedge \tilde{\bm\theta}^{3},
 \end{cases}
\end{equation}
more complex conjugates of them form a base of the bivectorial space.

Let us use a coordinate system to guide the obtaining of the tensor components of the bivectors with aid of equations (\ref{dual_basis}), therefore we have for ${\bm Z}^{1} = \tilde{\bm\theta}^{0} \wedge \tilde{\bm\theta}^{2}$,
\begin{equation}
 \tilde{\bm\theta}^{0} \wedge \tilde{\bm\theta}^{2} = (-l_{\mu} dx^{\mu})\wedge(\bar{m}_{\nu} dx^{\nu}) = -l_{\mu}\bar{m}_{\nu}\cdot \frac{1}{2}\left(dx^{\mu}\wedge dx^{\nu} - dx^{\nu}\wedge dx^{\mu}\right) = \frac{1}{2}\left(-l_{\mu}\bar{m}_{\nu}+ l_{\nu}\bar{m}_{\mu}\right)dx^{\mu}\wedge dx^{\nu}, \nonumber
\end{equation}
so that we can define
\begin{equation}
\label{U1}
 U_{\mu\nu} = -l_{\mu}\bar{m}_{\nu}+ l_{\nu}\bar{m}_{\mu},
\end{equation}
and we have
\begin{equation}
 \tilde{\bm\theta}^{0} \wedge \tilde{\bm\theta}^{2} = \frac{1}{2} U_{\mu\nu}\, dx^{\mu}\wedge dx^{\nu}.
\end{equation}
The same work goes for $\tilde{\bm\theta}^{3} \wedge \tilde{\bm\theta}^{1}$,
\begin{equation}
 \tilde{\bm\theta}^{3} \wedge \tilde{\bm\theta}^{1} =  \frac{1}{2}\left(k_{\mu}m_{\nu} - k_{\nu}m_{\mu}\right)dx^{\mu}\wedge dx^{\nu},
\end{equation}
with the definition
\begin{equation}
\label{V1}
 V_{\mu\nu} = k_{\mu}m_{\nu} - k_{\nu}m_{\mu},
\end{equation}
and with equality,
\begin{equation}
 \tilde{\bm\theta}^{3} \wedge \tilde{\bm\theta}^{1} = \frac{1}{2} V_{\mu\nu}\, dx^{\mu}\wedge dx^{\nu}.
\end{equation}
Finally we calculate $\tilde{\bm\theta}^{0} \wedge \tilde{\bm\theta}^{1} -\tilde{\bm\theta}^{2} \wedge \tilde{\bm\theta}^{3}$,
\begin{equation}
 \tilde{\bm\theta}^{0} \wedge \tilde{\bm\theta}^{1} -\tilde{\bm\theta}^{2} \wedge \tilde{\bm\theta}^{3} = \frac{1}{2} \left(l_{\mu}k_{\nu} - l_{\nu}k_{\mu} + m_{\mu}\bar{m}_{\nu} - m_{\nu}\bar{m}_{\mu} \right)dx^{\mu}\wedge dx^{\nu},
\end{equation}
with the definition
\begin{equation}
\label{W1}
 W_{\mu\nu} = l_{\mu}k_{\nu} - l_{\nu}k_{\mu} + m_{\mu}\bar{m}_{\nu} - m_{\nu}\bar{m}_{\mu}
\end{equation}
and with equality,
\begin{equation}
 \tilde{\bm\theta}^{0} \wedge \tilde{\bm\theta}^{1} -\tilde{\bm\theta}^{2} \wedge \tilde{\bm\theta}^{3} = \frac{1}{2} W_{\mu\nu}\,dx^{\mu}\wedge dx^{\nu}.
\end{equation}

%%%%%%%%%%%%%%%%%%%%%%%%%%%%%%%%%%

Let us represent the bivector $U_{\mu\nu}$ from equation (\ref{U1}) in Newman-Penrose coordinate system, where $U_{\alpha\beta} = -l_{\alpha}\bar{m}_{\beta}+ l_{\beta}\bar{m}_{\alpha}$, with accordance with (\ref{NP_basis_dual_vector}) we have that the only two components non-nulls are, with $l_{0}=-1$ and $\bar{m}_{2} = 1$,
\begin{equation}
 U_{02} = 1 \hspace*{1cm} \mbox{and} \hspace*{1cm} U_{20} = -1
\end{equation}
such that we have in matrix form,
\begin{equation}
 (U_{\alpha\beta}) = \begin{pmatrix}
                     0 & 0 & 1 & 0 \cr
                     0 & 0 & 0 & 0 \cr
                     -1 & 0 & 0 & 0 \cr
                     0 & 0 & 0 & 0
                    \end{pmatrix}
\end{equation}
The contravariant tensor in Newman-Penrose coordinate system is given by $U^{\alpha\beta} = -l^{\alpha}\bar{m}^{\beta}+ l^{\beta}\bar{m}^{\alpha} $ with only two components non-nulls,
\begin{equation}
 U^{13} = -1 \hspace*{1cm} \mbox{and} \hspace*{1cm} U^{31} = 1
\end{equation}
such that we have this contravariant tensor in matrix shape,
\begin{equation}
 (U^{\alpha\beta}) = \begin{pmatrix}
                     0 & 0 & 0 & 0 \cr
                     0 & 0 & 0 & -1 \cr
                     0 & 0 & 0 & 0 \cr
                     0 & 1 & 0 & 0
                    \end{pmatrix}.
\end{equation}
It is important to notice that,
\begin{equation}
\label{contracao UU}
 U_{\alpha\beta}U^{\alpha\beta} = 0.
\end{equation}
The dual tensor of the $U_{\alpha\beta}$, given by $\widetilde{U}_{\alpha\beta} = \dfrac{1}{2}\eta_{\alpha\beta\gamma\delta} U^{\gamma\delta}$ yields two nonzero components,
\begin{equation}
 \widetilde{U}_{\alpha\beta} = \frac{1}{2} \eta_{\alpha\beta\gamma\delta} \left(-l^{\gamma}\bar{m}^{\delta}+ l^{\delta}\bar{m}^{\gamma}\right) = \frac{1}{2}\left[ \eta_{\alpha\beta 13}(-l^{1}\bar{m}^{3}) + \eta_{\alpha\beta 31}(l^{1}\bar{m}^{3})\right] =   \eta_{\alpha 1\beta 3},\nonumber
\end{equation}
that are
\begin{equation}
 \widetilde{U}_{02} = -i \hspace*{1cm} \mbox{and} \hspace*{1cm} \widetilde{U}_{20} = i,
\end{equation}
where the matrix is given by,
\begin{equation}
 (\widetilde{U}_{\alpha\beta}) = -i \begin{pmatrix}
                     0 & 0 & 1 & 0 \cr
                     0 & 0 & 0 & 0 \cr
                     -1 & 0 & 0 & 0 \cr
                     0 & 0 & 0 & 0
                    \end{pmatrix}.
\end{equation}
We have seen that the tensor $U_{\alpha\beta}$ obeys the auto-duality condition from equation(\ref{sel-dual_property_1}),
\begin{equation}
 \widetilde{U}_{\alpha\beta} = -i U_{\alpha\beta}.
\end{equation}

%%%%%%%%%%%%%%%%%%%%%%%%%%%%%%%%%%

In the same way, let us represent the bivector $V_{\mu\nu}$ from equation (\ref{V1}) in Newman-Penrose coordinate system, where $V_{\alpha\beta} = k_{\alpha}m_{\beta} - k_{\beta}m_{\alpha}$, with accordance with (\ref{NP_basis_dual_vector}) we have that the only two components non-nulls are, with $k_{1}=-1$ and $m_{3} = 1$,
\begin{equation}
 V_{13} = -1 \hspace*{1cm} \mbox{and} \hspace*{1cm} V_{31} = 1
\end{equation}
such that we have in matrix form,
\begin{equation}
 (V_{\alpha\beta}) = \begin{pmatrix}
                     0 & 0 & 0 & 0 \cr
                     0 & 0 & 0 & -1 \cr
                     0 & 0 & 0 & 0 \cr
                     0 & 1 & 0 & 0
                    \end{pmatrix}
\end{equation}
The contravariant tensor in Newman-Penrose coordinate system is given by $V^{\alpha\beta} = k^{\alpha} m^{\beta} - k^{\beta} m^{\alpha} $ with only two components non-nulls,
\begin{equation}
 V^{02} = 1 \hspace*{1cm} \mbox{and} \hspace*{1cm} V^{20} = -1
\end{equation}
such that we have this contravariant tensor in matrix shape,
\begin{equation}
 (V^{\alpha\beta}) = \begin{pmatrix}
                     0 & 0 & 1 & 0 \cr
                     0 & 0 & 0 & 0 \cr
                     -1 & 0 & 0 & 0 \cr
                     0 & 0 & 0 & 0
                    \end{pmatrix}.
\end{equation}
It is important to notice that,
\begin{equation}
 V_{\alpha\beta}V^{\alpha\beta} = 0,
\end{equation}
but the below contractions are non-nulls,
\begin{equation}
\label{contracao UV}
 U_{\alpha\beta}V^{\alpha\beta} = U^{\alpha\beta}V_{\alpha\beta} = 2.
\end{equation}
The dual tensor of the $V_{\alpha\beta}$, given by $\widetilde{V}_{\alpha\beta} = \dfrac{1}{2}\eta_{\alpha\beta\gamma\delta} V^{\gamma\delta}$ yields two nonzero components,
\begin{equation}
 \widetilde{V}_{\alpha\beta} = \frac{1}{2} \eta_{\alpha\beta\gamma\delta} \left(k^{\gamma} m^{\delta} - k^{\delta} m^{\gamma}\right) = \frac{1}{2}\left[ \eta_{0\alpha\beta 2}(k^{0} m^{2}) + \eta_{2\alpha\beta 0}(-k^{0} m^{2})\right] = \eta_{0 \alpha \beta 2},\nonumber
\end{equation}
that are
\begin{equation}
 \widetilde{V}_{13} = i \hspace*{1cm} \mbox{and} \hspace*{1cm} \widetilde{V}_{31} = -i,
\end{equation}
where the matrix is given by,
\begin{equation}
 (\widetilde{V}_{\alpha\beta}) = -i \begin{pmatrix}
                     0 & 0 & 0 & 0 \cr
                     0 & 0 & 0 & -1 \cr
                     0 & 0 & 0 & 0 \cr
                     0 & 1 & 0 & 0
                    \end{pmatrix}.
\end{equation}
We have seen also that the tensor $V_{\alpha\beta}$ obeys the auto-duality condition from equation(\ref{sel-dual_property_1}),
\begin{equation}
 \widetilde{V}_{\alpha\beta} = -i V_{\alpha\beta}.
\end{equation}

%%%%%%%%%%%%%%%%%%%%%%%%%%%%%%%%%%%%%%%%%%%%%%%%

Finally let us see the tensor $W_{\alpha\beta} = l_{\alpha}k_{\beta} - l_{\beta}k_{\alpha} + m_{\alpha}\bar{m}_{\beta} - m_{\beta}\bar{m}_{\alpha}$ in Newman-Penrose coordinate system. In accordance with (\ref{NP_basis_dual_vector}) we have that the only two components non-nulls are, with $l_{0}=-1$, $k_{1} = -1$, $\bar{m}_{2} = 1$,
and $m_{3} =1$, such that,
\begin{equation}
 W_{01} = - W_{10} =1 \hspace*{1cm} \mbox{and} \hspace*{1cm} W_{23} = -W_{32} = -1
\end{equation}
such that we have in matrix form,
\begin{equation}
 (W_{\alpha\beta}) = \begin{pmatrix}
                     0 & 1 & 0 & 0 \cr
                     -1 & 0 & 0 & 0 \cr
                     0 & 0 & 0 & -1 \cr
                     0 & 0 & 1 & 0
                    \end{pmatrix}
\end{equation}
The contravariant tensor in Newman-Penrose coordinate system is given by 
$W^{\alpha\beta} = l^{\alpha}k^{\beta} - l^{\beta}k^{\alpha} + m^{\alpha}\bar{m}^{\beta} - m^{\beta}\bar{m}^{\alpha}$ with only two components non-nulls,
\begin{equation}
 W^{01} = - W^{10} = -1 \hspace*{1cm} \mbox{and} \hspace*{1cm} W^{23} = -W^{32} = 1
\end{equation}
such that we have this contravariant tensor in matrix shape,
\begin{equation}
 (W^{\alpha\beta}) = \begin{pmatrix}
                     0 & -1 & 0 & 0 \cr
                     1 & 0 & 0 & 0 \cr
                     0 & 0 & 0 & 1 \cr
                     0 & 0 & -1 & 0
                    \end{pmatrix}
\end{equation}
It is important to notice that,
\begin{equation}
\label{contracao UW VW}
 U^{\alpha\beta} W_{\alpha\beta} = V^{\alpha\beta}W_{\alpha\beta} = 0
\end{equation}
but the below contractions are non-nulls,
\begin{equation}
\label{contracao WW}
 W_{\alpha\beta}W^{\alpha\beta} = -4.
\end{equation}
The dual tensor $\widetilde{W}_{\alpha\beta} = \dfrac{1}{2}\eta_{\alpha\beta\gamma\delta} W^{\gamma\delta}$ yields four nonzero components,
\begin{equation}
 \widetilde{W}_{\alpha\beta} = \eta_{01\alpha\beta} W^{01} + \eta_{\alpha\beta 23} W^{23} = \eta_{01\alpha\beta}(-1) +  \eta_{\alpha\beta 23}(1) ,\nonumber
\end{equation}
that are
\begin{equation}
 \widetilde{W}_{01} = -\widetilde{W}_{10} = -i \hspace*{1cm} \mbox{and} \hspace*{1cm} \widetilde{W}_{23} =  -\widetilde{W}_{32} = i,
\end{equation}
where the matrix is given by,
\begin{equation}
 (\widetilde{W}_{\alpha\beta}) = -i \begin{pmatrix}
                     0 & 1 & 0 & 0 \cr
                     -1 & 0 & 0 & 0 \cr
                     0 & 0 & 0 & -1 \cr
                     0 & 0 & 1 & 0
                    \end{pmatrix}.
\end{equation}
We have seen also that the tensor $W_{\alpha\beta}$ obeys the auto-duality condition from equation(\ref{sel-dual_property_1}),
\begin{equation}
 \widetilde{W}_{\alpha\beta} = -i W_{\alpha\beta}.
\end{equation}
The tensors $U_{\alpha\beta}$, $V_{\alpha\beta}$ and $W_{\alpha\beta}$ form a base, where we can write any bivector as a linear combination of them. Thus the complex electromagnetic tensor ${\cal F}_{\alpha\beta}$ from equation (\ref{matrix_complex_electomagnetic_tensor}) can be write as,
\begin{equation}
\label{matrix_complex_electomagnetic_tensor_2}
 {\cal F}_{\alpha\beta} = 2 F_{02} U_{\alpha\beta} - 2 F_{13}V_{\alpha\beta} + (F_{01}-F_{23})W_{\alpha\beta}.
\end{equation}
With this idea we can form a self-dual base of the complex bivectors  by writing
\begin{equation}
\begin{cases}
 {Z^{1}}_{\alpha\beta} = U_{\alpha\beta} \cr
 {Z^{2}}_{\alpha\beta} = V_{\alpha\beta} \cr
 {Z^{3}}_{\alpha\beta} = W_{\alpha\beta}
\end{cases}.
\end{equation}
Like the equation (\ref{identidade_I}) we can introduce the tensor $I_{\alpha\beta\gamma\delta} = g_{\alpha\beta\gamma\delta} + i\eta_{\alpha\beta\gamma\delta}$ in Newman-Penrose coordinates and  similarly to equation (\ref{identidade_I_F_2}), with the self-duality, ${\widetilde{Z}^{A}}_{\alpha\beta} = - i {{Z}^{A}}_{\alpha\beta}$ with $A=1,2,3$, we have that,
\begin{equation}
 I_{\alpha\beta\gamma\delta}{Z}^{A\,\gamma\delta} = {Z^{A}}_{\alpha\beta},
\end{equation}
or then
\begin{equation}
\label{identidade_I_1}
 I_{\alpha\beta\gamma\delta}{U}^{\gamma\delta} = U_{\alpha\beta}, \hspace*{1cm}I_{\alpha\beta\gamma\delta}{V}^{\gamma\delta} = V_{\alpha\beta}\hspace*{1cm} \mbox{and}\hspace*{1cm} I_{\alpha\beta\gamma\delta}{W}^{\gamma\delta} = W_{\alpha\beta}.
\end{equation}
With these three equations and with results of equations (\ref{contracao UV}) and (\ref{contracao WW}), where the only non-null contractions that we can obtain is wrote as,
\begin{equation}
 V^{\alpha\beta}I_{\alpha\beta\gamma\delta}{U}^{\gamma\delta} = V^{\alpha\beta}U_{\alpha\beta} = 2, \hspace*{1cm}
 U^{\alpha\beta}I_{\alpha\beta\gamma\delta}{V}^{\gamma\delta} = U^{\alpha\beta}V_{\alpha\beta} =2 \hspace*{1cm} 
 \mbox{and}\hspace*{1cm}W^{\alpha\beta} I_{\alpha\beta\gamma\delta}{W}^{\gamma\delta} = W^{\alpha\beta}W_{\alpha\beta} = -4. \nonumber
\end{equation}
We can  rearrange these three equations as follows
\begin{equation}
 \frac{1}{2} I_{\alpha\beta\gamma\delta}V^{\alpha\beta}U^{\gamma\delta} = 1 \hspace*{1cm}
 \frac{1}{2} I_{\alpha\beta\gamma\delta}U^{\alpha\beta}V^{\gamma\delta} = 1 \hspace*{1cm} \mbox{and}\hspace*{1cm} 
 -\frac{1}{4} I_{\alpha\beta\gamma\delta}W^{\alpha\beta}W^{\gamma\delta} = 1, \nonumber
\end{equation}
and when we add the three equations we have
\begin{equation}
 \left[\frac{1}{2}\left( V^{\alpha\beta}U^{\gamma\delta} + U^{\alpha\beta}V^{\gamma\delta}\right) -\frac{1}{4}W^{\alpha\beta}W^{\gamma\delta}\right] I_{\alpha\beta\gamma\delta} = 3.
\end{equation}
By comparing the above equation with the equation (\ref{contracao II}) we find that,
\begin{equation}
\label{identidade_I_2}
 I^{\alpha\beta\gamma\delta} = \frac{1}{2}\left( V^{\alpha\beta}U^{\gamma\delta} + U^{\alpha\beta}V^{\gamma\delta}\right) -\frac{1}{4}W^{\alpha\beta}W^{\gamma\delta}
 \hspace*{1cm} \mbox{and}\hspace*{1cm}
 I_{\alpha\beta\gamma\delta} = \frac{1}{2}\left( V_{\alpha\beta}U_{\gamma\delta} + U_{\alpha\beta}V_{\gamma\delta}\right) -\frac{1}{4}W_{\alpha\beta}W_{\gamma\delta}.
\end{equation}
Note that using the above identity we can obtain the three equations (\ref{identidade_I_1}).

The complex electromagnetic bivector (\ref{matrix_complex_electomagnetic_tensor_2}) can be write as 
\begin{equation}
\label{complex_electomagnetic_tensor_3}
 \frac{1}{2}{\cal F}_{\alpha\beta} = F_{A}{Z^{A}}_{\alpha\beta} = F_{1}{Z^{1}}_{\alpha\beta} + F_{2}{Z^{2}}_{\alpha\beta} + F_{3}{Z^{3}}_{\alpha\beta}.
\end{equation}
It is important to note that the electromagnetic tensor $F_{\alpha\beta}$ is given by,
\begin{equation}
 F_{\alpha\beta} = F_{A}{Z^{A}}_{\alpha\beta} + \overline{F}_{A}{\overline{Z}\,^{A}}_{\alpha\beta},
\end{equation}
where $\overline{F}_{A}{\overline{Z}\,^{A}}_{\alpha\beta} $ is the complex conjugate. The dual $\widetilde{F}_{\alpha\beta}$ follows as 
\begin{equation}
 \widetilde{F}_{\alpha\beta} = F_{A}{\widetilde{Z}^{A}}_{\alpha\beta} + \overline{F}_{A}{\overline{\widetilde Z}\,^{A}}_{\alpha\beta}, \nonumber
\end{equation}
with self-duality ${\widetilde{Z}\,^{A}}_{\alpha\beta} = -i {Z^{A}}_{\alpha\beta}$ and more ${\overline{\widetilde Z}\,^{A}}_{\alpha\beta} = i {\overline{Z}\,^{A}}_{\alpha\beta}$, 
we have that 
\begin{equation}
 \widetilde{F}_{\alpha\beta} = -i F_{A}{Z^{A}}_{\alpha\beta} + i \overline{F}_{A}{\overline{Z}\,^{A}}_{\alpha\beta}, \nonumber
\end{equation}
we can write the complex electromagnetic bivector ${\cal F}_{\alpha\beta} = F_{\alpha\beta} + i \widetilde{F}_{\alpha\beta}$ as,
\begin{equation}
 {\cal F}_{\alpha\beta} = F_{A}{Z^{A}}_{\alpha\beta} + \overline{F}_{A}{\overline{Z}\,^{A}}_{\alpha\beta} +i \left(-i F_{A}{Z^{A}}_{\alpha\beta} +i \overline{F}_{A}{\overline{Z}\,^{A}}_{\alpha\beta} \right) = 
 2 F_{A}{Z^{A}}_{\alpha\beta}.
\end{equation}
The above equation confirms the equation(\ref{complex_electomagnetic_tensor_3}) in two ways:
\begin{equation}
\label{bivector_F_1}
 \frac{1}{2}{\cal F}_{\alpha\beta} = F_{A}{Z^{A}}_{\alpha\beta} =
 F_{1} {Z^{1}}_{\alpha\beta} + F_{2}{Z^{2}}_{\alpha\beta} + F_{3}{Z^{3}}_{\alpha\beta} = 
 F_{02} U_{\alpha\beta} - F_{13}V_{\alpha\beta} + (F_{01}-F_{23})\frac{W_{\alpha\beta}}{2}.
\end{equation}
or
\begin{equation}
\label{bivector_F_2}
 \frac{1}{2}{\cal F}_{\alpha\beta} = F_{A}{Z^{A}}_{\alpha\beta} = F_{02} U_{\alpha\beta} - F_{13}V_{\alpha\beta} + \frac{F_{01}-F_{23}}{2} W_{\alpha\beta}.
\end{equation}
In the equation (\ref{bivector_F_1}), R. Debever in reference \cite{Debever}, Israel in reference \cite{Israel} and M.Cahen, R. Debever and L.Defrise in the refernece \cite{Cahen}, choose the term $F_{3} = F_{01}-F_{23}$ and the bivector components $ {Z^{3}}_{\alpha\beta} = \dfrac{W_{\alpha\beta}}{2}$. Thus we can write the bivector from equation (\ref{electromagnetic_2-form}) as,
\begin{equation}
 \frac{1}{4} {\cal F}_{\alpha\beta}~\tilde{\bm\theta}^{\alpha} \wedge\tilde{\bm \theta}^{\beta} = F_{1} \frac{1}{2} U_{\alpha\beta}~\tilde{\bm\theta}^{\alpha} \wedge\tilde{\bm \theta}^{\beta}  + F_{2} \frac{1}{2} V_{\alpha\beta}~\tilde{\bm\theta}^{\alpha} \wedge\tilde{\bm \theta}^{\beta} + F_{3} \frac{1}{2} \left(\frac{1}{2} W_{\alpha\beta}~\tilde{\bm\theta}^{\alpha} \wedge\tilde{\bm \theta}^{\beta}\right)
\end{equation}
where we have:
\begin{equation}
\label{Z_1}
 \frac{1}{2} U_{\alpha\beta}~\tilde{\bm\theta}^{\alpha} \wedge\tilde{\bm \theta}^{\beta} =\tilde{\bm \theta}^{0} \wedge\tilde{\bm \theta}^{2} = {\bm Z}^1,
\end{equation}
\begin{equation}
\label{Z_2}
 \frac{1}{2} V_{\alpha\beta}~\tilde{\bm\theta}^{\alpha} \wedge\tilde{\bm \theta}^{\beta} =\tilde{\bm \theta}^{3} \wedge\tilde{\bm \theta}^{1} =  {\bm Z}^2
\end{equation}
and
\begin{equation}
\label{Z_3}
 \frac{1}{2} \left(\frac{1}{2} W_{\alpha\beta}~\tilde{\bm\theta}^{\alpha} \wedge\tilde{\bm \theta}^{\beta}\right) = \frac{1}{2}\left(\tilde{\bm\theta}^{0} \wedge\tilde{\bm \theta}^{1} -\tilde{\bm \theta}^{2} \wedge\tilde{\bm \theta}^{3}\right) =  {\bm Z}^3.
\end{equation}
and we can write the electromagnetic bivector from definition (\ref{electromagnetic_2-form}) as follows,
\begin{equation}
 {\bm F} = F_{1}{\bm Z}^{1} + F_{2}{\bm Z}^{2} + F_{3}{\bm Z}^{3},
\end{equation}
or reduced
\begin{equation}
\label{bivector_F_3}
 {\bm F} = F_{A}{\bm Z}^{A} \hspace*{1cm} \mbox{with}~A=1,2,3.
\end{equation}

We can return to the equation (\ref{Invariant_3}) where we have the invariant,
\begin{equation}
\frac{1}{8} {\cal F}_{\alpha\beta}{\cal F}^{\alpha\beta} = -2 F_{02}F_{13} -\frac{1}{2}(F_{01} - F_{23})^2 ,
\end{equation}
where the choice of R. Debever, Israel and M.Cahen, R. Debever and L.Defrise in the refernece \cite{Debever,Israel,Cahen} is $F_{1}=F_{02} $, $F_{2} = -F_{13} $ and $F_{3} = F_{01}-F_{23}$, such as,
\begin{equation}
\frac{1}{8} {\cal F}_{\alpha\beta}{\cal F}^{\alpha\beta} = 2 F_{1}F_{2} -\frac{1}{2}(F_{3})^2 ,
\end{equation}
or similar to a scalar products,
\begin{equation}
\label{Invariant_4}
\frac{1}{8} {\cal F}_{\alpha\beta}{\cal F}^{\alpha\beta} = \gamma^{AB}F_{A}F_{B},
\end{equation}
where the metric tensor is given by,
\begin{equation}
\gamma^{AB} = \begin{pmatrix}
               0 & 1 & 0 \cr
               1 & 0 & 0 \cr
               0 & 0 & -\frac{1}{2}
              \end{pmatrix}.
\end{equation}

Let us see the exterior product of the set $\{{\bm Z}^1,{\bm Z}^2,{\bm Z}^3 \}$, where the only nonzero exterior products are seen as,
\begin{equation}
 {\bm Z}^1 \wedge {\bm Z}^2 =  (\tilde{\bm \theta}^{0} \wedge\tilde{\bm \theta}^{2})\wedge (\tilde{\bm\theta}^{3} \wedge\tilde{\bm \theta}^{1}) = 
 \tilde{\bm \theta}^{0} \wedge\tilde{\bm \theta}^{1}\wedge\tilde{\bm \theta}^{2} \wedge\tilde{\bm \theta}^{3}, \nonumber
\end{equation}
or
\begin{equation}
\label{Z_1 ^ Z_2}
 {\bm Z}^1 \wedge {\bm Z}^2 = \gamma^{12}~~\tilde{\bm \theta}^{0} \wedge\tilde{\bm \theta}^{1}\wedge\tilde{\bm \theta}^{2} \wedge\tilde{\bm \theta}^{3}.
\end{equation}
The same reasoning holds true for,
\begin{equation}
\label{Z_2 ^ Z_1}
 {\bm Z}^2 \wedge {\bm Z}^1 = \gamma^{21}~~\tilde{\bm \theta}^{0} \wedge\tilde{\bm \theta}^{1}\wedge\tilde{\bm \theta}^{2} \wedge\tilde{\bm \theta}^{3}.
\end{equation}
And the only other nonzero exterior product is given by,
\begin{eqnarray}
 {\bm Z}^3 \wedge {\bm Z}^3 &=& \frac{1}{2}  \left(\tilde{\bm\theta}^{0} \wedge\tilde{\bm \theta}^{1} -\tilde{\bm \theta}^{2} \wedge\tilde{\bm \theta}^{3}\right) \wedge \frac{1}{2}  \left(\tilde{\bm\theta}^{0} \wedge\tilde{\bm \theta}^{1} -\tilde{\bm \theta}^{2} \wedge\tilde{\bm \theta}^{3}\right)\cr
 &=& -\frac{1}{4}\left(\tilde{\bm\theta}^{0} \wedge\tilde{\bm \theta}^{1}\wedge\tilde{\bm \theta}^{2} \wedge\tilde{\bm \theta}^{3} +\tilde{\bm \theta}^{2} \wedge\tilde{\bm \theta}^{3} \wedge\tilde{\bm \theta}^{0} \wedge\tilde{\bm \theta}^{1}\right)\cr
 &=& -\frac{1}{2}\tilde{\bm\theta}^{0} \wedge\tilde{\bm \theta}^{1}\wedge\tilde{\bm \theta}^{2} \wedge\tilde{\bm \theta}^{3}\nonumber
\end{eqnarray}
or
\begin{equation}
\label{Z_3 ^ Z_3}
 {\bm Z}^3 \wedge {\bm Z}^3 =  \gamma^{33}~~\tilde{\bm\theta}^{0} \wedge\tilde{\bm \theta}^{1}\wedge\tilde{\bm \theta}^{2} \wedge\tilde{\bm \theta}^{3}.
\end{equation}
With these three above equations (\ref{Z_1 ^ Z_2}), (\ref{Z_2 ^ Z_1}) and (\ref{Z_3 ^ Z_3}) we can resume the exterior product as,
\begin{equation}
 \label{Z_{A} ^ Z_{B}}
 {\bm Z}^{A} \wedge {\bm Z}^{B} =  \gamma^{AB}~~\tilde{\bm\theta}^{0} \wedge\tilde{\bm \theta}^{1}\wedge\tilde{\bm \theta}^{2} \wedge\tilde{\bm \theta}^{3}.
\end{equation}

The components of the conjugate set $\{\overline{\bm Z}\,^1,\overline{\bm Z}\,^2,\overline{\bm Z}\,^3 \}$ are given by,
\begin{equation}
 \overline{\bm Z}^1 = \overline{\tilde{\bm\theta}}\,^{0} \wedge \overline{\tilde{\bm\theta}}\,^{2} =\tilde{\bm \theta}^{0} \wedge\tilde{\bm \theta}^{3},
\end{equation}
\begin{equation}
 \overline{\bm Z}\,^2 = \overline{\tilde{\bm\theta}}\,^{3} \wedge \overline{\tilde{\bm\theta}}\,^{1} =\tilde{\bm \theta}^{2} \wedge\tilde{\bm \theta}^{1},
\end{equation}
and
\begin{equation}
 \overline{\bm Z}\,^3 = \frac{1}{2} \left( \overline{\tilde{\bm\theta}}\,^{0} \wedge \overline{\tilde{\bm\theta}}\,^{1} -  \overline{\tilde{\bm\theta}}\,^{2} \wedge \overline{\tilde{\bm\theta}}\,^{3} \right) =  \frac{1}{2}\left(\tilde{\bm\theta}^{0} \wedge\tilde{\bm \theta}^{1} +\tilde{\bm \theta}^{2} \wedge\tilde{\bm \theta}^{3}\right) 
\end{equation}
with the exterior products given by,
\begin{equation}
 \label{Z_m ^ Z_n conjugado}
 \overline{\bm Z}\,^{A} \wedge \overline{\bm Z}\,^{B} =  - \gamma^{AB}~~\tilde{\bm\theta}^{0} \wedge\tilde{\bm \theta}^{1}\wedge\tilde{\bm \theta}^{2} \wedge\tilde{\bm \theta}^{3}.
\end{equation}
We can also verify that,
\begin{equation}
  {\bm Z}^{A} \wedge \overline{\bm Z}\,^{B} = 0.
\end{equation}

Let us return to the equation  (\ref{bivector_F_3}), where we have ${\bm F} = F_{A}{\bm Z}^{A}$ with $A=1,2,3$, we have that,
\begin{equation}
 {\bm F} \wedge {\bm F} = F_{A}{\bm Z}^{A} \wedge F_{B}{\bm Z}^{B} = F_{A}F_{B}~{\bm Z}^{A} \wedge {\bm Z}^{B} = F_{A}F_{B}~\gamma^{AB}~\tilde{\bm\theta}^{0} \wedge\tilde{\bm \theta}^{1}\wedge\tilde{\bm \theta}^{2} \wedge\tilde{\bm \theta}^{3}, \nonumber
\end{equation}
with aid of equation (\ref{Invariant_4}) we have,
\begin{equation}
\label{produto externo F^F_1}
 {\bm F} \wedge {\bm F} = \frac{1}{8}{\cal F}_{\alpha\beta}{\cal F}^{\alpha\beta} ~ \tilde{\bm \theta}^{0} \wedge\tilde{\bm \theta}^{1}\wedge\tilde{\bm \theta}^{2} \wedge\tilde{\bm \theta}^{3}.
\end{equation}
Also from equation (\ref{electromagnetic_2-form}) where we have
${\bm F} = \frac{1}{4} {\cal F}_{\alpha\beta} ~ \tilde{\bm\theta}^{\alpha} \wedge \tilde{\bm\theta}^{\beta}$, we can calculate,
\begin{equation}
\label{produto externo F^F_2}
 {\bm F} \wedge {\bm F} = \frac{1}{16}{\cal F}_{\alpha\beta} {\cal F}_{\gamma\delta} ~ \tilde{\bm\theta}^{\alpha} \wedge \tilde{\bm\theta}^{\beta}\wedge \tilde{\bm\theta}^{\gamma} \wedge \tilde{\bm\theta}^{\delta}.
\end{equation}
Comparing the two equations above (\ref{produto externo F^F_1}) and (\ref{produto externo F^F_2}),  we that,
\begin{equation}
\label{produto externo F^F_3}
 {\cal F}_{\alpha\beta} {\cal F}_{\gamma\delta} ~ \tilde{\bm\theta}^{\alpha} \wedge \tilde{\bm\theta}^{\beta}\wedge \tilde{\bm\theta}^{\gamma} \wedge \tilde{\bm\theta}^{\delta} = 2{\cal F}_{\alpha\beta}{\cal F}^{\alpha\beta} ~ \tilde{\bm \theta}^{0} \wedge\tilde{\bm \theta}^{1}\wedge\tilde{\bm \theta}^{2} \wedge\tilde{\bm \theta}^{3}.
\end{equation}

If we recall the equation (\ref{Invariant_3}) and rewrite it as,
\begin{equation}
 \frac{1}{4} {\cal F}_{\alpha\beta}{\cal F}^{\alpha\beta} = - (F_{01} - F_{23})^2 -4 F_{02}F_{13}.
\end{equation}
In the above equation, Newman and Penrose \cite{Newman}, H. Stephani et al. in the reference \cite{Kramer} and  
J.Griffiths and J. Podolský in the reference \cite{Griffiths}, they use,
\begin{equation}
\label{electromagnetic_components_NP}
 \begin{cases}
  \Phi_{0} = F_{02} \cr
  \Phi_{1} = \dfrac{F_{01}-F_{23}}{2} \cr
  \Phi_{2} = -F_{13},
 \end{cases}
\end{equation}
where we can identify fron equation (\ref{electromagnetic_field-strenght_NP}) that,
\begin{equation}
\label{dyad_Phi_0}
 \Phi_{0} = \frac{1}{2}\left(-F_y + i F_z \right) = -\frac{1}{2}(E_y-iB_y) + \frac{i}{2}(E_z-iB_z), 
\end{equation}
\begin{equation}
\label{dyad_Phi_1}
 \Phi_{1} = F_x = (E_x-iB_x) 
\end{equation}
and
\begin{equation}
\label{dyad_Phi_2}
 \Phi_{2} = \frac{1}{2}\left(F_y + i F_z \right) = \frac{1}{2}(E_y-iB_y) + \frac{i}{2}(E_z-iB_z),
\end{equation}
so that we have the Lorentz invariant,
\begin{equation}
\label{Invariant_5}
 {\cal F}_{\alpha\beta}{\cal F}^{\alpha\beta} = 16\left(\Phi_{0}\Phi_{2} - 
\Phi_{1}^2\right).
\end{equation}
We can rewrite the above equation,
\begin{equation}
 \frac{1}{8}~{\cal F}_{\alpha\beta}{\cal F}^{\alpha\beta} = 2 \Phi_{0}\Phi_{2} - 2 \Phi_{1}^2, \nonumber
\end{equation}
or
\begin{equation}
 \frac{1}{8}~{\cal F}_{\alpha\beta}{\cal F}^{\alpha\beta} = \gamma^{AB}~\Phi_{A}\phi_{B},
\end{equation}
where, $A,~B =0,1,2$ and the metric tensor is,
\begin{equation}
\gamma^{AB} = 
 \begin{pmatrix}
               0 & 0 & 1 \cr
               0 & -2 & 0 \cr
               1 & 0 & 0
              \end{pmatrix}.
\end{equation}

We recall the equation (\ref{matrix_complex_electomagnetic_tensor_2}) and rewrite it as below,
\begin{equation}
\label{matrix_complex_electomagnetic_tensor_3}
 \frac{1}{2}{\cal F}_{\alpha\beta} =  F_{02} U_{\alpha\beta} -  F_{13}V_{\alpha\beta} + \frac{(F_{01}-F_{23})}{2}W_{\alpha\beta} = \Phi_{0}U_{\alpha\beta} + \Phi_{1}W_{\alpha\beta} + \Phi_{2} V_{\alpha\beta}.
\end{equation}
We obtain the electromagnetic bivector,
\begin{equation}
 \label{bivector_F_4}
 \frac{1}{4}{\cal F}_{\alpha\beta}~\tilde{\bm\theta}^{\alpha} \wedge \tilde{\bm\theta}^{\beta} = \Phi_{0} \left(\frac{1}{2}~ U_{\alpha\beta}~\tilde{\bm\theta}^{\alpha} \wedge \tilde{\bm\theta}^{\beta}\right) + \Phi_{1}\left(\frac{1}{2}~W_{\alpha\beta}~\tilde{\bm\theta}^{\alpha} \wedge \tilde{\bm\theta}^{\beta}\right) + \Phi_{2}\left(\frac{1}{2}~ V_{\alpha\beta}~\tilde{\bm\theta}^{\alpha} \wedge \tilde{\bm\theta}^{\beta}\right),
\end{equation}
where we define the bivector basis,
\begin{equation}
\label{base_U_V_W}
 \begin{cases}
  \dfrac{1}{2}~U_{\alpha\beta}~\tilde{\bm\theta}^{\alpha} \wedge \tilde{\bm\theta}^{\beta} = \tilde{\bm\theta}^{0} \wedge \tilde{\bm\theta}^{2} = \bm{U}, \\[6pt]
  \dfrac{1}{2}~W_{\alpha\beta} ~\tilde{\bm\theta}^{\alpha} \wedge \tilde{\bm\theta}^{\beta}= \tilde{\bm\theta}^{0} \wedge \tilde{\bm\theta}^{1} - \tilde{\bm\theta}^{2} \wedge \tilde{\bm\theta}^{3} = \bm{W}, \\[6pt]
  \dfrac{1}{2}~V_{\alpha\beta} ~\tilde{\bm\theta}^{\alpha} \wedge \tilde{\bm\theta}^{\beta}= \tilde{\bm\theta}^{3} \wedge \tilde{\bm\theta}^{1} = \bm{V} ,
 \end{cases}
\end{equation}
where the basis of bivector space in $V^4$ space is formed by six bivectors
$\{\bm{U}, \bm{V}, \bm{W}, \overline{\bm{U}}, \overline{\bm{V}}, \overline{\bm{W}}  \}$.
With definition (\ref{electromagnetic_2-form}) where ${\bm F} = \frac{1}{4} {\cal F}_{\alpha\beta} ~ \tilde{\bm\theta}^{\alpha} \wedge \tilde{\bm\theta}^{\beta}$ we write the electromagnetic bivector from equation (\ref{bivector_F_4}) as
\begin{equation}
 \label{bivector_F_5}
 \bm{F} = \Phi_{0}\bm{U} + \Phi_{1}\bm{W} + \Phi_{2}\bm{V}.
\end{equation}
The terms $\Phi_{0}$, $\Phi_{1}$ and $\Phi_{2}$ are called dyad components of the electromagnetic bivector $\bm{F}$ \cite{Carmeli}.

The nonzero external products of basis (\ref{base_U_V_W}) result in,
\begin{equation}
 \bm{U} \wedge \bm{V} = \bm{V} \wedge \bm{U} = \tilde{\bm \theta}^{0} \wedge\tilde{\bm \theta}^{1}\wedge\tilde{\bm \theta}^{2} \wedge\tilde{\bm \theta}^{3},
\end{equation}
and
\begin{equation}
 \bm{W} \wedge \bm{W} = - 2 ~\tilde{\bm \theta}^{0} \wedge\tilde{\bm \theta}^{1}\wedge\tilde{\bm \theta}^{2} \wedge\tilde{\bm \theta}^{3}.
\end{equation}
For conjugate complex we have that similar results seen in the equation (\ref{Z_m ^ Z_n conjugado}), such as
\begin{equation}
 \overline{\bm U} \wedge \overline{\bm V} = \overline{\bm V} \wedge \overline{\bm U} = -\frac{1}{2}~\overline{\bm W} \wedge \overline{\bm W} = -\tilde{\bm\theta}^{0} \wedge\tilde{\bm \theta}^{1}\wedge\tilde{\bm \theta}^{2} \wedge\tilde{\bm \theta}^{3}.
\end{equation}
For the other exterior products we have that $\bm{U} \wedge \overline{\bm{V}} = 0$, $\bm{V} \wedge \overline{\bm{U}} = 0$ and $\bm{W} \wedge \overline{\bm{W}} = 0$. With these exterior products we obtain for $ {\bm F} = \frac{1}{4} {\cal F}_{\alpha\beta} ~ \tilde{\bm\theta}^{\alpha} \wedge \tilde{\bm\theta}^{\beta}$  that,
\begin{equation}
 \bm{F} \wedge \bm{F}  = \frac{1}{16}{\cal F}_{\alpha\beta} {\cal F}_{\gamma\delta} ~ \tilde{\bm\theta}^{\alpha} \wedge \tilde{\bm\theta}^{\beta}\wedge \tilde{\bm\theta}^{\gamma} \wedge \tilde{\bm\theta}^{\delta}, \nonumber 
\end{equation}
and for $\bm{F} = \Phi_{0}\bm{U} + \Phi_{1}\bm{W} + \Phi_{2}\bm{V}$ from equation (\ref{bivector_F_5}) and with aid of equation (\ref{Invariant_5}) we have that,
\begin{equation}
 \bm{F} \wedge \bm{F}  = -2(\Phi_{0}\Phi_{2} + \Phi_{1}^2)~\tilde{\bm\theta}^{0} \wedge\tilde{\bm \theta}^{1}\wedge\tilde{\bm \theta}^{2} \wedge\tilde{\bm \theta}^{3} = \frac{1}{8} {\cal F}_{\alpha\beta}{\cal F}^{\alpha\beta} ~\tilde{\bm\theta}^{0} \wedge\tilde{\bm \theta}^{1}\wedge\tilde{\bm \theta}^{2} \wedge\tilde{\bm \theta}^{3} , \nonumber 
\end{equation}
where the two above equations yield in the identity obtained in the equation (\ref{produto externo F^F_3}).

The exterior products of eletromagnetic bivector $\bm{F} = \Phi_{0}\bm{U} + \Phi_{1}\bm{W} + \Phi_{2}\bm{V}$ with anyone bivector of the basis 
$\{\bm{U}, \bm{V}, \bm{W}\}$, result in 4-forms as follow,
\begin{equation}
 \bm{F} \wedge \bm{U} =  \Phi_{2} ~\tilde{\bm\theta}^{0} \wedge\tilde{\bm \theta}^{1}\wedge\tilde{\bm \theta}^{2} \wedge\tilde{\bm \theta}^{3},
\end{equation}
\begin{equation}
 \bm{F} \wedge \bm{V} =  \Phi_{0} ~\tilde{\bm\theta}^{0} \wedge\tilde{\bm \theta}^{1}\wedge\tilde{\bm \theta}^{2} \wedge\tilde{\bm \theta}^{3},
\end{equation}
\begin{equation}
 \bm{F} \wedge \bm{W} = - 2 \Phi_{1} ~\tilde{\bm\theta}^{0} \wedge\tilde{\bm \theta}^{1}\wedge\tilde{\bm \theta}^{2} \wedge\tilde{\bm \theta}^{3}.
\end{equation}
In term of contractions, we can use the equation (\ref{matrix_complex_electomagnetic_tensor_3}), where we have 
\begin{equation}
\label{bivector_F_6}
 \frac{1}{2}{\cal F}_{\alpha\beta} = \Phi_{0}U_{\alpha\beta} + \Phi_{1}W_{\alpha\beta} + \Phi_{2} V_{\alpha\beta}, 
\end{equation}
and the contraction of this equation with $U^{\alpha\beta}$ results in
\begin{equation}
 \frac{1}{2}{\cal F}_{\alpha\beta}U^{\alpha\beta} = \Phi_{2} V_{\alpha\beta}U^{\alpha\beta} = 2\Phi_{2},\nonumber
\end{equation}
where we have used $V_{\alpha\beta}U^{\alpha\beta} = 2$ from equation (\ref{contracao UV}) and nulls for $U_{\alpha\beta}U^{\alpha\beta} = W_{\alpha\beta}U^{\alpha\beta} = 0$ in accordance with (\ref{contracao UU}) and (\ref{contracao UW VW}). The above equation summarizes in, 
\begin{equation}
\label{F ab U ab}
 \Phi_{2} = \frac{1}{4}{\cal F}_{\alpha\beta}U^{\alpha\beta}.
\end{equation}
The contraction of electromagnetic bivector ${\cal F}_{\alpha\beta}$ with $V^{\alpha\beta}$ results in,
\begin{equation}
\label{F ab V ab}
 \Phi_{0} = \frac{1}{4}{\cal F}_{\alpha\beta}V^{\alpha\beta}.
\end{equation}
Then, we calculate the contraction of electromagnetic bivector ${\cal F}_{\alpha\beta}$ with $W^{\alpha\beta}$, that it results in,
\begin{equation}
 \frac{1}{2}{\cal F}_{\alpha\beta}W^{\alpha\beta} = \Phi_{1} W_{\alpha\beta}W^{\alpha\beta} = -4\Phi_{1},\nonumber
\end{equation}
where we have used the equation (\ref{contracao WW}) and the above equation summarizes in, 
\begin{equation}
\label{F ab W ab}
 \Phi_{1} = -\frac{1}{8}{\cal F}_{\alpha\beta}W^{\alpha\beta}.
\end{equation}

We can also use the identity (\ref{identidade_I_2}) to write $\Phi_{0}$,$\Phi_{1}$ and $\Phi_{2}$ in terms of the electromagnetic bivector $F_{\alpha\beta}$ from equation (\ref{electromagnetic_tensor}) by using the identity (\ref{identidade_I_F_1}), where we have that $I_{\alpha\beta\gamma\delta}F^{\gamma\delta} =
 \frac{1}{2} {\cal F}_{\alpha\beta}$ and for making a contraction with $V^{\alpha\beta}$ we obtain,
\begin{equation}
  V^{\alpha\beta}I_{\alpha\beta\gamma\delta}F^{\gamma\delta} =
 \frac{1}{2} {\cal F}_{\alpha\beta}V^{\alpha\beta}, \nonumber
\end{equation}
with aid of the equations (\ref{identidade_I_1}) and  (\ref{F ab V ab}) we have,
\begin{equation}
 V_{\gamma\delta}F^{\gamma\delta}= 2\Phi_{0},\nonumber
\end{equation}
or then
\begin{equation}
 \Phi_{0} = \frac{1}{2}V^{\alpha\beta}F_{\alpha\beta}.\nonumber
\end{equation}
With definition in the equation (\ref{V1}), 
$V^{\alpha\beta} = k^{\alpha}m^{\beta} - k^{\beta}m^{\alpha}$, 
the above equation becomes 
\begin{equation}
 \Phi_{0} = \frac{1}{2}\left(k^{\alpha}m^{\beta} + k^{\beta}m^{\alpha}\right)F_{\alpha\beta} = k^{\alpha}m^{\beta}F_{\alpha\beta},
\end{equation}
in Newman-Penrose coordinates, with equation (\ref{NP_basis_vector}) and (\ref{electromagnetic_tensor}) we must verify that $\Phi_{0} = F_{02}$.

In the same way for $\Phi_{1}$, we put ${\cal F}_{\alpha\beta} = 2~I_{\alpha\beta\gamma\delta}F^{\gamma\delta} $ in equation (\ref{F ab W ab})
\begin{equation}
 \Phi_{1} = -\frac{1}{8}\left(2~I_{\alpha\beta\gamma\delta}F^{\gamma\delta}\right)W^{\alpha\beta} = -\frac{1}{4}\left(I_{\alpha\beta\gamma\delta}W^{\alpha\beta}\right)F^{\gamma\delta} = -\frac{1}{4} W_{\alpha\beta} F^{\alpha\beta}. \nonumber
\end{equation}
With definition in the equation (\ref{W1}), $W_{\alpha\beta} = l_{\alpha}k_{\beta} - l_{\beta}k_{\alpha} + m_{\alpha}\bar{m}_{\beta} - m_{\beta}\bar{m}_{\alpha}$, the above equation becomes 
\begin{equation}
 \Phi_{1} =  -\frac{1}{4}\left(l_{\alpha}k_{\beta} - l_{\beta}k_{\alpha} + m_{\alpha}\bar{m}_{\beta} - m_{\beta}\bar{m}_{\alpha} \right)F^{\alpha\beta} = -\frac{1}{2}\left(l_{\alpha}k_{\beta}+m_{\alpha}\bar{m}_{\beta} \right)F^{\alpha\beta},\nonumber
\end{equation}
or then
\begin{equation}
 \Phi_{1} = \frac{1}{2}\left(k^{\alpha}l^{\beta}+\bar{m}^{\alpha}m^{\beta} \right)F_{\alpha\beta}
\end{equation}
where in Newman-Penrose coordinates, with equation (\ref{NP_basis_vector}) and (\ref{electromagnetic_tensor}) we must verify that $\Phi_{1} = \dfrac{1}{2}\left(F_{01} -F_{23} \right)$.

Finally for for $\Phi_{2}$, we can put ${\cal F}_{\alpha\beta} = 2~I_{\alpha\beta\gamma\delta}F^{\gamma\delta} $ in equation (\ref{F ab U ab})
\begin{equation}
 \Phi_{2} = \frac{1}{4}\left(2~I_{\alpha\beta\gamma\delta}F^{\gamma\delta}\right)U^{\alpha\beta} = \frac{1}{2}\left(I_{\alpha\beta\gamma\delta}U^{\alpha\beta}\right)F^{\gamma\delta} = \frac{1}{2} U^{\alpha\beta} F_{\alpha\beta}. \nonumber
\end{equation}
With definition in the equation (\ref{U1}), $U^{\alpha\beta} = -l^{\alpha}\bar{m}^{\beta} + l^{\beta}\bar{m}^{\alpha}$, the above equation becomes 
\begin{equation}
 \Phi_{2} = - l^{\alpha}\bar{m}^{\beta}F_{\alpha\beta} = \bar{m}^{\alpha}l^{\beta} F_{\alpha\beta},
\end{equation}
where in Newman-Penrose coordinates, with equation (\ref{NP_basis_vector}) and (\ref{electromagnetic_tensor}) we must verify that $\Phi_{2} = -F_{13}$ in accordance with the definition (\ref{electromagnetic_components_NP}).

%%%%%%%%%%%%%%%%%%%%%%%%%%%%%%%%%%%%%%%%%%%%%%%%%%%%%%%%%%%%%%
%%%%%%%%%%%%%%%%%%%%%%%%%%%%%%%%%%%%%%%%%%%%%%%%%%%%%%%%%%%%%%
%%%%%%%%%%%%%%%%%%%%%%%%%%%%%%%%%%%%%%%%%%%%%%%%%%%%%%%%%%%%%%
%%%%%%%%%%%%%%%%%%%%%%%%%%%%%%%%%%%%%%%%%%%%%%%%%%%%%%%%%%%%%%
%%%%%%%%%%%%%%%%%%%%%%%%%%%%%%%%%%%%%%%%%%%%%%%%%%%%%%%%%%%%%%
%%%%%%%%%%%%%%%%%%%%%%%%%%%%%%%%%%%%%%%%%%%%%%%%%%%%%%%%%%%%%%

\subsection{The Maxwell equations in the Newman-Penrose formalism}

When we refer the electromagnetic field tensor $F_{\alpha\beta}$ to a standard
Minkowski frame, its components are related by definition to
the components of the electric and magnetic 3-vector fields $\bm E$ and $\bm B$ \cite{Wytler2}. In Newman-Penrose formalism the Maxwell's source-free equations is described in terms of dyad components of the electromagnetic bivector, $\Phi_{0}$, $\Phi_{1}$ and $\Phi_{2}$.
For this, let us return to the equation (\ref{bivector_F_6}), where we have
\begin{equation}
 \frac{1}{2}{\cal F}_{\alpha\beta} = \Phi_{0}U_{\alpha\beta} + \Phi_{1}W_{\alpha\beta} + \Phi_{2} V_{\alpha\beta}. \nonumber
\end{equation}
If we calculate the Maxwell equations in vacuum, where the electromagnetic current is null, we have that $ \nabla_{\alpha} {\cal F}^{\alpha\beta} = 0$.
Thus, we have that,
\begin{equation}
\label{Newman_Penrose_Maxwell_equations_1}
 \frac{1}{2}\nabla_{\alpha}{\cal F}^{\alpha\beta} = \left(\nabla_{\alpha}\Phi_{0} \right)U^{\alpha\beta} +\Phi_{0}\nabla_{\alpha}  U^{\alpha\beta} + \left(\nabla_{\alpha}\Phi_{1} \right)W^{\alpha\beta} + \Phi_{1}\nabla_{\alpha} W^{\alpha\beta} + \left(\nabla_{\alpha}\Phi_{2} \right) V^{\alpha\beta} +\Phi_{2} \nabla_{\alpha}V^{\alpha\beta} = 0 .
\end{equation}
We must obtain the four Maxwell equations in the four null directions $\bm k$, $\bm l$, $\bm m$ and $\bar{\bm m}$ by 
\begin{equation}
 k_{\beta}\nabla_{\alpha} {\cal F}^{\alpha\beta}, \hspace*{1cm}
 l_{\beta}\nabla_{\alpha} {\cal F}^{\alpha\beta}, \hspace*{1cm}
 m_{\beta}\nabla_{\alpha} {\cal F}^{\alpha\beta}, \hspace*{1cm}
 \mbox{and} \hspace*{1cm}
 \bar{m}_{\beta}\nabla_{\alpha} {\cal F}^{\alpha\beta}.
\end{equation}
To obtain the Maxwell equations (\ref{Newman_Penrose_Maxwell_equations_1}) in null directions, it is necessary to calculate the contractions with aid of equation (\ref{contraction_tetrad}),
\begin{eqnarray}
\label{contraction_1}
 k_{\beta} U^{\alpha\beta} &=& k_{\beta}\left(-l^{\alpha}\bar{m}^{\beta} + l^{\beta}\bar{m}^{\alpha}\right) = -\bar{m}^{\alpha},
\cr
 k_{\beta} V^{\alpha\beta} &=& k_{\beta}\left( k^{\alpha}m^{\beta} - k^{\beta}m^{\alpha}\right) = 0,
\cr
k_{\beta} W^{\alpha\beta} &=& k_{\beta}\left(l^{\alpha}k^{\beta} - l^{\beta}k^{\alpha} + m^{\alpha}\bar{m}^{\beta} - m^{\beta}\bar{m}^{\alpha}\right) = k^{\alpha},
\end{eqnarray}
\begin{eqnarray}
\label{contraction_2}
 l_{\beta} U^{\alpha\beta} &=& l_{\beta}\left(-l^{\alpha}\bar{m}^{\beta} + l^{\beta}\bar{m}^{\alpha}\right) = 0,
\cr
 l_{\beta} V^{\alpha\beta} &=& l_{\beta}\left( k^{\alpha}m^{\beta} - k^{\beta}m^{\alpha}\right) = m^{\alpha},
\cr
l_{\beta} W^{\alpha\beta} &=& l_{\beta}\left(l^{\alpha}k^{\beta} - l^{\beta}k^{\alpha} + m^{\alpha}\bar{m}^{\beta} - m^{\beta}\bar{m}^{\alpha}\right) = -l^{\alpha},
\end{eqnarray}
\begin{eqnarray}
\label{contraction_3}
 m_{\beta} U^{\alpha\beta} &=& m_{\beta}\left(-l^{\alpha}\bar{m}^{\beta} + l^{\beta}\bar{m}^{\alpha}\right) = -l^{\alpha},
\cr
 m_{\beta} V^{\alpha\beta} &=& m_{\beta}\left( k^{\alpha}m^{\beta} - k^{\beta}m^{\alpha}\right) = 0,
\cr
m_{\beta} W^{\alpha\beta} &=& m_{\beta}\left(l^{\alpha}k^{\beta} - l^{\beta}k^{\alpha} + m^{\alpha}\bar{m}^{\beta} - m^{\beta}\bar{m}^{\alpha}\right) = m^{\alpha},
\end{eqnarray}
\begin{eqnarray}
\label{contraction_4}
 \bar{m}_{\beta} U^{\alpha\beta} &=& \bar{m}_{\beta}\left(-l^{\alpha}\bar{m}^{\beta} + l^{\beta}\bar{m}^{\alpha}\right) = 0,
\cr
 \bar{m}_{\beta} V^{\alpha\beta} &=& \bar{m}_{\beta}\left( k^{\alpha}m^{\beta} - k^{\beta}m^{\alpha}\right) = k^{\alpha},
\cr
\bar{m}_{\beta} W^{\alpha\beta} &=& \bar{m}_{\beta}\left(l^{\alpha}k^{\beta} - l^{\beta}k^{\alpha} + m^{\alpha}\bar{m}^{\beta} - m^{\beta}\bar{m}^{\alpha}\right) = -\bar{m}^{\alpha}.
\end{eqnarray}
Thus, let us  make a projection of equation (\ref{Newman_Penrose_Maxwell_equations_1})  on the direction $k_{\beta}$, where we have,
\begin{equation}
 k_{\beta}\nabla_{\alpha}{\cal F}^{\alpha\beta} = \left(\nabla_{\alpha}\Phi_{0} \right)k_{\beta}U^{\alpha\beta} +\Phi_{0}k_{\beta}\nabla_{\alpha}  U^{\alpha\beta} + \left(\nabla_{\alpha}\Phi_{1} \right)k_{\beta}W^{\alpha\beta} + \Phi_{1}k_{\beta}\nabla_{\alpha} W^{\alpha\beta} + \left(\nabla_{\alpha}\Phi_{2} \right)k_{\beta} V^{\alpha\beta} +\Phi_{2}k_{\beta} \nabla_{\alpha}V^{\alpha\beta} = 0, \nonumber
\end{equation}
where we can use the equations (\ref{contraction_1}) to reduce the above equation to,
\begin{equation}
k^{\alpha}\nabla_{\alpha}\Phi_{1} - \bar{m}^{\alpha} \nabla_{\alpha}\Phi_{0}  +\Phi_{0}k_{\beta}\nabla_{\alpha}  U^{\alpha\beta}   + \Phi_{1}k_{\beta}\nabla_{\alpha} W^{\alpha\beta} +\Phi_{2}k_{\beta} \nabla_{\alpha}V^{\alpha\beta} = 0. \nonumber
\end{equation}
We should note that the projections of the covariant derivatives in the null directions seen above equation, they can be defined as directional derivatives
\begin{equation}
\label{directional_derivatives}
 D=k^{\alpha}\nabla_{\alpha}, \hspace*{0.5cm}
 \Delta = l^{\alpha}\nabla_{\alpha}, \hspace*{0.5cm}
 \delta = m^{\alpha}\nabla_{\alpha} \hspace*{0.5cm}
 \hspace*{0.5cm} \mbox{and} \hspace*{0.5cm} \bar{\delta} = \bar{m}^{\alpha}\nabla_{\alpha},
\end{equation}
so that the above equation is rewritten as,
\begin{equation}
\label{Newman_Penrose_Maxwell_equations_2}
D\Phi_{1} - \bar{\delta} \Phi_{0}  +\Phi_{0}k_{\beta}\nabla_{\alpha}  U^{\alpha\beta}   + \Phi_{1}k_{\beta}\nabla_{\alpha} W^{\alpha\beta} +\Phi_{2}k_{\beta} \nabla_{\alpha}V^{\alpha\beta} = 0. %\nonumber
\end{equation}
Now we can work with the term $k_{\beta}\nabla_{\alpha}  U^{\alpha\beta}$ to reduce it in to spin coefficients, noting that,
\begin{equation}
 \nabla_{\alpha}(k_{\beta}U^{\alpha\beta}) =  (\nabla_{\alpha}k_{\beta}) U^{\alpha\beta} + k_{\beta}\nabla_{\alpha}U^{\alpha\beta}, \nonumber
\end{equation}
where we can use the identity (\ref{contraction_1}), $k_{\beta} U^{\alpha\beta} = - \bar{m}^{\alpha}$, and we can isolate the term $k_{\beta}\nabla_{\alpha}  U^{\alpha\beta}$, such as
\begin{equation}
\label{1a_identidade_Maxwell_1}
 k_{\beta}\nabla_{\alpha}  U^{\alpha\beta} = -\nabla_{\alpha}\bar{m}^{\alpha} - (-l^{\alpha}\bar{m}^{\beta} + l^{\beta}\bar{m}^{\alpha})\nabla_{\alpha}k_{\beta}.
\end{equation}
The term $l^{\beta}\bar{m}^{\alpha}\nabla_{\alpha}k_{\beta}$ can be simplified by,
\begin{equation}
 l^{\beta}\bar{m}^{\alpha}\nabla_{\alpha}k_{\beta} = \bar{m}^{\alpha}\nabla_{\alpha}(l_{\beta}k^{\beta}) - \bar{m}^{\alpha}k^{\beta}\nabla_{\alpha}l_{\beta} = -\bar{m}^{\alpha}k^{\beta}\nabla_{\alpha}l_{\beta}\nonumber
\end{equation}
and put it in the equation (\ref{1a_identidade_Maxwell_1}), we have that
\begin{equation}
\label{1a_identidade_Maxwell_2}
 k_{\beta}\nabla_{\alpha}  U^{\alpha\beta} = -\nabla_{\alpha}\bar{m}^{\alpha} + l^{\alpha}\bar{m}^{\beta}\nabla_{\alpha}k_{\beta} + k^{\beta} \bar{m}^{\alpha}\nabla_{\alpha}l_{\beta}.
\end{equation}
Now we can get another view of the term  $\nabla_{\alpha}\bar{m}^{\alpha}$ using the fact that,
\begin{equation}
\nabla_{\alpha}\bar{m}^{\alpha} =  \nabla_{\alpha}(\gamma^{\alpha\beta}\bar{m}_{\beta}) = \gamma^{\alpha\beta}\nabla_{\alpha} \bar{m}_{\beta} = (-k^{\alpha}l^{\beta}-k^{\beta}l^{\alpha} +  m^{\alpha}\bar{m}^{\beta} + m^{\beta}\bar{m}^{\alpha})\nabla_{\alpha}\bar{m}_{\beta} , \nonumber
\end{equation}
or
\begin{equation}
\nabla_{\alpha}\bar{m}^{\alpha} =  -k^{\alpha}l^{\beta}\nabla_{\alpha}\bar{m}_{\beta}-k^{\beta}l^{\alpha}\nabla_{\alpha}\bar{m}_{\beta} +  m^{\alpha}\bar{m}^{\beta}\nabla_{\alpha}\bar{m}_{\beta} + m^{\beta}\bar{m}^{\alpha}\nabla_{\alpha}\bar{m}_{\beta}. \nonumber
\end{equation}
We have from $\nabla_{\alpha}(\bar{m}^{\beta}\bar{m}_{\beta}) = 0$, where it results in $  \bar{m}^{\beta}\nabla_{\alpha}\bar{m}_{\beta}=0$ and from $\nabla_{\alpha}(\bar{m}^{\beta}k_{\beta})=0$ that results in $k^{\beta}\nabla_{\alpha}\bar{m}_{\beta} = - \bar{m}^{\beta}\nabla_{\alpha}k_{\beta} $, and so on, we have that the above equation reduces to, 
\begin{equation}
\label{divergencia m bar}
\nabla_{\alpha}\bar{m}^{\alpha} =  \bar{m}^{\beta}k^{\alpha}\nabla_{\alpha}l_{\beta} + \bar{m}^{\beta}l^{\alpha}\nabla_{\alpha}k_{\beta} + m^{\beta}\bar{m}^{\alpha}\nabla_{\alpha}\bar{m}_{\beta}. 
\end{equation}
Replacing this term in the equation in the equation (\ref{1a_identidade_Maxwell_2}) we have,
\begin{equation}
 k_{\beta}\nabla_{\alpha}  U^{\alpha\beta} = -(\bar{m}^{\beta}k^{\alpha}\nabla_{\alpha}l_{\beta} + \bar{m}^{\beta}l^{\alpha}\nabla_{\alpha}k_{\beta} + m^{\beta}\bar{m}^{\alpha}\nabla_{\alpha}\bar{m}_{\beta}) + \bar{m}^{\beta}l^{\alpha}\nabla_{\alpha}k_{\beta} + k^{\beta} \bar{m}^{\alpha}\nabla_{\alpha}l_{\beta}, \nonumber
\end{equation}
or
\begin{equation}
 k_{\beta}\nabla_{\alpha}  U^{\alpha\beta} = k^{\beta} \bar{m}^{\alpha}\nabla_{\alpha}l_{\beta} - m^{\beta}\bar{m}^{\alpha} \nabla_{\alpha}\bar{m}_{\beta} - \bar{m}^{\beta}k^{\alpha}\nabla_{\alpha} l_{\beta}  .
\end{equation}
We can verify the spin coefficient displayed in the equation (\ref{spin_coefficient_12}), we have that $2\alpha =  k^{\alpha} \bar{m}^{\beta}\nabla_{\beta}l_{\alpha} -  m^{\alpha}\bar{m}^{\beta} \nabla_{\beta}\bar{m}_{\alpha}$ and from equation (\ref{spin_coefficient_08}) we have $\pi = \bar{m}^{\alpha}k^{\beta}\nabla_{\beta} l_{\alpha}$. Thus the above equation results in
\begin{equation}
\label{1a_identidade_Maxwell_5}
k_{\beta}\nabla_{\alpha}  U^{\alpha\beta} = 2\alpha -\pi.
\end{equation}
Now we have to work with the term  $k_{\beta}\nabla_{\alpha}  W^{\alpha\beta}$ to reduce it in to spin coefficient, where we have that,
\begin{equation}
\label{1a_identidade_Maxwell_6}
   k_{\beta}\nabla_{\alpha}W^{\alpha\beta} =  \nabla_{\alpha}(k_{\beta}W^{\alpha\beta}) -  (\nabla_{\alpha}k_{\beta}) W^{\alpha\beta} = \nabla_{\alpha}k^{\alpha} - (l^{\alpha}k^{\beta} - l^{\beta}k^{\alpha} + m^{\alpha}\bar{m}^{\beta} - m^{\beta}\bar{m}^{\alpha}) \nabla_{\alpha}k_{\beta}. 
\end{equation}
We must simplify the above equation obtaining an identity for $\nabla_{\alpha}k^{\alpha}$ since $ \nabla_{\alpha}k^{\alpha} = \nabla_{\alpha}(\gamma^{\alpha\beta} k_{\beta})= \gamma^{\alpha\beta}\nabla_{\alpha}k_{\beta} $, then we obtain,
\begin{equation}
\label{divergencia k}
 \nabla_{\alpha}k^{\alpha} = (-k^{\alpha}l^{\beta}-k^{\beta}l^{\alpha} +  m^{\alpha}\bar{m}^{\beta} + m^{\beta}\bar{m}^{\alpha})\nabla_{\alpha}k_{\beta} , 
\end{equation}
and replacing the above equation in the equation (\ref{1a_identidade_Maxwell_6}) we obtain,
\begin{equation}
   k_{\beta}\nabla_{\alpha}W^{\alpha\beta} = - 2 l^{\alpha}k^{\beta}\nabla_{\alpha}k_{\beta} + 2m^{\beta}\bar{m}^{\alpha} \nabla_{\alpha}k_{\beta}. 
\end{equation}
We must note that $\nabla_{\alpha}(k_{\beta}k^{\beta})= 2k^{\beta}\nabla_{\alpha}k_{\beta} = 0$ such as that the term $l^{\alpha}k^{\beta} \nabla_{\alpha}k_{\beta} = 0$, and the above equation becomes
\begin{equation}
   k_{\beta}\nabla_{\alpha}W^{\alpha\beta} =   2m^{\beta}\bar{m}^{\alpha} \nabla_{\alpha}k_{\beta}. \nonumber
\end{equation}
We can verify the spin coefficient displayed in the equation (\ref{spin_coefficient_02}), where we have that $m^{\beta}\bar{m}^{\alpha} \nabla_{\alpha}k_{\beta} = -\rho$, and then
\begin{equation}
\label{1a_identidade_Maxwell_7}
   k_{\beta}\nabla_{\alpha}W^{\alpha\beta} = -2\rho.
\end{equation}
Now we have to work with the term  $k_{\beta}\nabla_{\alpha}  V^{\alpha\beta}$ to reduce it in to spin coefficient, where we have that,
\begin{equation}
\label{1a_identidade_Maxwell_8}
   k_{\beta}\nabla_{\alpha}V^{\alpha\beta} =  \nabla_{\alpha}(k_{\beta}V^{\alpha\beta}) - V^{\alpha\beta} (\nabla_{\alpha}k_{\beta})  =  -(k^{\alpha}m^{\beta} - k^{\beta}m^{\alpha}) \nabla_{\alpha}k_{\beta} = -m^{\beta}k^{\alpha}\nabla_{\alpha}k_{\beta} + k^{\beta}m^{\alpha}\nabla_{\alpha}k_{\beta}. 
\end{equation}
Again, we must note that $\nabla_{\alpha}(k_{\beta}k^{\beta})= 2k^{\beta}\nabla_{\alpha}k_{\beta} = 0$ such as that the term $k^{\beta}m^{\alpha} \nabla_{\alpha}k_{\beta} = 0$, and the above equation becomes
\begin{equation}
   k_{\beta}\nabla_{\alpha}V^{\alpha\beta} = -m^{\beta}k^{\alpha}\nabla_{\alpha}k_{\beta}. \nonumber
\end{equation}
We can verify the spin coefficient displayed in the equation (\ref{spin_coefficient_01}), where we have that $ m^{\beta}k^{\alpha}\nabla_{\alpha}k_{\beta} = -\kappa$, and then
\begin{equation}
\label{1a_identidade_Maxwell_9}
   k_{\beta}\nabla_{\alpha}V^{\alpha\beta} = \kappa.
\end{equation}
Thus, replacing the equations (\ref{1a_identidade_Maxwell_5}), (\ref{1a_identidade_Maxwell_7}) and (\ref{1a_identidade_Maxwell_9}) in the first Maxwell equation (\ref{Newman_Penrose_Maxwell_equations_2}) in Newman-Penrose formalism, 
\begin{equation}
D\Phi_{1} - \bar{\delta} \Phi_{0}  +\Phi_{0}(2\alpha-\pi) + \Phi_{1}(-2\rho) +\Phi_{2}(\kappa) = 0. \nonumber
\end{equation}
or then,
\begin{equation}
 \label{Newman_Penrose_Maxwell_equations_3}
 D\Phi_{1} - \bar{\delta} \Phi_{0} = (\pi - 2\alpha)\Phi_{0} + 2\rho \Phi_{1} - \kappa \Phi_{2}.
\end{equation}

%%%%%%%%%%%%%%%%%%%%%%%%%%%%%%%%%%%%%%%%%%%%%%%%%%%%%%%%%%%
%%%%%%%%%%%%%%%%%%%%%%%%%%%%%%%%%%%%%%%%%%%%%%%%%%%%%%%%%%%

For the second Maxwell equation let us  make a projection of equation (\ref{Newman_Penrose_Maxwell_equations_1})  on the direction $l_{\beta}$, where we have,
\begin{equation}
 l_{\beta}\nabla_{\alpha}{\cal F}^{\alpha\beta} = \left(\nabla_{\alpha}\Phi_{0} \right)l_{\beta}U^{\alpha\beta} +\Phi_{0}l_{\beta}\nabla_{\alpha}  U^{\alpha\beta} + \left(\nabla_{\alpha}\Phi_{1} \right)l_{\beta}W^{\alpha\beta} + \Phi_{1}l_{\beta}\nabla_{\alpha} W^{\alpha\beta} + \left(\nabla_{\alpha}\Phi_{2} \right)l_{\beta} V^{\alpha\beta} +\Phi_{2}l_{\beta} \nabla_{\alpha}V^{\alpha\beta} = 0, \nonumber
\end{equation}
where we can use the equations (\ref{contraction_2}) to reduce the above equation to,
\begin{equation}
-l^{\alpha}\nabla_{\alpha}\Phi_{1} + m^{\alpha} \nabla_{\alpha}\Phi_{2}  +\Phi_{0}l_{\beta}\nabla_{\alpha}  U^{\alpha\beta}   + \Phi_{1}l_{\beta}\nabla_{\alpha} W^{\alpha\beta} +\Phi_{2}l_{\beta} \nabla_{\alpha}V^{\alpha\beta} = 0. \nonumber
\end{equation}
We can use the definition of the covariant derivatives in the null directions (\ref{directional_derivatives}) in the above equation where it follows,
\begin{equation}
\label{2a_identidade_Maxwell_1}
-\Delta\Phi_{1} + \delta\Phi_{2}  +\Phi_{0}l_{\beta}\nabla_{\alpha}  U^{\alpha\beta}   + \Phi_{1}l_{\beta}\nabla_{\alpha} W^{\alpha\beta} +\Phi_{2}l_{\beta} \nabla_{\alpha}V^{\alpha\beta} = 0. 
\end{equation}
Now we can work the term $l_{\beta}\nabla_{\alpha}  U^{\alpha\beta}$ to reduce it in to spin coefficient, using the equation
\begin{equation}
 \nabla_{\alpha}(l_{\beta}U^{\alpha\beta}) =  (\nabla_{\alpha}l_{\beta}) U^{\alpha\beta} + l_{\beta}\nabla_{\alpha}U^{\alpha\beta}, \nonumber
\end{equation}
where we can use the identity (\ref{contraction_2}), $l_{\beta} U^{\alpha\beta} = 0$, and we can isolate the term $l_{\beta}\nabla_{\alpha}  U^{\alpha\beta}$, such as
\begin{equation}
  l_{\beta}\nabla_{\alpha}U^{\alpha\beta} = - (-l^{\alpha}\bar{m}^{\beta} + l^{\beta}\bar{m}^{\alpha}) (\nabla_{\alpha}l_{\beta}). \nonumber
\end{equation}
Again, we have that $\bar{m}^{\alpha}\nabla_{\alpha}(l_{\beta}l^{\beta}) = 2 l^{\beta} \bar{m}^{\alpha}\nabla_{\alpha}l_{\beta} =0$. Thus, the above equation reduces to 
\begin{equation}
  l_{\beta}\nabla_{\alpha}U^{\alpha\beta} = l^{\alpha}\bar{m}^{\beta}\nabla_{\alpha}l_{\beta} , \nonumber
\end{equation}
therefore we can verify the spin coefficient displayed in the (\ref{spin_coefficient_05}), where we have that $l^{\alpha}\bar{m}^{\beta}\nabla_{\alpha}l_{\beta} = \nu$, and then,
\begin{equation}
\label{2a_identidade_Maxwell_2}
  l_{\beta}\nabla_{\alpha}U^{\alpha\beta} = \nu. 
\end{equation}
The next term of equation (\ref{2a_identidade_Maxwell_1}) that we can reduce in spin coefficient is $l_{\beta}\nabla_{\alpha}  W^{\alpha\beta}$ by,
\begin{equation}
 \nabla_{\alpha}(l_{\beta}W^{\alpha\beta}) =  (\nabla_{\alpha}l_{\beta}) W^{\alpha\beta} + l_{\beta}\nabla_{\alpha}W^{\alpha\beta}, \nonumber
\end{equation}
with aid of equation (\ref{contraction_2}) and definition (\ref{W1}) we have,
\begin{equation}
 \nabla_{\alpha}(-l^{\alpha}) =  (l^{\alpha}k^{\beta} -l^{\beta}k^{\alpha} + m^{\alpha}\bar{m}^{\beta} - m^{\beta}\bar{m}^{\alpha})\nabla_{\alpha}l_{\beta}  + l_{\beta}\nabla_{\alpha}W^{\alpha\beta}, \nonumber
\end{equation}
or then,
\begin{equation}
\label{2a_identidade_Maxwell_3}
 l_{\beta}\nabla_{\alpha}W^{\alpha\beta} =  -\nabla_{\alpha}l^{\alpha} -  l^{\alpha}k^{\beta}\nabla_{\alpha}l_{\beta} - m^{\alpha}\bar{m}^{\beta}\nabla_{\alpha}l_{\beta} + m^{\beta}\bar{m}^{\alpha}\nabla_{\alpha}l_{\beta}. 
\end{equation}
We can replace the term $\nabla_{\alpha}l^{\alpha}$ by $\nabla_{\alpha}(\gamma^{\alpha\beta}l_{\beta})$ and obtain,
\begin{equation}
 \nabla_{\alpha}l^{\alpha} = \gamma^{\alpha\beta}\nabla_{\alpha}l_{\beta} = (-k^{\alpha}l^{\beta}-k^{\beta}l^{\alpha} +  m^{\alpha}\bar{m}^{\beta} + m^{\beta}\bar{m}^{\alpha})\nabla_{\alpha}l_{\beta}  , \nonumber 
\end{equation}
where we obtain
\begin{equation}
\label{divergencia l}
 \nabla_{\alpha}l^{\alpha} = -k^{\beta}l^{\alpha}\nabla_{\alpha}l_{\beta} +  m^{\alpha}\bar{m}^{\beta}\nabla_{\alpha}l_{\beta} + m^{\beta}\bar{m}^{\alpha}\nabla_{\alpha}l_{\beta}.
 \end{equation}
 Replacing this result in to equation (\ref{2a_identidade_Maxwell_3}) we  have,
\begin{equation}
 l_{\beta}\nabla_{\alpha}W^{\alpha\beta} = -2 \bar{m}^{\beta}m^{\alpha}\nabla_{\alpha}l_{\beta}, \nonumber 
\end{equation}
where this written in terms of spin coefficient (\ref{spin_coefficient_06}), it is  given by,
\begin{equation}
\label{2a_identidade_Maxwell_4}
 l_{\beta}\nabla_{\alpha}W^{\alpha\beta} = -2\mu.
\end{equation} 
The last term of equation (\ref{2a_identidade_Maxwell_1}) that we can reduce in spin coefficients is $l_{\beta}\nabla_{\alpha}  V^{\alpha\beta}$ by,
\begin{equation}
 \nabla_{\alpha}(l_{\beta}V^{\alpha\beta}) =  (\nabla_{\alpha}l_{\beta}) V^{\alpha\beta} + l_{\beta}\nabla_{\alpha}V^{\alpha\beta}, \nonumber
\end{equation}
with aid of equation (\ref{contraction_2}) and definition (\ref{V1}) we have,
\begin{equation}
 \nabla_{\alpha}m^{\alpha} =  (k^{\alpha}m^{\beta} - k^{\beta}m^{\alpha})\nabla_{\alpha}l_{\beta}  + l_{\beta}\nabla_{\alpha}V^{\alpha\beta}, \nonumber
\end{equation}
or then,
\begin{equation}
\label{2a_identidade_Maxwell_5}
 l_{\beta}\nabla_{\alpha}V^{\alpha\beta} =  \nabla_{\alpha}m^{\alpha} -  k^{\alpha}m^{\beta}\nabla_{\alpha}l_{\beta} +  k^{\beta}m^{\alpha}\nabla_{\alpha}l_{\beta}. 
\end{equation}
We can replace the term $\nabla_{\alpha}m^{\alpha}$ by $\nabla_{\alpha}(\gamma^{\alpha\beta}m_{\beta})$ and obtain,
\begin{equation}
 \nabla_{\alpha}m^{\alpha} = \gamma^{\alpha\beta}\nabla_{\alpha}m_{\beta} = (-k^{\alpha}l^{\beta}-k^{\beta}l^{\alpha} +  m^{\alpha}\bar{m}^{\beta} + m^{\beta}\bar{m}^{\alpha})\nabla_{\alpha}m_{\beta}  , \nonumber 
\end{equation}
where we obtain
\begin{equation}
\label{divergencia m}
 \nabla_{\alpha}m^{\alpha} = -k^{\beta}l^{\alpha}\nabla_{\alpha}m_{\beta} -  l^{\beta}k^{\alpha}\nabla_{\alpha}m_{\beta} + \bar{m}^{\beta}m^{\alpha}\nabla_{\alpha}m_{\beta}.
\end{equation}
Replacing this result in to equation (\ref{2a_identidade_Maxwell_5}) we  have,
\begin{eqnarray}
 l_{\beta}\nabla_{\alpha}V^{\alpha\beta} &=& -k^{\beta}l^{\alpha}\nabla_{\alpha}m_{\beta} -  l^{\beta}k^{\alpha}\nabla_{\alpha}m_{\beta} + \bar{m}^{\beta}m^{\alpha}\nabla_{\alpha}m_{\beta} -  k^{\alpha}m^{\beta}\nabla_{\alpha}l_{\beta} +  k^{\beta}m^{\alpha}\nabla_{\alpha}l_{\beta} \cr
  &=&   m^{\beta}l^{\alpha}\nabla_{\alpha}k_{\beta} + m^{\beta}k^{\alpha}\nabla_{\alpha} l_{\beta} + \bar{m}^{\beta}m^{\alpha}\nabla_{\alpha}m_{\beta} - m^{\beta}k^{\alpha}\nabla_{\alpha}l_{\beta}  - l^{\beta}m^{\alpha}\nabla_{\alpha}k_{\beta}\cr
   &=& (\bar{m}^{\beta}m^{\alpha}\nabla_{\alpha}m_{\beta} -l^{\beta}m^{\alpha}\nabla_{\alpha}k_{\beta}) + m^{\beta}l^{\alpha}\nabla_{\alpha}k_{\beta}.  \nonumber 
\end{eqnarray}
The above equation in terms of spin coefficients of the equations (\ref{spin_coefficient_04}) and (\ref{spin_coefficient_10}) is  given by,
\begin{equation}
\label{2a_identidade_Maxwell_6}
 l_{\beta}\nabla_{\alpha}V^{\alpha\beta} = 2\beta-\tau.
\end{equation} 
Thus, replacing the equations (\ref{2a_identidade_Maxwell_2}),  (\ref{2a_identidade_Maxwell_4}) and (\ref{2a_identidade_Maxwell_6}) in the second Maxwell equation (\ref{2a_identidade_Maxwell_1}), we obtain in Newman-Penrose formalism, the below equation
\begin{equation}
 \label{2a_identidade_Maxwell_7}
 \Delta\Phi_{1} - \delta\Phi_{2} = \nu\Phi_{0} - 2\mu\Phi_{1} + (2\beta-\tau)\Phi_{2}.
\end{equation}

%%%%%%%%%%%%%%%%%%%%%%%%%%%%%%%%%%%%%%%%%%%%%%%%%%%%%%%%
%%%%%%%%%%%%%%%%%%%%%%%%%%%%%%%%%%%%%%%%%%%%%%%%%%%%%%%%
%%%%%%%%%%%%%%%%%%%%%%%%%%%%%%%%%%%%%%%%%%%%%%%%%%%%%%%%
%%%%%%%%%%%%%%%%%%%%%%%%%%%%%%%%%%%%%%%%%%%%%%%%%%%%%%%%

For the third Maxwell equation let us  make a projection of equation (\ref{Newman_Penrose_Maxwell_equations_1})  on the direction $m_{\beta}$, where we have,
\begin{equation}
 m_{\beta}\nabla_{\alpha}{\cal F}^{\alpha\beta} = \left(\nabla_{\alpha}\Phi_{0} \right)m_{\beta}U^{\alpha\beta} +\Phi_{0}m_{\beta}\nabla_{\alpha}  U^{\alpha\beta} + \left(\nabla_{\alpha}\Phi_{1} \right)m_{\beta}W^{\alpha\beta} + \Phi_{1}m_{\beta}\nabla_{\alpha} W^{\alpha\beta} + \left(\nabla_{\alpha}\Phi_{2} \right)m_{\beta} V^{\alpha\beta} +\Phi_{2}m_{\beta} \nabla_{\alpha}V^{\alpha\beta} = 0, \nonumber
\end{equation}
where we can use the equations (\ref{contraction_3}) to reduce the above equation to,
\begin{equation}
-l^{\alpha}\nabla_{\alpha}\Phi_{0} + m^{\alpha} \nabla_{\alpha}\Phi_{1}  +\Phi_{0}m_{\beta}\nabla_{\alpha}  U^{\alpha\beta}   + \Phi_{1}m_{\beta}\nabla_{\alpha} W^{\alpha\beta} +\Phi_{2}m_{\beta} \nabla_{\alpha}V^{\alpha\beta} = 0. \nonumber
\end{equation}
We can use the definition of the covariant derivatives in the null directions (\ref{directional_derivatives}) in the above equation where it follows,
\begin{equation}
\label{3a_identidade_Maxwell_1}
\Delta\Phi_{0} - \delta\Phi_{1}   = \Phi_{0}m_{\beta}\nabla_{\alpha}  U^{\alpha\beta}   + \Phi_{1}m_{\beta}\nabla_{\alpha} W^{\alpha\beta} +\Phi_{2}m_{\beta} \nabla_{\alpha}V^{\alpha\beta}. 
\end{equation}
The term $m_{\beta}\nabla_{\alpha}  U^{\alpha\beta}$ can reduced it in to spin coefficients, using the equation
\begin{equation}
 \nabla_{\alpha}(m_{\beta}U^{\alpha\beta}) =  (\nabla_{\alpha}m_{\beta}) U^{\alpha\beta} + m_{\beta}\nabla_{\alpha}U^{\alpha\beta}, \nonumber
\end{equation}
where we can use the identity (\ref{contraction_3}), $m_{\beta} U^{\alpha\beta} = -l^{\alpha}$, and we can isolate the term $m_{\beta}\nabla_{\alpha}  U^{\alpha\beta}$, such as
\begin{equation}
  m_{\beta}\nabla_{\alpha}U^{\alpha\beta} = -\nabla_{\alpha}l^{\alpha} - (-l^{\alpha}\bar{m}^{\beta} + l^{\beta}\bar{m}^{\alpha}) \nabla_{\alpha}m_{\beta}. \nonumber
\end{equation}
We can use equation (\ref{divergencia l}) and the above equation follows 
\begin{eqnarray}
  m_{\beta}\nabla_{\alpha}U^{\alpha\beta} &=& k^{\beta}l^{\alpha}\nabla_{\alpha}l_{\beta} -  m^{\alpha}\bar{m}^{\beta}\nabla_{\alpha}l_{\beta} - m^{\beta}\bar{m}^{\alpha}\nabla_{\alpha}l_{\beta}+  
  l^{\alpha}\bar{m}^{\beta}\nabla_{\alpha}m_{\beta} - l^{\beta}\bar{m}^{\alpha}\nabla_{\alpha}m_{\beta} \cr
  &=& k^{\beta}l^{\alpha}\nabla_{\alpha}l_{\beta} -  \bar{m}^{\beta}m^{\alpha}\nabla_{\alpha}l_{\beta} + l^{\beta}\bar{m}^{\alpha}\nabla_{\alpha}m_{\beta} -  
  m^{\beta}l^{\alpha}\nabla_{\alpha}\bar{m}_{\beta} - l^{\beta}\bar{m}^{\alpha}\nabla_{\alpha}m_{\beta} \cr
  &=& (k^{\beta}l^{\alpha}\nabla_{\alpha}l_{\beta} - m^{\beta}l^{\alpha}\nabla_{\alpha}\bar{m}_{\beta}) -  \bar{m}^{\beta}m^{\alpha}\nabla_{\alpha}l_{\beta} , \nonumber
\end{eqnarray}
therefore the above equation in terms of the spin coefficients of the equations (\ref{spin_coefficient_06}) and (\ref{spin_coefficient_11}) is written as,
\begin{equation}
\label{3a_identidade_Maxwell_2}
  m_{\beta}\nabla_{\alpha}U^{\alpha\beta} = 2\gamma - \mu. 
\end{equation}
The next term of equation (\ref{3a_identidade_Maxwell_1}) that we can reduce in spin coefficient is $m_{\beta}\nabla_{\alpha}  W^{\alpha\beta}$ by,
\begin{equation}
 \nabla_{\alpha}(m_{\beta}W^{\alpha\beta}) =  (\nabla_{\alpha}m_{\beta}) W^{\alpha\beta} + m_{\beta}\nabla_{\alpha}W^{\alpha\beta}, \nonumber
\end{equation}
with aid of equation (\ref{contraction_3}) and definition (\ref{W1}) we have,
\begin{equation}
 \nabla_{\alpha}m^{\alpha} =  (l^{\alpha}k^{\beta} -l^{\beta}k^{\alpha} + m^{\alpha}\bar{m}^{\beta} - m^{\beta}\bar{m}^{\alpha})\nabla_{\alpha}m_{\beta}  + l_{\beta}\nabla_{\alpha}W^{\alpha\beta}, \nonumber
\end{equation}
we can use equation (\ref{divergencia m}) in to above equation and obtain,
\begin{eqnarray}
 m_{\beta}\nabla_{\alpha}W^{\alpha\beta} &=& -k^{\beta}l^{\alpha}\nabla_{\alpha}m_{\beta} -  l^{\beta}k^{\alpha}\nabla_{\alpha}m_{\beta} + \bar{m}^{\beta}m^{\alpha}\nabla_{\alpha}m_{\beta} -  l^{\alpha}k^{\beta}\nabla_{\alpha}l_{\beta} + l^{\beta}k^{\alpha}\nabla_{\alpha}m_{\beta} - m^{\alpha}\bar{m}^{\beta}\nabla_{\alpha}m_{\beta}\cr
 &=& -2k^{\beta}l^{\alpha}\nabla_{\alpha}m_{\beta} ~ = ~ 2m^{\beta}l^{\alpha}\nabla_{\alpha}k_{\beta},
\end{eqnarray}
where this written in terms of spin coefficient (\ref{spin_coefficient_04}) is  given by,
\begin{equation}
\label{3a_identidade_Maxwell_4}
 m_{\beta}\nabla_{\alpha}W^{\alpha\beta} = -2\tau.
\end{equation} 
The last term of equation (\ref{3a_identidade_Maxwell_1}) that we can reduce in spin coefficient is $m_{\beta}\nabla_{\alpha}  V^{\alpha\beta}$ by,
\begin{equation}
 \nabla_{\alpha}(m_{\beta}V^{\alpha\beta}) =  (\nabla_{\alpha}m_{\beta}) V^{\alpha\beta} + m_{\beta}\nabla_{\alpha}V^{\alpha\beta}, \nonumber
\end{equation}
with aid of equation (\ref{contraction_2}) and definition (\ref{V1}) we have,
\begin{equation}
 0 =  (k^{\alpha}m^{\beta} - k^{\beta}m^{\alpha})\nabla_{\alpha}m_{\beta}  + m_{\beta}\nabla_{\alpha}V^{\alpha\beta}, \nonumber
\end{equation}
or then,
\begin{equation}
 m_{\beta}\nabla_{\alpha}V^{\alpha\beta} = - k^{\alpha}m^{\beta}\nabla_{\alpha}m_{\beta} +  k^{\beta}m^{\alpha}\nabla_{\alpha}m_{\beta} =   k^{\beta}m^{\alpha}\nabla_{\alpha}m_{\beta} = -m^{\beta}m^{\alpha}\nabla_{\alpha}k_{\beta} \nonumber
\end{equation}
The above equation written in terms of spin coefficient (\ref{spin_coefficient_03}) is  given by,
\begin{equation}
\label{3a_identidade_Maxwell_6}
 m_{\beta}\nabla_{\alpha}V^{\alpha\beta} = \sigma.
\end{equation} 
Thus, replacing the equations (\ref{3a_identidade_Maxwell_2}),  (\ref{3a_identidade_Maxwell_4}) and (\ref{3a_identidade_Maxwell_6}) in the second Maxwell equation (\ref{3a_identidade_Maxwell_1}), we obtain in Newman-Penrose formalism, the below equation
\begin{equation}
 \label{3a_identidade_Maxwell_7}
 \Delta\Phi_{0} - \delta\Phi_{1} = (2\gamma - \mu)\Phi_{0} - 2\tau\Phi_{1} + \sigma \Phi_{2}.
\end{equation}

%%%%%%%%%%%%%%%%%%%%%%%%%%%%%%%%%%%%%%%%%%%%%%%%%%%%%%%%%%%%%%
%%%%%%%%%%%%%%%%%%%%%%%%%%%%%%%%%%%%%%%%%%%%%%%%%%%%%%%%%%%%%%
%%%%%%%%%%%%%%%%%%%%%%%%%%%%%%%%%%%%%%%%%%%%%%%%%%%%%%%%%%%%%%

For the fourth  Maxwell equation let us  make a projection of equation (\ref{Newman_Penrose_Maxwell_equations_1})  on the direction $\bar{m}_{\beta}$, where we have,
\begin{equation}
 \bar{m}_{\beta}\nabla_{\alpha}{\cal F}^{\alpha\beta} = \left(\nabla_{\alpha}\Phi_{0} \right)\bar{m}_{\beta}U^{\alpha\beta} +\Phi_{0}\bar{m}_{\beta}\nabla_{\alpha}  U^{\alpha\beta} + \left(\nabla_{\alpha}\Phi_{1} \right)\bar{m}_{\beta}W^{\alpha\beta} + \Phi_{1}\bar{m}_{\beta}\nabla_{\alpha} W^{\alpha\beta} + \left(\nabla_{\alpha}\Phi_{2} \right)\bar{m}_{\beta} V^{\alpha\beta} +\Phi_{2}m_{\beta} \nabla_{\alpha}V^{\alpha\beta} = 0, \nonumber
\end{equation}
where we can use the equations (\ref{contraction_4}) to reduce the above equation to,
\begin{equation}
-\bar{m}^{\alpha}\nabla_{\alpha}\Phi_{1} + k^{\alpha} \nabla_{\alpha}\Phi_{2}  +\Phi_{0}\bar{m}_{\beta}\nabla_{\alpha}  U^{\alpha\beta}   + \Phi_{1}\bar{m}_{\beta}\nabla_{\alpha} W^{\alpha\beta} +\Phi_{2}\bar{m}_{\beta} \nabla_{\alpha}V^{\alpha\beta} = 0. \nonumber
\end{equation}
We can use the definition of the covariant derivatives in the null directions (\ref{directional_derivatives}) in the above equation where it follows,
\begin{equation}
\label{4a_identidade_Maxwell_1}
-D\Phi_{2} + \bar{\delta}\Phi_{1}   = \Phi_{0}\bar{m}_{\beta}\nabla_{\alpha}  U^{\alpha\beta}   + \Phi_{1}\bar{m}_{\beta}\nabla_{\alpha} W^{\alpha\beta} +\Phi_{2}\bar{m}_{\beta} \nabla_{\alpha}V^{\alpha\beta}. 
\end{equation}
The term $\bar{m}_{\beta}\nabla_{\alpha}  U^{\alpha\beta}$ can reduced it in to spin coefficient, using the equation
\begin{equation}
 \nabla_{\alpha}(\bar{m}_{\beta}U^{\alpha\beta}) =  (\nabla_{\alpha}\bar{m}_{\beta}) U^{\alpha\beta} + \bar{m}_{\beta}\nabla_{\alpha}U^{\alpha\beta}, \nonumber
\end{equation}
where we can use the identity (\ref{contraction_4}), $\bar{m}_{\beta} U^{\alpha\beta} = 0$, and we can isolate the term $\bar{m}_{\beta}\nabla_{\alpha}  U^{\alpha\beta}$, such as
\begin{equation}
  \bar{m}_{\beta}\nabla_{\alpha}U^{\alpha\beta} =  - (-l^{\alpha}\bar{m}^{\beta} + l^{\beta}\bar{m}^{\alpha}) \nabla_{\alpha}\bar{m}_{\beta} = - l^{\beta}\bar{m}^{\alpha} \nabla_{\alpha}\bar{m}_{\beta} = \bar{m}^{\beta}\bar{m}^{\alpha} \nabla_{\alpha}l_{\beta}, \nonumber
\end{equation}
therefore we can see the spin coefficient of the equation (\ref{spin_coefficient_07}) and write the above equation as,
\begin{equation}
\label{4a_identidade_Maxwell_2}
  \bar{m}_{\beta}\nabla_{\alpha}U^{\alpha\beta} = \lambda. 
\end{equation}
The next term of equation (\ref{4a_identidade_Maxwell_1}) that we can reduce in spin coefficient is $\bar{m}_{\beta}\nabla_{\alpha}  W^{\alpha\beta}$ by,
\begin{equation}
 \nabla_{\alpha}(\bar{m}_{\beta}W^{\alpha\beta}) =  (\nabla_{\alpha}\bar{m}_{\beta}) W^{\alpha\beta} + \bar{m}_{\beta}\nabla_{\alpha}W^{\alpha\beta}, \nonumber
\end{equation}
with aid of equation (\ref{contraction_4}) and definition (\ref{W1}) we have,
\begin{equation}
 -\nabla_{\alpha}\bar{m}^{\alpha} =  (l^{\alpha}k^{\beta} -l^{\beta}k^{\alpha} + m^{\alpha}\bar{m}^{\beta} - m^{\beta}\bar{m}^{\alpha})\nabla_{\alpha}\bar{m}_{\beta}  + \bar{m}_{\beta}\nabla_{\alpha}W^{\alpha\beta}, \nonumber
\end{equation}
we can use equation (\ref{divergencia m bar}) in to above equation and obtain,
\begin{eqnarray}
 \bar{m}_{\beta}\nabla_{\alpha}W^{\alpha\beta} &=& k^{\beta}l^{\alpha}\nabla_{\alpha}\bar{m}_{\beta} +  l^{\beta}k^{\alpha}\nabla_{\alpha}\bar{m}_{\beta} - m^{\beta}\bar{m}^{\alpha}\nabla_{\alpha}\bar{m}_{\beta} -  l^{\alpha}k^{\beta}\nabla_{\alpha}\bar{m}_{\beta} + l^{\beta}k^{\alpha}\nabla_{\alpha}\bar{m}_{\beta} + m^{\beta}\bar{m}^{\alpha}\nabla_{\alpha}\bar{m}_{\beta}\cr
 &=& 2 l^{\beta}k^{\alpha}\nabla_{\alpha}\bar{m}_{\beta} ~ = ~ -2\bar{m}^{\beta}k^{\alpha}\nabla_{\alpha}l_{\beta}  . 
\end{eqnarray}
where this written in terms of spin coefficient (\ref{spin_coefficient_08}) is  given by,
\begin{equation}
\label{4a_identidade_Maxwell_4}
 \bar{m}_{\beta}\nabla_{\alpha}W^{\alpha\beta} = -2\pi.
\end{equation} 
The last term of equation (\ref{4a_identidade_Maxwell_1}) that we can reduce in spin coefficients is $\bar{m}_{\beta}\nabla_{\alpha}  V^{\alpha\beta}$ by,
\begin{equation}
 \nabla_{\alpha}(\bar{m}_{\beta}V^{\alpha\beta}) =  (\nabla_{\alpha}\bar{m}_{\beta}) V^{\alpha\beta} + \bar{m}_{\beta}\nabla_{\alpha}V^{\alpha\beta}, \nonumber
\end{equation}
with aid of equation (\ref{contraction_4}) and definition (\ref{V1}) we have,
\begin{equation}
 \nabla_{\alpha}k^{\alpha} =  (k^{\alpha}m^{\beta} - k^{\beta}m^{\alpha})\nabla_{\alpha}\bar{m}_{\beta}  + \bar{m}_{\beta}\nabla_{\alpha}V^{\alpha\beta}, \nonumber
\end{equation}
with equation (\ref{divergencia k}) we have that,
\begin{eqnarray}
 \bar{m}_{\beta}\nabla_{\alpha}V^{\alpha\beta} &=& - k^{\alpha}l^{\beta}\nabla_{\alpha}k_{\beta}  +  m^{\alpha}\bar{m}^{\beta}\nabla_{\alpha}k_{\beta} + m^{\beta}\bar{m}^{\alpha}\nabla_{\alpha}k_{\beta} -  
 k^{\alpha}m^{\beta}\nabla_{\alpha}\bar{m}_{\beta} +  k^{\beta}m^{\alpha}\nabla_{\alpha}\bar{m}_{\beta} \cr 
 &=& -l^{\beta}k^{\alpha}\nabla_{\alpha}k_{\beta}  -  k^{\beta}m^{\alpha}\nabla_{\alpha}\bar{m}_{\beta} + m^{\beta}\bar{m}^{\alpha}\nabla_{\alpha}k_{\beta} -  
 m^{\beta}k^{\alpha}\nabla_{\alpha}\bar{m}_{\beta} +  k^{\beta}m^{\alpha}\nabla_{\alpha}\bar{m}_{\beta} \cr
 &=& m^{\beta}\bar{m}^{\alpha}\nabla_{\alpha}k_{\beta} - (l^{\beta}k^{\alpha}\nabla_{\alpha}k_{\beta} - \bar{m}^{\beta}k^{\alpha}\nabla_{\alpha}m_{\beta})  \nonumber
\end{eqnarray}
The above equation written in terms of spin coefficients of the equations (\ref{spin_coefficient_02}) and (\ref{spin_coefficient_09}) is  given by,
\begin{equation}
\label{4a_identidade_Maxwell_6}
 \bar{m}_{\beta}\nabla_{\alpha}V^{\alpha\beta} =  - \rho + 2\epsilon.
\end{equation} 
Thus, replacing the equations (\ref{4a_identidade_Maxwell_2}),  (\ref{4a_identidade_Maxwell_4}) and (\ref{4a_identidade_Maxwell_6}) in the fourth Maxwell equation (\ref{4a_identidade_Maxwell_1}), we obtain in Newman-Penrose formalism, the below equation
\begin{equation}
 \label{4a_identidade_Maxwell_7}
  D\Phi_{2} - \bar{\delta}\Phi_{1} = -\lambda\Phi_{0} + 2\pi\Phi_{1} +(\rho-2\epsilon)\Phi_{2}.
\end{equation}

Finally, we can display the four Maxwell equations in vacuum  (\ref{Newman_Penrose_Maxwell_equations_3}), (\ref{2a_identidade_Maxwell_7}), (\ref{3a_identidade_Maxwell_7}) and (\ref{4a_identidade_Maxwell_7}), in terms of the spin coefficients as \cite{Newman},
\begin{equation}
 \begin{cases}
  D\Phi_{1} - \bar{\delta} \Phi_{0} = (\pi - 2\alpha)\Phi_{0} + 2\rho \Phi_{1} - \kappa \Phi_{2},\cr
  D\Phi_{2} - \bar{\delta}\Phi_{1} = -\lambda\Phi_{0} + 2\pi\Phi_{1} +(\rho-2\epsilon)\Phi_{2},\cr
   \Delta\Phi_{0} - \delta\Phi_{1} = (2\gamma - \mu)\Phi_{0} - 2\tau\Phi_{1} + \sigma \Phi_{2},\cr  
    \Delta\Phi_{1} - \delta\Phi_{2} = \nu\Phi_{0} - 2\mu\Phi_{1} + (2\beta-\tau)\Phi_{2}.
 \end{cases}
\end{equation}
For a spacetime to have as its source an electromagnetic  field  satisfying  Maxwell's  source-free  equations, this spacetime satisfies an exact solution of General Relativity called electrovacuum solution of the Einstein field equation or called source-free Einstein-Maxwell solutions \cite{Wytler2}. We will see further how the Einstein-Maxwell equations can be expressed in terms of the dyad components of the electromagnetic field $\Phi_{0}$, $\Phi_{1}$ and $\Phi_{2}$.

%%%%%%%%%%%%%%%%%%%%%%%%%%%%%%%%%%%%%%%%%%%%%%%%%%%%%%%%%%%%%%
%%%%%%%%%%%%%%%%%%%%%%%%%%%%%%%%%%%%%%%%%%%%%%%%%%%%%%%%%%%%%%
%%%%%%%%%%%%%%%%%%%%%%%%%%%%%%%%%%%%%%%%%%%%%%%%%%%%%%%%%%%%%%

%%%%%%%%%%%%%%%%%%%%%%%%%%%%%%%%%%%%%%%%%%%%%%%%%%%%%%%%%%%%%%
%%%%%%%%%%%%%%%%%%%%%%%%%%%%%%%%%%%%%%%%%%%%%%%%%%%%%%%%%%%%%%
%%%%%%%%%%%%%%%%%%%%%%%%%%%%%%%%%%%%%%%%%%%%%%%%%%%%%%%%%%%%%%
%%%%%%%%%%%%%%%%%%%%%%%%%%%%%%%%%%%%%%%%%%%%%%%%%%%%%%%%%%%%%%
%%%%%%%%%%%%%%%%%%%%%%%%%%%%%%%%%%%%%%%%%%%%%%%%%%%%%%%%%%%%%%
%%%%%%%%%%%%%%%%%%%%%%%%%%%%%%%%%%%%%%%%%%%%%%%%%%%%%%%%%%%%%%

\subsection{Energy-momentum tensor of the electromagnetic field}
The energy-momentum tensor is a source of the gravitational field. A gravitational field produced by a source-free electromagnetic field is a exact solution of Einstein's field equations \cite{Kramer,Wytler2}.
The energy-momentum tensor of the electromagnetic field is given by,
\begin{equation}
 T_{\alpha\beta} = \frac{1}{2}~{\cal F}_{\alpha\gamma}{\overline{\cal F}_{\beta}}\,^{\gamma},
\end{equation}
where we have from equation (\ref{bivector_F_6}) that
\begin{equation}
 {\cal F}_{\alpha\gamma} = 2\left(\Phi_{0}U_{\alpha\gamma} + \Phi_{1}W_{\alpha\gamma} + \Phi_{2} V_{\alpha\gamma}\right), \nonumber
\end{equation}
and
\begin{equation}
 {\overline{\cal F}_{\beta}}\,^{\gamma} = 2\left(\overline\Phi_{0}\overline{U}_{\beta}\,^{\gamma} + \overline\Phi_{1}\overline{W}_{\beta}\,^{\gamma} + \overline\Phi_{2} \overline{V}_{\beta}\,^{\gamma}\right), \nonumber
\end{equation}
where we write the complex conjugate as,
\begin{equation}
 \overline{U}_{\beta}\,^{\gamma} = -l_{\beta} m^{\gamma} + l^{\gamma}m_{\beta}, \nonumber
\end{equation}
\begin{equation}
 \overline{V}_{\beta}\,^{\gamma} = k_{\beta} \bar{m}^{\gamma} - k^{\gamma}\bar{m}_{\beta} \nonumber
\end{equation}
and
\begin{equation}
 \overline{W}_{\beta}\,^{\gamma} = l_{\beta} k^{\gamma} - l^{\gamma}k_{\beta} + \bar{m}_{\beta}m^{\gamma} - \bar{m}^{\gamma}m_{\beta}. \nonumber
\end{equation}
So that the energy-momentum tensor tensor can be written as
\begin{eqnarray}
\label{Energy_momentum_tensor_1}
 T_{\alpha\beta} &=& 2\,\bigg(\Phi_{0}\overline\Phi_{0} U_{\alpha\gamma}\overline{U}_{\beta}\,^{\gamma}  + \Phi_{0}\overline\Phi_{1}U_{\alpha\gamma}\overline{W}_{\beta}\,^{\gamma} + \Phi_{0}\overline\Phi_{2}U_{\alpha\gamma}\overline{V}_{\beta}\,^{\gamma}\nonumber \\[2pt]
 & & + \Phi_{1}\overline\Phi_{0}W_{\alpha\gamma}\overline{U}_{\beta}\,^{\gamma} + \Phi_{1}\overline\Phi_{1}W_{\alpha\gamma}\overline{W}_{\beta}\,^{\gamma} + \Phi_{1}\overline\Phi_{2}W_{\alpha\gamma}\overline{V}_{\beta}\,^{\gamma} \nonumber \\[2pt]
 & & +  \Phi_{2}\overline\Phi_{0} V_{\alpha\gamma}\overline{U}_{\beta}\,^{\gamma} +  \Phi_{2}\overline\Phi_{1} V_{\alpha\gamma}\overline{W}_{\beta}\,^{\gamma} + \Phi_{2}\overline\Phi_{2} V_{\alpha\gamma} \overline{V}_{\beta}\,^{\gamma}
 \bigg).
\end{eqnarray}
Let us calculate each term of the energy-momentum tensor tensor. With the only non-null contractions $k_{\alpha}l^{\alpha} =-1$ and $m_{\alpha}\bar{m}^{\alpha}=1$, the terms of the energy-momentum tensor tensor are,
\begin{equation}
  U_{\alpha\gamma}\overline{U}_{\beta}\,^{\gamma} = \left(-l_{\alpha}\bar{m}_{\gamma} + l_{\gamma}\bar{m}_{\alpha} \right)\left( -l_{\beta} m^{\gamma} + l^{\gamma}m_{\beta}\right) = l_{\alpha}l_{\beta}, \nonumber
\end{equation}
\begin{equation}
  U_{\alpha\gamma}\overline{W}_{\beta}\,^{\gamma} = \left(-l_{\alpha}\bar{m}_{\gamma} + l_{\gamma}\bar{m}_{\alpha} \right)\left( l_{\beta} k^{\gamma} - l^{\gamma}k_{\beta} + \bar{m}_{\beta}m^{\gamma} - \bar{m}^{\gamma}m_{\beta}\right) = -l_{\alpha}\bar{m}_{\beta} - \bar{m}_{\alpha}l_{\beta}, \nonumber
\end{equation}
\begin{equation}
  U_{\alpha\gamma}\overline{V}_{\beta}\,^{\gamma} = \left(-l_{\alpha}\bar{m}_{\gamma} + l_{\gamma}\bar{m}_{\alpha} \right)\left(k_{\beta} \bar{m}^{\gamma} - k^{\gamma}\bar{m}_{\beta}\right) = \bar{m}_{\alpha}\bar{m}_{\beta}, \nonumber
\end{equation}
\begin{equation}
 W_{\alpha\gamma}\overline{U}_{\beta}\,^{\gamma} = \left(l_{\alpha}k_{\gamma} - l_{\gamma}k_{\alpha} + m_{\alpha}\bar{m}_{\gamma} - m_{\gamma}\bar{m}_{\alpha}\right)\left(-l_{\beta} m^{\gamma} + l^{\gamma}m_{\beta} \right) = - l_{\alpha} m_{\beta} - m_{\alpha}l_{\beta}, \nonumber
\end{equation}
\begin{equation}
 W_{\alpha\gamma}\overline{W}_{\beta}\,^{\gamma} = \left(l_{\alpha}k_{\gamma} - l_{\gamma}k_{\alpha} + m_{\alpha}\bar{m}_{\gamma} - m_{\gamma}\bar{m}_{\alpha}\right)\left(l_{\beta} k^{\gamma} - l^{\gamma}k_{\beta} + \bar{m}_{\beta}m^{\gamma} - \bar{m}^{\gamma}m_{\beta} \right) = l_{\alpha} k_{\beta} + k_{\alpha} l_{\beta} +m_{\alpha}\bar{m}_{\beta} + \bar{m}_{\alpha}m_{\beta}, \nonumber
\end{equation}
\begin{equation}
 W_{\alpha\gamma}\overline{V}_{\beta}\,^{\gamma} = \left(l_{\alpha}k_{\gamma} - l_{\gamma}k_{\alpha} + m_{\alpha}\bar{m}_{\gamma} - m_{\gamma}\bar{m}_{\alpha}\right)\left(k_{\beta} \bar{m}^{\gamma} - k^{\gamma}\bar{m}_{\beta} \right) = -k_{\alpha} \bar{m}_{\beta} - \bar{m}_{\alpha}k_{\beta}, \nonumber
\end{equation}
\begin{equation}
 V_{\alpha\gamma}\overline{U}_{\beta}\,^{\gamma} = \left(k_{\alpha}m_{\gamma} - k_{\gamma}m_{\alpha}\right)\left(-l_{\beta} m^{\gamma} + l^{\gamma}m_{\beta} \right) =  m_{\alpha} m_{\beta}, \nonumber
\end{equation}
\begin{equation}
 V_{\alpha\gamma}\overline{W}_{\beta}\,^{\gamma} = \left(k_{\alpha}m_{\gamma} - k_{\gamma}m_{\alpha}\right)\left(l_{\beta} k^{\gamma} - l^{\gamma}k_{\beta} + \bar{m}_{\beta}m^{\gamma} - \bar{m}^{\gamma}m_{\beta} \right) = - k_{\alpha} m_{\beta} - k_{\beta}m_{\alpha}, \nonumber
\end{equation}
\begin{equation}
 V_{\alpha\gamma}\overline{V}_{\beta}\,^{\gamma} = \left(k_{\alpha}m_{\gamma} - k_{\gamma}m_{\alpha}\right)\left(k_{\beta} \bar{m}^{\gamma} - k^{\gamma}\bar{m}_{\beta} \right) = k_{\alpha} k_{\beta}. \nonumber
\end{equation}
So returning these values to the equation  (\ref{Energy_momentum_tensor_1}) we have,
\begin{eqnarray}
\label{Energy_momentum_tensor_2}
 T_{\alpha\beta} &=& 2\,\bigg[\Phi_{0}\overline\Phi_{0} l_{\alpha}l_{\beta}  - \Phi_{0}\overline\Phi_{1}\left(l_{\alpha}\bar{m}_{\beta} + \bar{m}_{\alpha}l_{\beta}\right) + \Phi_{0}\overline\Phi_{2}\bar{m}_{\alpha}\bar{m}_{\beta}\nonumber \\[2pt]
 & & - \Phi_{1}\overline\Phi_{0}\left( l_{\alpha} m_{\beta} + m_{\alpha}l_{\beta}\right) + \Phi_{1}\overline\Phi_{1}\left( l_{\alpha} k_{\beta} + k_{\alpha} l_{\beta} +m_{\alpha}\bar{m}_{\beta} + \bar{m}_{\alpha}m_{\beta}\right) - \Phi_{1}\overline\Phi_{2}\left( k_{\alpha} \bar{m}_{\beta} + \bar{m}_{\alpha}k_{\beta}\right) \nonumber \\[2pt]
 & & +  \Phi_{2}\overline\Phi_{0} m_{\alpha} m_{\beta} -  \Phi_{2}\overline\Phi_{1}\left( k_{\alpha} m_{\beta} + k_{\beta}m_{\alpha} \right) + \Phi_{2}\overline\Phi_{2} k_{\alpha} k_{\beta}
 \bigg].
\end{eqnarray}

We should note that for the electromagnetic field that:
\begin{equation}
 {T_{\alpha}}^{\alpha} = 0 \hspace*{1cm}\Longrightarrow \hspace*{1cm}
 {\cal F}_{\alpha\beta} \bar{\cal F}^{\alpha\beta} = 0.
\end{equation}

Now we will see that the energy-momentum tensor of the non-null electromagnetic field can be obtained with $\Phi_{0}=\Phi_{2}=0$, where the electromagnetic field is ${\cal F}_{\alpha\beta} = 2~\Phi_{1}W_{\alpha\beta} $ and the energy-momentum tensor of the electromagnetic field of equation (\ref{Energy_momentum_tensor_2}) reduces to
\begin{equation}
 \label{Energy_momentum_tensor_3}
 T_{\alpha\beta} = 2~\Phi_{1}\overline\Phi_{1}\left( k_{\alpha} l_{\beta}+ l_{\alpha} k_{\beta} + m_{\alpha}\bar{m}_{\beta} + \bar{m}_{\alpha}m_{\beta}\right).
\end{equation}
Let us verify that the  energy-momentum tensor of the non-null electromagnetic field in coordinate basis
\begin{equation}
 \label{Energy_momentum_tensor_4}
 T_{\mu\nu} = 2~\Phi_{1}\overline\Phi_{1}\left( k_{\mu} l_{\nu} + l_{\mu} k_{\nu} + m_{\mu}\bar{m}_{\nu} + \bar{m}_{\mu}m_{\nu}\right),
\end{equation}
where in Minkowski coordinate basis the null tetrad is given by (\ref{coordinate_basis_dual_vector}),
\begin{equation}
 (k_{\mu})= \frac{1}{\sqrt{2}} \begin{pmatrix}
              -1 \cr 1 \cr 0 \cr 0
             \end{pmatrix},
\hspace*{1cm}
 (l_{\mu})= \frac{1}{\sqrt{2}} \begin{pmatrix}
              -1 \cr -1 \cr 0 \cr 0
             \end{pmatrix},
\hspace*{1cm}
 (m_{\mu})= \frac{1}{\sqrt{2}} \begin{pmatrix}
              0 \cr 0 \cr 1 \cr -i
             \end{pmatrix}  
\hspace*{1cm} \mbox{and} \hspace*{1cm} 
 (\bar{m}_{\mu})= \frac{1}{\sqrt{2}} \begin{pmatrix}
              0 \cr 0 \cr 1 \cr i
             \end{pmatrix}, \nonumber
\end{equation}
such that the energy-momentum tensor in the equation (\ref{Energy_momentum_tensor_4}) results in,
\begin{equation}
 \label{Energy_momentum_tensor_5}
 T_{\mu\nu} = 2~\Phi_{1}\overline\Phi_{1}\begin{pmatrix}
                                     1 & 0 & 0 & 0 \cr
                                     0 & -1 & 0 & 0 \cr
                                     0 & 0 & 1 & 0 \cr
                                     0 & 0 & 0 & 1
                                    \end{pmatrix}
     \hspace*{1cm}\Longrightarrow \hspace*{1cm}
    {T_{\mu}}^{\nu} = 2~\Phi_{1}\overline\Phi_{1}\begin{pmatrix}
                                     -1 & 0 & 0 & 0 \cr
                                     0 & -1 & 0 & 0 \cr
                                     0 & 0 & 1 & 0 \cr
                                     0 & 0 & 0 & 1
                                    \end{pmatrix} .
\end{equation}
We must compare this result with the result obtained in references \cite{Wytler2}, where we have in a inertial frame ${\cal O}$,
\begin{equation}
\label{Energy_momentum_tensor_6}
 {T_{\mu}}^{\nu} = \frac{1}{2}\left(E_x^{2} + B_x^{2} \right)
 \begin{pmatrix}
                                     -1 & 0 & 0 & 0 \cr
                                     0 & -1 & 0 & 0 \cr
                                     0 & 0 & 1 & 0 \cr
                                     0 & 0 & 0 & 1
                                    \end{pmatrix} .
\end{equation}
In accordance with reference \cite{Wytler2}, we can obtain an  electric field 3-vector $\bm{E}$  parallel to  magnetic field 3-vector $\bm{B}$  in any inertial frame ${\cal O}$, so that we can identify and confirm with the equation (\ref{dyad_Phi_1}) that,
\begin{equation}
 2~\Phi_{1}\overline\Phi_{1} = \frac{1}{2}\left(E_x^{2} + B_x^{2} \right), 
 \hspace*{1cm} \Phi_{1} = \frac{1}{2}\left(E_x - i B_x \right), 
 \hspace*{1cm} \mbox{and} \hspace*{1cm} \overline\Phi_{1} = \frac{1}{2}\left(E_x + i B_x \right).
\end{equation}
We can use the Lorentz transformations to rotate the complex electromagnetic filed vector {\bf F}  in any desired coordinate system of an inertial frame ${\cal O}'$.

The null electromagnetic field (pure electromagnetic radiation) is obtained when $\Phi_{0} = \Phi_{1} = 0$ implying from (\ref{dyad_Phi_0}) that $E_y = B_z$ and $E_z = -B_y$, and  consequently we have $ \Phi_2 = E_y$. The complex self-dual electromagnetic field is 
${\cal F}_{\alpha\beta} = 2~\Phi_{2} V_{\alpha\beta} $ and the energy-momentum tensor (\ref{Energy_momentum_tensor_2}) results in,
\begin{equation}
 \label{Energy_momentum_tensor_7}
 T_{\alpha\beta} = 2~\Phi_{2}\overline\Phi_{2}~ k_{\alpha} k_{\beta},
\end{equation}
that is a energy-momentum tensor of the null dust with $k_{\alpha}k^{\alpha}=0$ \cite{Kramer}. In accordance with reference \cite{Wytler2}, the solution for null electromagnetic field has electric and magnetic 3-vectors given by $\bm{E} = (0.E_y,0)$ and $\bm{B} = (0,0,B_z)$, with $|\bm{E}| = |\bm{B}|$ therefore we have the energy-momentum tensor of pure electromagnetic radiation (photons) in coordinate basis given by,
\begin{equation}
\label{Energy_momentum_tensor_8}
 T_{\mu\nu} =  \begin{pmatrix}
                        |\bm{E}|^2 & -|\bm{E}|^2 & 0 & 0 \cr
                        -|\bm{E}|^2 &  |\bm{E}|^2 & 0 & 0 \cr
                        0 & 0 & 0 & 0 \cr
                        0 & 0 & 0 & 0
                                    \end{pmatrix} .
\end{equation}

%%%%%%%%%%%%%%%%%%%%%%%%%%%%%%%%%%%%%%%%%%%%%%%%%%%%%%%%%%%%%%
%%%%%%%%%%%%%%%%%%%%%%%%%%%%%%%%%%%%%%%%%%%%%%%%%%%%%%%%%%%%%%
%%%%%%%%%%%%%%%%%%%%%%%%%%%%%%%%%%%%%%%%%%%%%%%%%%%%%%%%%%%%%%
%%%%%%%%%%%%%%%%%%%%%%%%%%%%%%%%%%%%%%%%%%%%%%%%%%%%%%%%%%%%%%
%%%%%%%%%%%%%%%%%%%%%%%%%%%%%%%%%%%%%%%%%%%%%%%%%%%%%%%%%%%%%%
%%%%%%%%%%%%%%%%%%%%%%%%%%%%%%%%%%%%%%%%%%%%%%%%%%%%%%%%%%%%%%

\subsection{The Einstein-Maxwell field equations}

The Ricci tensor from non-coordinate Newman-Penrose basis for coordinate basis with aid of tetrad field is given by,
\begin{equation}
 R_{\mu\nu} = R_{\alpha\beta} {\omega^{\alpha}}_{\mu}{\omega^{\beta}}_{\nu}.
\end{equation}
where we can explicit the  components fo Ricci tensor as,
\begin{eqnarray}
 R_{\mu\nu} = R_{00}{\omega^{0}}_{\mu}{\omega^{0}}_{\nu} + R_{01}{\omega^{0}}_{\mu}{\omega^{1}}_{\nu} + R_{02}{\omega^{0}}_{\mu}{\omega^{2}}_{\nu} + R_{03}{\omega^{0}}_{\mu}{\omega^{3}}_{\nu}\cr
 + R_{10}{\omega^{1}}_{\mu}{\omega^{0}}_{\nu} + R_{11}{\omega^{1}}_{\mu}{\omega^{1}}_{\nu} + R_{12}{\omega^{1}}_{\mu}{\omega^{2}}_{\nu} + R_{13}{\omega^{1}}_{\mu}{\omega^{3}}_{\nu}\cr
  + R_{20}{\omega^{2}}_{\mu}{\omega^{0}}_{\nu} + R_{21}{\omega^{2}}_{\mu}{\omega^{1}}_{\nu} + R_{22}{\omega^{2}}_{\mu}{\omega^{2}}_{\nu} + R_{23}{\omega^{2}}_{\mu}{\omega^{3}}_{\nu} \cr 
  + R_{30}{\omega^{3}}_{\mu}{\omega^{0}}_{\nu} + R_{31}{\omega^{3}}_{\mu}{\omega^{1}}_{\nu} + R_{32}{\omega^{3}}_{\mu}{\omega^{2}}_{\nu} + R_{33}{\omega^{3}}_{\mu}{\omega^{3}}_{\nu}.\nonumber
\end{eqnarray}
We can use,
\begin{equation}
 {\omega^{0}}_{\mu} = -l_{\mu},  \hspace*{1cm} {\omega^{1}}_{\mu} = -k_{\mu},  \hspace*{1cm} {\omega^{2}}_{\mu} = m_{\mu},  \hspace*{1cm} \mbox{and}  \hspace*{1cm} {\omega^{3}}_{\mu} = \bar{m}_{\mu},
\end{equation}
to simplify the Ricci tensor equation as
\begin{eqnarray}
\label{Ricci_tensor_NP_1}
 R_{\mu\nu} &=& R_{00}l_{\mu}l_{\nu} + R_{01}(l_{\mu}k_{\nu} + l_{\nu}k_{\mu}) - R_{02}(l_{\mu}\bar{m}_{\nu} + l_{\nu}\bar{m}_{\mu}) - R_{03}(l_{\mu}m_{\nu}+l_{\nu}m_{\mu}) + R_{11}k_{\mu}k_{\nu}\cr
 & & - R_{12}(k_{\mu}\bar{m}_{\nu} + k_{\nu}\bar{m}_{\mu}) - R_{13}(k_{\mu}m_{\nu}+k_{\nu}m_{\mu}) + R_{22}\bar{m}_{\mu}\bar{m}_{\nu} + R_{23}(\bar{m}_{\mu}m_{\nu}+\bar{m}_{\nu}m_{\mu}) + R_{33}m_{\mu}m_{\nu}.
\end{eqnarray}
The curvature scalar $R$ is obtained by $R=g^{\mu\nu}R_{\mu\nu} = 2R_{01}l_{\mu}k^{\mu} + 2 R_{23}\bar{m}_{\mu}m^{\mu} = 2(R_{23}-R_{01})$.

Through the reduced tensor $S_{\mu\nu}$   defined by \cite{Wytler3},
\begin{equation}
 S_{\mu\nu} = R_{\mu\nu} - \frac{1}{4}g_{\mu\nu}R,
\end{equation}
and with aid of the equation (\ref{Ricci_tensor_NP_1}), we can define the ten independent components of the Ricci tensor terms by the scalar quantities defined as
\begin{equation}
\label{Ricci_Phi_00}
 \Phi_{00} = \frac{1}{2}S_{\mu\nu}k^{\mu}k^{\nu} = \frac{1}{2}\left(R_{\mu\nu}k^{\mu}k^{\nu} - \frac{1}{4}R g_{\mu\nu} k^{\mu}k^{\nu} \right) = \frac{1}{2}R_{00}l_{\mu}l_{\nu}k^{\mu}k^{\nu} = \frac{1}{2} R_{00},
\end{equation}
with
\begin{equation}
 \overline{\Phi}_{00} = \frac{1}{2} R_{00} = \Phi_{00}.
\end{equation}
\begin{equation}
\label{Ricci_Phi_01}
 \Phi_{01} = \frac{1}{2}S_{\mu\nu}k^{\mu}m^{\nu} = \frac{1}{2}\left(R_{\mu\nu}k^{\mu}m^{\nu} - \frac{1}{4}R g_{\mu\nu} k^{\mu}m^{\nu} \right) = \frac{1}{2}R_{02}l_{\mu}\bar{m}_{\nu}k^{\mu}m^{\nu} = \frac{1}{2} R_{02},
\end{equation}
\begin{equation}
 \Phi_{10} = \frac{1}{2}S_{\mu\nu}k^{\mu}\bar{m}^{\nu} = \frac{1}{2}\left(R_{\mu\nu}k^{\mu}\bar{m}^{\nu} - \frac{1}{4}R g_{\mu\nu} k^{\mu}\bar{m}^{\nu} \right) = \frac{1}{2}R_{02}l_{\mu}m_{\nu}k^{\mu}\bar{m}^{\nu} = \frac{1}{2} R_{03},
\end{equation}
with
\begin{equation}
 \overline{\Phi}_{01} = \frac{1}{2} R_{03} = \Phi_{10}.
\end{equation}
\begin{equation}
\label{Ricci_Phi_02}
 \Phi_{02} = \frac{1}{2}S_{\mu\nu}m^{\mu}m^{\nu} = \frac{1}{2}\left(R_{\mu\nu}m^{\mu}m^{\nu} - \frac{1}{4}R g_{\mu\nu} m^{\mu}m^{\nu} \right) = \frac{1}{2}R_{22}\bar{m}_{\mu}\bar{m}_{\nu}m^{\mu}m^{\nu} = \frac{1}{2} R_{22},
\end{equation}
\begin{equation}
 \Phi_{20} = \frac{1}{2}S_{\mu\nu}\bar{m}^{\mu}\bar{m}^{\nu} = \frac{1}{2}\left(R_{\mu\nu}\bar{m}^{\mu}\bar{m}^{\nu} - \frac{1}{4}R g_{\mu\nu} \bar{m}^{\mu}\bar{m}^{\nu} \right) = \frac{1}{2}R_{33}m_{\mu}m_{\nu}\bar{m}^{\mu}\bar{m}^{\nu} = \frac{1}{2} R_{33},
\end{equation}
with
\begin{equation}
 \overline{\Phi}_{02} = \frac{1}{2} R_{33} = \Phi_{20}.
\end{equation}
\begin{equation}
\label{Ricci_Phi_11}
 \Phi_{11} = \frac{1}{4}S_{\mu\nu}(k^{\mu}l^{\nu}+  m^{\mu}\bar{m}^{\nu}) = \frac{1}{4}\left[R_{\mu\nu}(k^{\mu}l^{\nu}+  m^{\mu}\bar{m}^{\nu}) - \frac{1}{4}R g_{\mu\nu} (k^{\mu}l^{\nu}+  m^{\mu}\bar{m}^{\nu}) \right] = \frac{1}{4}\left(R_{01} + R_{23}\right).
\end{equation}
with
\begin{equation}
 \overline{\Phi}_{11} = \frac{1}{4}\left(\bar{R}_{01} + \bar{R}_{23}\right)  =  \left(R_{01} + R_{32}\right) = \Phi_{11}.
\end{equation}
\begin{equation}
\label{Ricci_Phi_12}
 \Phi_{12} = \frac{1}{2}S_{\mu\nu}l^{\mu}m^{\nu} = \frac{1}{2}\left(R_{\mu\nu}l^{\mu}m^{\nu} - \frac{1}{4}R g_{\mu\nu} l^{\mu}m^{\nu} \right) = \frac{1}{2}R_{12}k_{\mu}\bar{m}_{\nu}l^{\mu}m^{\nu} = \frac{1}{2} R_{12},
\end{equation}
\begin{equation}
 \Phi_{21} = \frac{1}{2}S_{\mu\nu}l^{\mu}\bar{m}^{\nu} = \frac{1}{2}\left(R_{\mu\nu}l^{\mu}\bar{m}^{\nu} - \frac{1}{4}R g_{\mu\nu} l^{\mu}\bar{m}^{\nu} \right) = \frac{1}{2}R_{13}k_{\mu}m_{\nu}l^{\mu}\bar{m}^{\nu} = \frac{1}{2} R_{13},
\end{equation}
with
\begin{equation}
 \overline{\Phi}_{12} = \frac{1}{2} R_{13} = \Phi_{21}.
\end{equation}
\begin{equation}
\label{Ricci_Phi_22}
 \Phi_{22} = \frac{1}{4}S_{\mu\nu}l^{\mu}l^{\nu} = \frac{1}{2}\left(R_{\mu\nu}l^{\mu}l^{\nu} - \frac{1}{4}R g_{\mu\nu} l^{\mu}l^{\nu} \right) = \frac{1}{2} R_{11}k_{\mu}k_{\nu}l^{\mu}l^{\nu} = \frac{1}{2} R_{11}.
\end{equation}
with
\begin{equation}
 \overline{\Phi}_{22} = \frac{1}{2} R_{11}  =  \Phi_{22}.
\end{equation}
Thus, we can display these components in matrix form as follows,
\begin{equation}
 (\Phi_{AB}) = \begin{pmatrix}
                \Phi_{00} & \Phi_{01} & \Phi_{02} \cr
                \Phi_{10} & \Phi_{11} & \Phi_{12} \cr
                \Phi_{20} & \Phi_{21} & \Phi_{22} 
               \end{pmatrix} = 
               \begin{pmatrix}
                \Phi_{00} & \Phi_{01} & \Phi_{02} \cr
                \overline{\Phi}_{01} & \Phi_{11} & \Phi_{12} \cr
                \overline{\Phi}_{02} & \overline{\Phi}_{12} & \Phi_{22} 
               \end{pmatrix},
\end{equation}
with the complex components of Ricci tensor satisfying 
$\overline{\Phi}_{AB} = \Phi_{BA}$.

The Einstein-Maxwell field equations is written as \cite{Kramer, Wytler2},
\begin{equation}
\label{Einstein_equation_2}
 R_{\alpha\beta} = 8\pi G T_{\alpha\beta}.
\end{equation}
The electromagnetic energy-momentum tensor obtained in the equation (\ref{Energy_momentum_tensor_2}) is the source of curvature spacetime,  
\begin{eqnarray}
 T_{\alpha\beta} &=& 2\,\bigg[\Phi_{0}\overline\Phi_{0} l_{\alpha}l_{\beta}  - \Phi_{0}\overline\Phi_{1}\left(l_{\alpha}\bar{m}_{\beta} + \bar{m}_{\alpha}l_{\beta}\right) + \Phi_{0}\overline\Phi_{2}\bar{m}_{\alpha}\bar{m}_{\beta}\nonumber \\[2pt]
 & & - \Phi_{1}\overline\Phi_{0}\left( l_{\alpha} m_{\beta} + m_{\alpha}l_{\beta}\right) + \Phi_{1}\overline\Phi_{1}\left( l_{\alpha} k_{\beta} + k_{\alpha} l_{\beta} +m_{\alpha}\bar{m}_{\beta} + \bar{m}_{\alpha}m_{\beta}\right) - \Phi_{1}\overline\Phi_{2}\left( k_{\alpha} \bar{m}_{\beta} + \bar{m}_{\alpha}k_{\beta}\right) \nonumber \\[2pt]
 & & +  \Phi_{2}\overline\Phi_{0} m_{\alpha} m_{\beta} -  \Phi_{2}\overline\Phi_{1}\left( k_{\alpha} m_{\beta} + k_{\beta}m_{\alpha} \right) + \Phi_{2}\overline\Phi_{2} k_{\alpha} k_{\beta}
 \bigg]. \nonumber
\end{eqnarray}

Let us make a contraction $T_{\alpha\beta}k^{\alpha}k^{\beta}$ in the equation (\ref{Einstein_equation_2}), and we have that the only non-null contraction term in the equation of electromagnetic energy-momentum tensor is $T_{\alpha\beta} k^{\alpha}k^{\beta} = 2 \Phi_{0}\overline{\Phi}_{0}$. From  equation (\ref{Ricci_Phi_00}) we have that $R_{\alpha\beta}k^{\alpha}k^{\beta} = 2\Phi_{00}$. Thus we have the first Einstein-Maxwell field equation,
\begin{equation}
 \Phi_{00} = 8\pi G \Phi_{0}\overline{\Phi}_{0}.
\end{equation}

The second contraction is $T_{\alpha\beta}k^{\alpha}m^{\beta}$ in the equation (\ref{Einstein_equation_2}), so that the only non-null contraction term in the equation of electromagnetic energy-momentum tensor is $T_{\alpha\beta} k^{\alpha}m^{\beta} = 2 \Phi_{0}\overline{\Phi}_{1}$. From  equation (\ref{Ricci_Phi_01}) we have that $R_{\alpha\beta}k^{\alpha}m^{\beta} = 2\Phi_{01}$. Thus we have the second Einstein-Maxwell field equation,
\begin{equation}
 \Phi_{01} = 8\pi G \Phi_{0}\overline{\Phi}_{1}.
\end{equation}

The third contraction is $T_{\alpha\beta}m^{\alpha}m^{\beta}$ in the equation (\ref{Einstein_equation_2}), so that the only non-null contraction term in the equation of electromagnetic energy-momentum tensor is $T_{\alpha\beta} m^{\alpha}m^{\beta} = 2 \Phi_{0}\overline{\Phi}_{2}$. From  equation (\ref{Ricci_Phi_02}) we have that $R_{\alpha\beta}m^{\alpha}m^{\beta} = 2\Phi_{02}$. Thus we have the third Einstein-Maxwell field equation,
\begin{equation}
 \Phi_{02} = 8\pi G \Phi_{0}\overline{\Phi}_{2}.
\end{equation}

The fourth contraction is $T_{\alpha\beta}(k^{\alpha}l^{\beta} +m^{\alpha}\bar{m}^{\beta})$ in the equation (\ref{Einstein_equation_2}), so that the only non-null contraction term in the equation of electromagnetic energy-momentum tensor is $T_{\alpha\beta}(k^{\alpha}l^{\beta} +m^{\alpha}\bar{m}^{\beta}) = 4 \Phi_{1}\overline{\Phi}_{1}$. From  equation (\ref{Ricci_Phi_11}) we have that $R_{\alpha\beta}(k^{\alpha}l^{\beta} + m^{\alpha}\bar{m}^{\beta}) = 4\Phi_{11}$. Thus we have the fourth Einstein-Maxwell field equation,
\begin{equation}
 \Phi_{11} = 8\pi G \Phi_{1}\overline{\Phi}_{1}.
\end{equation}

The fifth contraction is $T_{\alpha\beta}l^{\alpha}m^{\beta}$ in the equation (\ref{Einstein_equation_2}), so that the only non-null contraction term in the equation of electromagnetic energy-momentum tensor is $T_{\alpha\beta} l^{\alpha}m^{\beta} = 2 \Phi_{1}\overline{\Phi}_{2}$. From  equation (\ref{Ricci_Phi_12}) we have that $R_{\alpha\beta}l^{\alpha}m^{\beta} = 2\Phi_{12}$. Thus we have the fifth Einstein-Maxwell field equation,
\begin{equation}
 \Phi_{12} = 8\pi G \Phi_{1}\overline{\Phi}_{2}.
\end{equation}

And finally, the sixth contraction is $T_{\alpha\beta}l^{\alpha}l^{\beta}$ in the equation (\ref{Einstein_equation_2}), so that the only non-null contraction term in the equation of electromagnetic energy-momentum tensor is $T_{\alpha\beta} l^{\alpha}l^{\beta} = 2 \Phi_{2}\overline{\Phi}_{2}$. From  equation (\ref{Ricci_Phi_22}) we have that $R_{\alpha\beta}l^{\alpha}m^{\beta} = 2\Phi_{22}$. Thus the sixth Einstein-Maxwell field equation is,
\begin{equation}
 \label{Einstein_Maxwell_NP_Phi_22}
 \Phi_{22} = 8\pi G \Phi_{2}\overline{\Phi}_{2}.
\end{equation}

Summarizing, we have that the Einstein-Maxwell field equations written in Newman-Penrose formalism are given by,
\begin{equation}
 \label{Einstein_Maxwell_NP_1}
 \Phi_{AB} = 8\pi G \Phi_{A}\overline{\Phi}_{B},
\end{equation}
where $A,B=0,1,2$.

%%%%%%%%%%%%%%%%%%%%%%%%%%%%%%%%%%%%%%%%%%%%%
We have two conditions (i) non-null electromagnetic field and (ii) null electromagnetic field. For non-null electromagnetic field, we have seen in the equations (\ref{Energy_momentum_tensor_3}) and (\ref{Energy_momentum_tensor_6}) where we have only $\Phi_{1} \neq 0$, and for null electromagnetic field, we have seen in the equations (\ref{Energy_momentum_tensor_7}) and (\ref{Energy_momentum_tensor_8}) that the only non-zero electromagnetic field component is $\Phi_{2}$.

%%%%%%%%%%%%%%%%%%%%%%%%%%%%%%%%%%%%%%%%%%%%%%%%%%%%%%%%%%%%%%
%%%%%%%%%%%%%%%%%%%%%%%%%%%%%%%%%%%%%%%%%%%%%%%%%%%%%%%%%%%%%%
%%%%%%%%%%%%%%%%%%%%%%%%%%%%%%%%%%%%%%%%%%%%%%%%%%%%%%%%%%%%%%
%%%%%%%%%%%%%%%%%%%%%%%%%%%%%%%%%%%%%%%%%%%%%%%%%%%%%%%%%%%%%%
%%%%%%%%%%%%%%%%%%%%%%%%%%%%%%%%%%%%%%%%%%%%%%%%%%%%%%%%%%%%%%
%%%%%%%%%%%%%%%%%%%%%%%%%%%%%%%%%%%%%%%%%%%%%%%%%%%%%%%%%%%%%%

\section{The complex components of the Weyl curvature tensor in Newman-Penrose formalism}

The Weyl tensor is completely traceless, ${C^{\mu}}_{\lambda\mu\nu}=0$, where the contraction with respect to each pair of indices vanishes, and it has ten independent components.

In addition, similarly to Riemann curvature tensor, we have that Weyl curvature tensor obeys the first Bianchi identity \cite{Hall, Wald, Nakahara},
\begin{equation}
 \label{first_Bianchi_equation}
 C_{\kappa\lambda\mu\nu} +  C_{\kappa\mu\nu\lambda} + C_{\kappa\nu\lambda\mu} = 0.
\end{equation}

\subsection{Duality and bivectors}
Similarly to the case of the electromagnetic field we must introduce the duality concept for the  Weyl curvature tensor. Tensors with two pair of antisymmetric indices like  Riemann and Weyl curvature tensors have two duality operations, the left dual $^{\sim}C_{\kappa\lambda\mu\nu}$ and right $C^{\sim}_{\kappa\lambda\mu\nu}$, defined by \cite{Kramer,Wytler3,Hall},
\begin{equation}
 ^{\sim}C_{\kappa\lambda\mu\nu} = \frac{1}{2}\epsilon_{\kappa\lambda\rho\sigma}{C^{\rho\sigma}}_{\mu\nu} \hspace*{1cm} \mbox{and} \hspace*{1cm} C^{\sim}_{\kappa\lambda\mu\nu} = {C_{\kappa\lambda}}^{\rho\sigma} ~\frac{1}{2}\epsilon_{\mu\nu\rho\sigma}.
\end{equation}
In the same way as the electromagnetic field, we introduce the complex Weyl tensor,
\begin{equation}
 {{\cal C}^{*}}_{\kappa\lambda\mu\nu} = C_{\kappa\lambda\mu\nu} + iC^{\sim}_{\kappa\lambda\mu\nu},
\end{equation}
such that it is self-dual, 
\begin{equation}
  ^{\sim}{{\cal C}^{*}}_{\kappa\lambda\mu\nu} = -i{{\cal C}^{*}}_{\kappa\lambda\mu\nu}.
\end{equation}

We have described the complex electromagnetic field ${\cal F}_{\mu\nu}$ as a linear combination of the self-dual base of complex bivectors $U_{\alpha\beta},V_{\alpha\beta}$ and $W_{\alpha\beta}$ in accordance with (\ref{bivector_F_6}), where we have 
$\frac{1}{2}{\cal F}_{\alpha\beta} = \Phi_{0}U_{\alpha\beta} + \Phi_{1}W_{\alpha\beta} + \Phi_{2} V_{\alpha\beta}$. In the same way we can express the Weyl tensor as a linear combination of the self-dual base of complex bivectors $U_{\alpha\beta},V_{\alpha\beta}$ and $W_{\alpha\beta}$.
For this proposal it is necessary to note that the Weyl tensor is completely traceless, ${C^{\mu}}_{\lambda\mu\nu}=0$. 

We have to write double combinations of complex bivectors $U_{\alpha\beta},V_{\alpha\beta}$ and $W_{\alpha\beta}$ such that these combinations yield the traceless conditions of Weyl tensor. The first double combination that we can try is $U_{\alpha\beta}U_{\gamma\delta}$, where from equation (\ref{U1}) we have that $U_{\alpha\beta} = -l_{\alpha}\bar{m}_{\beta}+l_{\beta}\bar{m}_{\alpha}$ and 
the below contraction results in,
\begin{equation}
 {U^{\alpha}}_{\beta}U_{\alpha\delta} = (-l^{\alpha}\bar{m}_{\beta}+l_{\beta}\bar{m}^{\alpha} ) (-l_{\alpha}\bar{m}_{\delta}+l_{\delta}\bar{m}_{\alpha}) = 0,
\end{equation}
recalling that the only non-null contractions between the tetrad field are $k_{\alpha}l^{\alpha} = -1$ and $m_{\alpha}\bar{m}^{\alpha} =1$.

The second double combination that we can try is $V_{\alpha\beta}V_{\gamma\delta}$, where from equation (\ref{V1}) we have that $V_{\alpha\beta} = k_{\alpha}m_{\beta}-k_{\beta}m_{\alpha}$, 
the below contraction results in,
\begin{equation}
 {V^{\alpha}}_{\beta}V_{\alpha\delta} = (k^{\alpha}m_{\beta}-k_{\beta}m^{\alpha} ) (k_{\alpha}m_{\delta}-k_{\delta}m_{\alpha}) = 0.
\end{equation}

The third double combination that we can try is $W_{\alpha\beta}W_{\gamma\delta}$, where from equation (\ref{W1}) we have that $W_{\alpha\beta} = l_{\alpha}k_{\beta} - l_{\beta}k_{\alpha} + m_{\alpha}\bar{m}_{\beta} - m_{\beta}\bar{m}_{\alpha}$, 
the below contraction results in,
\begin{equation}
 {W^{\alpha}}_{\beta}W_{\alpha\delta} = (l^{\alpha}k_{\beta} - l_{\beta}k^{\alpha} + m^{\alpha}\bar{m}_{\beta} - m_{\beta}\bar{m}^{\alpha}) (l_{\alpha}k_{\delta} - l_{\delta}k_{\alpha} + m_{\alpha}\bar{m}_{\delta} - m_{\delta}\bar{m}_{\alpha}) = k_{\beta}l_{\delta} + k_{\delta}l_{\beta} -m_{\beta}\bar{m}_{\delta} - m_{\delta}\bar{m}_{\beta},
\end{equation}
but this contraction is not null. However we can combine with more two terms,
\begin{equation}
 {U^{\alpha}}_{\beta}V_{\alpha\delta} = (-l^{\alpha}\bar{m}_{\beta}+l_{\beta}\bar{m}^{\alpha} ) (k_{\alpha}m_{\delta}-k_{\delta}m_{\alpha}) = -k_{\delta}l_{\beta} + m_{\delta}\bar{m}_{\beta},
\end{equation}
and
\begin{equation}
 {V^{\alpha}}_{\beta}U_{\alpha\delta} = (k^{\alpha}m_{\beta}-k_{\beta}m^{\alpha} )(-l_{\alpha}\bar{m}_{\delta}+l_{\delta}\bar{m}_{\alpha}) = -k_{\beta}l_{\delta} + m_{\beta}\bar{m}_{\delta}.
\end{equation}
Thus we can combine the three above equations resulting in,
\begin{equation}
 {U^{\alpha}}_{\beta}V_{\alpha\delta} + {V^{\alpha}}_{\beta}U_{\alpha\delta} + {W^{\alpha}}_{\beta}W_{\alpha\delta} = 0,
\end{equation}
such that it is a base term to use in Weyl tensor combination.

The fourth double combination that we have is $U_{\alpha\beta}W_{\gamma\delta} + W_{\alpha\beta}U_{\gamma\delta}$, where the contraction results in,
\begin{eqnarray}
 {U^{\alpha}}_{\beta}W_{\alpha\delta} + {W^{\alpha}}_{\beta}U_{\alpha\delta} &=& (-l^{\alpha}\bar{m}_{\beta}+l_{\beta}\bar{m}^{\alpha})(l_{\alpha}k_{\delta} - l_{\delta}k_{\alpha} + m_{\alpha}\bar{m}_{\delta} - m_{\delta}\bar{m}_{\alpha})\cr 
 & & +  (l^{\alpha}k_{\beta} - l_{\beta}k^{\alpha} + m^{\alpha}\bar{m}_{\beta} - m_{\beta}\bar{m}^{\alpha})(-l_{\alpha}\bar{m}_{\delta}+l_{\delta}\bar{m}_{\alpha}) \cr
 &=& (l_{\beta}\bar{m}_{\delta} - l_{\delta}\bar{m}_{\beta}) + (-l_{\beta}\bar{m}_{\delta} + l_{\delta}\bar{m}_{\beta}) = 0.
\end{eqnarray}

The fifth and last independent double combination that we have is $V_{\alpha\beta}W_{\gamma\delta} + W_{\alpha\beta}V_{\gamma\delta}$, where the contraction results in,
\begin{eqnarray}
 {V^{\alpha}}_{\beta}W_{\alpha\delta} + {W^{\alpha}}_{\beta}V_{\alpha\delta} &=& (k^{\alpha}m_{\beta}-k_{\beta}m^{\alpha})(l_{\alpha}k_{\delta} - l_{\delta}k_{\alpha} + m_{\alpha}\bar{m}_{\delta} - m_{\delta}\bar{m}_{\alpha})\cr 
 & & +  (l^{\alpha}k_{\beta} - l_{\beta}k^{\alpha} + m^{\alpha}\bar{m}_{\beta} - m_{\beta}\bar{m}^{\alpha})(k_{\alpha}m_{\delta}-k_{\delta}m_{\alpha}) \cr
 &=& (k_{\beta}m_{\delta} - k_{\delta}m_{\beta}) - (k_{\beta}m_{\delta} - k_{\delta}m_{\beta}) = 0.
\end{eqnarray}

So, the complex Weyl tensor ${\cal C}^{*}_{\alpha\beta\gamma\delta}$ are written as a linear combination,
\begin{eqnarray}
\label{complex_Weyl_NP_1}
 \frac{1}{2}{\cal C}^{*}_{\alpha\beta\gamma\delta} &=& \Psi_{0} U_{\alpha\beta}U_{\gamma\delta} + \Psi_{1}(U_{\alpha\beta}W_{\gamma\delta} + W_{\alpha\beta}U_{\gamma\delta}) +\Psi_{2} (U_{\alpha\beta}V_{\alpha\delta} + V_{\alpha\beta}U_{\alpha\delta} + W_{\alpha\beta}W_{\alpha\delta}) \cr 
 & & +\Psi_{3}(V_{\alpha\beta}W_{\gamma\delta} + W_{\alpha\beta}V_{\gamma\delta}) + \Psi_{4} V_{\alpha\beta}V_{\gamma\delta}.
\end{eqnarray}
There are ten independent components in the five complex terms $\Psi_{0},\Psi_{1}, \Psi_{2}, \Psi_{3}$ and $\Psi_{4}$ and we can obtain these components by contractions with the respective bivectors of the base. 
To get the first component $\Psi_{0}$ we can contract above equation (\ref{complex_Weyl_NP_1}) with $V^{\alpha\beta}V^{\gamma\delta}$, where we have from equation (\ref{contracao UV}), $U_{\alpha\beta}V^{\alpha\beta} =2$, and the others contractions in equations of Weyl tensor (\ref{complex_Weyl_NP_1}) are zeros. Then we have,
\begin{equation}
 \frac{1}{2}{\cal C}^{*}_{\alpha\beta\gamma\delta}V^{\alpha\beta}V^{\gamma\delta} = \Psi_{0} U_{\alpha\beta}U_{\gamma\delta}V^{\alpha\beta}V^{\gamma\delta} = 4\Psi_{0}, \nonumber
\end{equation}
or
\begin{equation}
 \label{Psi_0_1}
 \Psi_{0} = \frac{1}{8} {\cal C}^{*}_{\alpha\beta\gamma\delta}V^{\alpha\beta}V^{\gamma\delta}.
\end{equation}
Similarly to the case of the electromagnetic field, we also have that 
\begin{equation}
\label{identidade_I_C_1}
 I_{\alpha\beta\gamma\delta} {C_{\epsilon\zeta}}^{\gamma\delta} = \frac{1}{4} \left( g_{\alpha\beta\gamma\delta} + i \eta_{\alpha\beta\gamma\delta} \right){C_{\epsilon\zeta}}^{\gamma\delta} = \frac{1}{2} \left( C_{\epsilon\zeta\alpha\beta} + i C^{\sim}_{\epsilon\zeta\alpha\beta}\right) = \frac{1}{2} {\cal C}^{*}_{\epsilon\zeta\alpha\beta}.
\end{equation}
We can contract the above equation with $V^{\epsilon\zeta}V^{\alpha\beta}$ such that,
\begin{equation}
 V^{\epsilon\zeta}V^{\alpha\beta}I_{\alpha\beta\gamma\delta} {C_{\epsilon\zeta}}^{\gamma\delta} =\frac{1}{2} {\cal C}^{*}_{\epsilon\zeta\alpha\beta}V^{\epsilon\zeta}V^{\alpha\beta} = 4\Psi_{0}, \nonumber
\end{equation}
with aid of equation (\ref{identidade_I_1}) where $V^{\alpha\beta}I_{\alpha\beta\gamma\delta} = V_{\gamma\delta}$, the above equation reduces to
\begin{equation}
 \label{Psi_0_2}
 \Psi_{0} = \frac{1}{4} C_{\alpha\beta\gamma\delta}V^{\alpha\beta}V^{\gamma\delta}.
\end{equation}
Thus we can rewrite it em terms of the complex null-tetrad basis as
\begin{equation}
 \Psi_{0} = \frac{1}{4} C_{\alpha\beta\gamma\delta}(-k^{\alpha}m^{\beta}+ k^{\beta}m^{\alpha}) (-k^{\gamma}m^{\delta}+ k^{\delta}m^{\gamma}) = \frac{1}{4} C_{\alpha\beta\gamma\delta}(-2k^{\alpha}m^{\beta}) (-2k^{\gamma}m^{\delta}), \nonumber
\end{equation}
such that
\begin{equation}
 \label{Psi_0_3}
 \Psi_{0} =  C_{\alpha\beta\gamma\delta} k^{\alpha}m^{\beta}k^{\gamma}m^{\delta}.
\end{equation}
For above Weyl component, in the pseudo-orthornormal Newman-Penrose non-coordinate basis (\ref{NP_basis_vector}), the non-null components become when $\alpha=0$, $\beta = 2$, $\gamma=0$ and $\delta =2$, resulting in
\begin{equation}
 \label{Psi_0_4}
 \Psi_{0} =  C_{0202} = C_{2020}.
\end{equation}
The conjugate complex yelds $\bar{\Psi}_{0} = C_{0303} = C_{3030}$.

Now to get the component $\Psi_{4}$ let us contract the equation (\ref{complex_Weyl_NP_1}) with $U^{\alpha\beta}U^{\gamma\delta}$, where we have from equation (\ref{contracao UV}), $U_{\alpha\beta}V^{\alpha\beta} = 2$, and the others contractions in equations of Weyl tensor (\ref{complex_Weyl_NP_1}) are zeros. Then we have,
\begin{equation}
 \frac{1}{2}{\cal C}^{*}_{\alpha\beta\gamma\delta}U^{\alpha\beta}U^{\gamma\delta} = \Psi_{4} V_{\alpha\beta}V_{\gamma\delta}U^{\alpha\beta}U^{\gamma\delta} = 4\Psi_{4}, \nonumber
\end{equation}
or 
\begin{equation}
 \label{Psi_4_1}
 \Psi_{4} = \frac{1}{8} {\cal C}^{*}_{\alpha\beta\gamma\delta}U^{\alpha\beta}U^{\gamma\delta}.
\end{equation}
In the same way, we obtain to Weyl component $\Phi_{4}$ the below equation,
\begin{equation}
 \label{Psi_4_2}
 \Psi_{4} =  C_{\alpha\beta\gamma\delta} l^{\alpha}\bar{m}^{\beta}l^{\gamma}\bar{m}^{\delta}.
\end{equation}
And for above Weyl component, in the pseudo-orthornormal Newman-Penrose non-coordinate basis (\ref{NP_basis_vector}), the non-null components become when $\alpha=1$, $\beta = 3$, $\gamma=1$ and $\delta =3$, resulting in
\begin{equation}
 \label{Psi_4_3}
 \Psi_{4} =  C_{1313} = C_{3131}.
\end{equation}
The conjugate complex yelds $\bar{\Psi}_{4} = C_{1212} = C_{2121}$.

%%%%%%%%%%%%%%%%%%%%%%%%%%%%%%%%%%%%%%%%%%%%%%%%%%%%%%%%%%

Now to get the component $\Psi_{1}$ let us contract the equation (\ref{complex_Weyl_NP_1}) with $V^{\alpha\beta}W^{\gamma\delta} + W^{\alpha\beta}V^{\gamma\delta}$, where we have from equations (\ref{contracao UV}), $U_{\alpha\beta}V^{\alpha\beta} = 2$, and (\ref{contracao WW}), $W_{\alpha\beta}W^{\alpha\beta} =-4$
and the others contractions in equations of Weyl tensor (\ref{complex_Weyl_NP_1}) are zeros, we have,
\begin{eqnarray}
 \frac{1}{2}{\cal C}^{*}_{\alpha\beta\gamma\delta}(V^{\alpha\beta}W^{\gamma\delta} + W^{\alpha\beta}V^{\gamma\delta}) = \Psi_{1}(U_{\alpha\beta}W_{\gamma\delta} + W_{\alpha\beta}U_{\gamma\delta})(V^{\alpha\beta}W^{\gamma\delta} + W^{\alpha\beta}V^{\gamma\delta}) = -16 \Psi_{1}, \nonumber
\end{eqnarray}
where it results in
\begin{equation}
 \label{Psi_1_1}
 \Psi_{1} = -\frac{1}{16}  {\cal C}^{*}_{\alpha\beta\gamma\delta}V^{\alpha\beta}W^{\gamma\delta}.
\end{equation}
We can use the identity of the equation (\ref{identidade_I_C_1}),  
${\cal C}^{*}_{\alpha\beta\gamma\delta} = 2 I_{\gamma\delta\epsilon\zeta} {C_{\alpha\beta}}^{\epsilon\zeta}$, into above equation (\ref{Psi_1_1}) where it results,
\begin{equation}
 \Psi_{1} = -\frac{1}{16} \left(2 I_{\gamma\delta\epsilon\zeta} {C_{\alpha\beta}}^{\epsilon\zeta}\right) V^{\alpha\beta}W^{\gamma\delta} = -\frac{1}{8} {C_{\alpha\beta}}^{\epsilon\zeta}V^{\alpha\beta}\left(I_{\gamma\delta\epsilon\zeta} W^{\gamma\delta}\right) = -\frac{1}{8} {C_{\alpha\beta}}^{\epsilon\zeta}V^{\alpha\beta}W_{\epsilon\zeta}, \nonumber
\end{equation}
or
\begin{equation}
 \label{Psi_1_2}
 \Psi_{1} = -\frac{1}{8}  C_{\alpha\beta\gamma\delta}V^{\alpha\beta}W^{\gamma\delta}. 
\end{equation}
We also have that the above equation can explicit in,
\begin{equation}
 \Psi_{1} = -\frac{1}{8}  C_{\alpha\beta\gamma\delta}\left(k^{\alpha}m^{\beta} - k^{\beta}m^{\alpha}\right)\left(l^{\gamma}k^{\delta} - l^{\delta}k^{\gamma} + m^{\gamma}\bar{m}^{\delta} - m^{\delta}\bar{m}^{\gamma}\right) = -\frac{1}{8}  C_{\alpha\beta\gamma\delta}\left(2k^{\alpha}m^{\beta} \right)\left(-2k^{\gamma}l^{\delta} - 2\bar{m}^{\gamma}m^{\delta} \right),\nonumber
\end{equation}
or 
\begin{equation}
 \label{Psi_1_3}
 \Psi_{1} = \frac{1}{2}  C_{\alpha\beta\gamma\delta}\left(k^{\alpha}m^{\beta} k^{\gamma}l^{\delta} + k^{\alpha}m^{\beta}\bar{m}^{\gamma}m^{\delta}\right).
\end{equation}
We can reduce the above equation, by traceless, $\gamma^{\alpha\gamma}C_{\alpha\beta\gamma\delta} = C_{\beta\delta}=0$ and with (\ref{components_g_1}),
\begin{equation}
\gamma^{\alpha\gamma}C_{\alpha\beta\gamma\delta} = \left[-\left(l^{\alpha}k^{\gamma} + l^{\gamma}k^{\alpha} \right)+
 \left(m^{\alpha}\bar{m}^{\gamma} + m^{\gamma}\bar{m}^{\alpha} \right)\right]C_{\alpha\beta\gamma\delta} = 0, \nonumber
\end{equation}
where it becomes
\begin{equation}
\left(l^{\alpha}k^{\gamma} + l^{\gamma}k^{\alpha} \right)C_{\alpha\beta\gamma\delta} =
 \left(m^{\alpha}\bar{m}^{\gamma} + m^{\gamma}\bar{m}^{\alpha} \right)C_{\alpha\beta\gamma\delta} , \nonumber
\end{equation}
with change of indices we obtain
\begin{equation}
\label{identidade_C_2}
l^{\alpha}k^{\gamma} \left(C_{\alpha\beta\gamma\delta} + C_{\gamma\beta\alpha\delta}\right) =
\bar{m}^{\alpha}m^{\gamma} \left(C_{\alpha\beta\gamma\delta} + C_{\gamma\beta\alpha\delta}\right)  ,
\end{equation}
Now, we can multiply the above equation by $k^{\beta}m^{\delta}$ where it results in
\begin{equation}
l^{\alpha}k^{\gamma}k^{\beta}m^{\delta} \left(C_{\alpha\beta\gamma\delta} + C_{\gamma\beta\alpha\delta}\right) =
\bar{m}^{\alpha}m^{\gamma}k^{\beta}m^{\delta} \left(C_{\alpha\beta\gamma\delta} + C_{\gamma\beta\alpha\delta}\right),\nonumber
\end{equation}
noting that $l^{\alpha}k^{\gamma}k^{\beta}m^{\delta}C_{\gamma\beta\alpha\delta} =0$ and $\bar{m}^{\alpha}m^{\gamma}k^{\beta}m^{\delta}C_{\alpha\beta\gamma\delta} = 0$ such as,
\begin{equation}
k^{\gamma}m^{\delta}l^{\alpha}k^{\beta} C_{\alpha\beta\gamma\delta}  =
k^{\beta}m^{\gamma} \bar{m}^{\alpha}m^{\delta} C_{\gamma\beta\alpha\delta},\nonumber
\end{equation}
with the properties $C_{\alpha\beta\gamma\delta}=C_{\gamma\delta\alpha\beta}=-C_{\gamma\delta\beta\alpha} $ and  $C_{\gamma\beta\alpha\delta}= - C_{\beta\gamma\alpha\delta}$ the above equation is rewriten as 
\begin{equation}
k^{\gamma}m^{\delta}k^{\beta}l^{\alpha} C_{\gamma\delta\beta\alpha}   =
k^{\beta}m^{\gamma} \bar{m}^{\alpha}m^{\delta} C_{\beta\gamma\alpha\delta}\nonumber
\end{equation}
On the left side of the above equation we choose the indices $\gamma \rightarrow \alpha$, $\delta \rightarrow \beta$, $\beta \rightarrow \gamma$ and $\alpha \rightarrow \delta$. On right side we choose $\beta \rightarrow \alpha$, $\gamma \rightarrow \beta$ and $\alpha \rightarrow \gamma$, where it follows
\begin{equation}
k^{\alpha}m^{\beta}k^{\gamma}l^{\delta} C_{\alpha\beta\gamma\delta}   =
k^{\alpha}m^{\beta} \bar{m}^{\gamma}m^{\delta} C_{\alpha\beta\gamma\delta}.
\end{equation}
Now we can return with this equation to equation (\ref{Psi_1_3}) to  write two results for component $\Psi_{1}$,
\begin{equation}
 \label{Psi_1_4}
 \Psi_{1} =  C_{\alpha\beta\gamma\delta} k^{\alpha}m^{\beta}k^{\gamma}l^{\delta}
\end{equation}
and
\begin{equation}
 \label{Psi_1_5}
 \Psi_{1} =  C_{\alpha\beta\gamma\delta} k^{\alpha}m^{\beta} \bar{m}^{\gamma}m^{\delta}.
\end{equation}
It is still possible to write the equation (\ref{Psi_1_4}) with the property $C_{\alpha\beta\gamma\delta} = C_{\gamma\delta\alpha\beta}$ as
\begin{equation}
  \Psi_{1} =  C_{\gamma\delta\alpha\beta} k^{\gamma}l^{\delta}k^{\alpha}m^{\beta}, \nonumber
\end{equation}
and we can choose the indices $\alpha \leftrightarrow \gamma$ and $\beta \leftrightarrow \delta$ to obtain,
\begin{equation}
 \label{Psi_1_6}
 \Psi_{1} =  C_{\alpha\beta\gamma\delta} k^{\alpha}l^{\beta}k^{\gamma}m^{\delta}.
\end{equation}
In Newman-Penrose complex null tetrads we from equations (\ref{Psi_1_4}),(\ref{Psi_1_5}) and (\ref{Psi_1_6}) that,
\begin{equation}
 \label{Psi_1_7}
 \Psi_{1} = C_{0201} =  C_{0102} = C_{0232} = C_{3202}.
\end{equation}

%%%%%%%%%%%%%%%%%%%%%%%%%%%%%%%%%%%
%%%%%%%%%%%%%%%%%%%%%%%%%%%%%%%%%%%

Now to get the component $\Psi_{3}$ let us contract the equation (\ref{complex_Weyl_NP_1}) with $U^{\alpha\beta}W^{\gamma\delta} + W^{\alpha\beta}U^{\gamma\delta}$, where we have from equations (\ref{contracao UV}), $U_{\alpha\beta}V^{\alpha\beta} = 2$, and (\ref{contracao WW}), $W_{\alpha\beta}W^{\alpha\beta} =-4$
and the others contractions in equations of Weyl tensor (\ref{complex_Weyl_NP_1}) are zeros, then we have,
\begin{eqnarray}
 \frac{1}{2}{\cal C}^{*}_{\alpha\beta\gamma\delta}(U^{\alpha\beta}W^{\gamma\delta} + W^{\alpha\beta}U^{\gamma\delta}) = \Psi_{3}(V_{\alpha\beta}W_{\gamma\delta} + W_{\alpha\beta}V_{\gamma\delta})(U^{\alpha\beta}W^{\gamma\delta} + W^{\alpha\beta}U^{\gamma\delta}) = -16 \Psi_{3}, \nonumber
\end{eqnarray}
where it results in
\begin{equation}
 \label{Psi_3_1}
 \Psi_{3} = -\frac{1}{16}  {\cal C}^{*}_{\alpha\beta\gamma\delta}U^{\alpha\beta}W^{\gamma\delta} 
\end{equation}
We can use the identity of the equation (\ref{identidade_I_C_1}),  
${\cal C}^{*}_{\alpha\beta\gamma\delta} = 2 I_{\gamma\delta\epsilon\zeta} {C_{\alpha\beta}}^{\epsilon\zeta}$, into above equation (\ref{Psi_3_1}) where it results,
\begin{equation}
 \Psi_{3} = -\frac{1}{16} \left(2 I_{\gamma\delta\epsilon\zeta} {C_{\alpha\beta}}^{\epsilon\zeta}\right) U^{\alpha\beta}W^{\gamma\delta} = -\frac{1}{8} {C_{\alpha\beta}}^{\epsilon\zeta}U^{\alpha\beta}\left(I_{\gamma\delta\epsilon\zeta} W^{\gamma\delta}\right) = -\frac{1}{8} {C_{\alpha\beta}}^{\epsilon\zeta}U^{\alpha\beta}W_{\epsilon\zeta}, \nonumber
\end{equation}
or
\begin{equation}
 \label{Psi_3_2}
 \Psi_{3} = -\frac{1}{8}  C_{\alpha\beta\gamma\delta}U^{\alpha\beta}W^{\gamma\delta}. 
\end{equation}
We also have that the above equation can explicit in,
\begin{equation}
 \Psi_{3} = -\frac{1}{8}  C_{\alpha\beta\gamma\delta}\left(-l^{\alpha}\bar{m}^{\beta} + l^{\beta}\bar{m}^{\alpha}\right)\left(l^{\gamma}k^{\delta} - l^{\delta}k^{\gamma} + m^{\gamma}\bar{m}^{\delta} - m^{\delta}\bar{m}^{\gamma}\right) = -\frac{1}{8}  C_{\alpha\beta\gamma\delta}\left(-2l^{\alpha}\bar{m}^{\beta} \right)\left(-2k^{\gamma}l^{\delta} - 2\bar{m}^{\gamma}m^{\delta} \right),\nonumber
\end{equation}
or 
\begin{equation}
 \label{Psi_3_3}
 \Psi_{3} = -\frac{1}{2}  C_{\alpha\beta\gamma\delta}\left(l^{\alpha}\bar{m}^{\beta} k^{\gamma}l^{\delta} + l^{\alpha}\bar{m}^{\beta}\bar{m}^{\gamma}m^{\delta}\right).
\end{equation}
Let us return to equation (\ref{identidade_C_2}) and we can multiply this  equation by $l^{\beta}\bar{m}^{\delta}$ where it results in
\begin{equation}
l^{\alpha}k^{\gamma}l^{\beta}\bar{m}^{\delta} \left(C_{\alpha\beta\gamma\delta} + C_{\gamma\beta\alpha\delta}\right) =
\bar{m}^{\alpha}m^{\gamma} l^{\beta}\bar{m}^{\delta} \left(C_{\alpha\beta\gamma\delta} + C_{\gamma\beta\alpha\delta}\right), \nonumber
\end{equation}
noting that $l^{\alpha}k^{\gamma}l^{\beta}\bar{m}^{\delta}C_{\alpha\beta\gamma\delta} =0$ and $\bar{m}^{\alpha}m^{\gamma} l^{\beta}\bar{m}^{\delta}C_{\gamma\beta\alpha\delta} = 0$ such as,
\begin{equation}
l^{\alpha}\bar{m}^{\delta} k^{\gamma}l^{\beta} C_{\gamma\beta\alpha\delta} =
l^{\beta}\bar{m}^{\alpha}\bar{m}^{\delta} m^{\gamma} C_{\alpha\beta\gamma\delta}, \nonumber
\end{equation}
with the properties $C_{\gamma\beta\alpha\delta} = C_{\alpha\delta\gamma\beta}$ and $C_{\alpha\beta\gamma\delta} = C_{\beta\alpha\delta\gamma}$ it results in
\begin{equation}
l^{\alpha}\bar{m}^{\delta} k^{\gamma}l^{\beta} C_{\alpha\delta\gamma\beta}= l^{\beta}\bar{m}^{\alpha}\bar{m}^{\delta} m^{\gamma} C_{\beta\alpha\delta\gamma}.\nonumber
\end{equation}
On the lef side of the above equation we choose the indices $\delta \leftrightarrow \beta$ and on right side we choose $\alpha \leftrightarrow \beta$ and $\delta \leftrightarrow \gamma$, where it follows
\begin{equation}
l^{\alpha}\bar{m}^{\beta} k^{\gamma}l^{\delta} C_{\alpha\beta\gamma\delta}= l^{\alpha}\bar{m}^{\beta}\bar{m}^{\gamma} m^{\delta} C_{\alpha\beta\gamma\delta}.
\end{equation}
Now we can return with this equation to equation (\ref{Psi_3_3}) to  write two results for component $\Psi_{3}$,
\begin{equation}
 \label{Psi_3_4}
 \Psi_{3} =  -C_{\alpha\beta\gamma\delta} l^{\alpha}\bar{m}^{\beta} k^{\gamma}l^{\delta}
\end{equation}
and
\begin{equation}
 \label{Psi_3_5}
 \Psi_{3} =  -C_{\alpha\beta\gamma\delta} l^{\alpha}\bar{m}^{\beta}\bar{m}^{\gamma} m^{\delta}.
\end{equation}
It is still possible to write the equation (\ref{Psi_3_4}) with the property $C_{\alpha\beta\gamma\delta} = C_{\gamma\delta\alpha\beta}$ as
\begin{equation}
  \Psi_{3} =  -C_{\gamma\delta\alpha\beta} k^{\gamma}l^{\delta}l^{\alpha}\bar{m}^{\beta}, \nonumber
\end{equation}
and we can choose the indices $\alpha \leftrightarrow \gamma$ and $\beta \leftrightarrow \delta$ to obtain,
\begin{equation}
 \label{Psi_3_6}
 \Psi_{3} =  -C_{\alpha\beta\gamma\delta} k^{\alpha}l^{\beta}l^{\gamma}\bar{m}^{\delta}.
\end{equation}
In Newman-Penrose complex null tetrads we from equation (\ref{Psi_3_4}), (\ref{Psi_3_5}) and (\ref{Psi_3_6}) that,
\begin{equation}
 \label{Psi_3_7}
 \Psi_{3} = -C_{1301} =  -C_{0113} = -C_{1332} = -C_{3213}.
\end{equation}

%%%%%%%%%%%%%%%%%%%%%%%%%%%%%%%%%%%%%%%%%%%%%%%%%%%%%%%%%%%%%%
%%%%%%%%%%%%%%%%%%%%%%%%%%%%%%%%%%%%%%%%%%%%%%%%%%%%%%%%%%%%%%
%%%%%%%%%%%%%%%%%%%%%%%%%%%%%%%%%%%%%%%%%%%%%%%%%%%%%%%%%%%%%%
%17 de abril de 2021

To get the component $\Psi_{2}$ we can contract above equation (\ref{complex_Weyl_NP_1}) with 
$U^{\alpha\beta}V^{\gamma\delta} + V^{\alpha\beta}U^{\gamma\delta} + W^{\alpha\beta}W^{\gamma\delta}$, where the only non zero component is 
\begin{equation}
 \frac{1}{2}{\cal C}^{*}_{\alpha\beta\gamma\delta}\left(U^{\alpha\beta}V^{\gamma\delta} + V^{\alpha\beta}U^{\gamma\delta} + W^{\alpha\beta}W^{\gamma\delta}\right) = \Psi_{2}\left(U_{\alpha\beta}V_{\gamma\delta} + V_{\alpha\beta}U_{\gamma\delta} + W_{\alpha\beta}W_{\gamma\delta}\right)\left( U^{\alpha\beta}V^{\gamma\delta} + V^{\alpha\beta}U^{\gamma\delta} + W^{\alpha\beta}W^{\gamma\delta}\right), \nonumber
\end{equation}
with non null contractions (\ref{contracao UV}) and (\ref{contracao WW}) it results 
\begin{equation}
 \frac{1}{2}{\cal C}^{*}_{\alpha\beta\gamma\delta}\left(U^{\alpha\beta}V^{\gamma\delta} + V^{\alpha\beta}U^{\gamma\delta} + W^{\alpha\beta}W^{\gamma\delta}\right)  = 24\Psi_{2} , \nonumber
\end{equation}
We can use the identity of the equation (\ref{identidade_I_C_1}),  
${\cal C}^{*}_{\alpha\beta\gamma\delta} = 2 I_{\gamma\delta\epsilon\zeta} {C_{\alpha\beta}}^{\epsilon\zeta}$, into above equation and rewrite it as,
\begin{equation}
I_{\gamma\delta\epsilon\zeta} {C_{\alpha\beta}}^{\epsilon\zeta}\left(U^{\alpha\beta}V^{\gamma\delta} + V^{\alpha\beta}U^{\gamma\delta} + W^{\alpha\beta}W^{\gamma\delta}\right) = 24\Psi_{2} , \nonumber
\end{equation}
or
\begin{equation}
{C_{\alpha\beta}}^{\epsilon\zeta}\left[U^{\alpha\beta}\left( I_{\gamma\delta\epsilon\zeta} V^{\gamma\delta}\right) + V^{\alpha\beta}\left( I_{\gamma\delta\epsilon\zeta} U^{\gamma\delta}\right) + W^{\alpha\beta}\left(I_{\gamma\delta\epsilon\zeta} W^{\gamma\delta}\right)\right] = 24\Psi_{2} , \nonumber
\end{equation}
where we can use the identities of equation (\ref{identidade_I_1}), such as
\begin{equation}
{C_{\alpha\beta}}^{\epsilon\zeta}\left[U^{\alpha\beta}V_{\epsilon\zeta} + V^{\alpha\beta}U_{\epsilon\zeta}  + W^{\alpha\beta}W_{\epsilon\zeta}\right] = 24\Psi_{2} . \nonumber
\end{equation}
Thus we have that above equation becomes
\begin{equation}
24\Psi_{2} = 2C_{\alpha\beta\gamma\delta}U^{\alpha\beta}V^{\gamma\delta} + C_{\alpha\beta\gamma\delta}W^{\alpha\beta}W^{\gamma\delta}. \nonumber
\end{equation}

With aid of equations (\ref{U1}), (\ref{V1}) and (\ref{W1}) we can write the above equation as,
\begin{equation}
24\Psi_{2} = 2C_{\alpha\beta\gamma\delta}\left(-l^{\alpha}\bar{m}^{\beta} + l^{\beta}\bar{m}^{\alpha} \right)\left(k^{\gamma}m^{\delta} - k^{\delta}m^{\gamma}\right) + C_{\alpha\beta\gamma\delta}\left(l^{\alpha}k^{\beta} - l^{\beta}k^{\alpha} + m^{\alpha}\bar{m}^{\beta} - m^{\beta}\bar{m}^{\alpha} \right)\left(l^{\gamma}k^{\delta} - l^{\delta}k^{\gamma} + m^{\gamma}\bar{m}^{\delta} - m^{\delta}\bar{m}^{\gamma} \right), \nonumber
\end{equation}
or
\begin{equation}
\Psi_{2} = \frac{1}{24}\left[2C_{\alpha\beta\gamma\delta}\left(-2l^{\alpha}\bar{m}^{\beta}\right)\left(2k^{\gamma}m^{\delta}\right) + C_{\alpha\beta\gamma\delta}\left(2l^{\alpha}k^{\beta} + 2m^{\alpha}\bar{m}^{\beta} \right)\left(2l^{\gamma}k^{\delta}  + 2m^{\gamma}\bar{m}^{\delta} \right)\right], \nonumber
\end{equation}
and it results in the below equation
\begin{equation}
\label{Psi_2_1}
 \Psi_{2} = \frac{1}{6}\left[-2C_{\alpha\beta\gamma\delta}~l^{\alpha}\bar{m}^{\beta}k^{\gamma}m^{\delta} + 2C_{\alpha\beta\gamma\delta}~l^{\alpha}k^{\beta}m^{\gamma}\bar{m}^{\delta} + C_{\alpha\beta\gamma\delta}~l^{\alpha}k^{\beta}l^{\gamma}k^{\delta}+ C_{\alpha\beta\gamma\delta}~m^{\alpha}\bar{m}^{\beta}m^{\gamma}\bar{m}^{\delta}\right].
\end{equation}

Finally, to simplify the component $\Psi_{2}$, we can start from the equation (\ref{identidade_C_2}) and we can multiply it by $\bar{m}^{\beta}m^{\delta}$ where it results in
\begin{equation}
 l^{\alpha}k^{\gamma}\bar{m}^{\beta}m^{\delta}\left(C_{\alpha\beta\gamma\delta} + C_{\gamma\beta\alpha\delta} \right) =  \bar{m}^{\alpha}m^{\gamma}\bar{m}^{\beta}m^{\delta}\left(C_{\alpha\beta\gamma\delta} + C_{\gamma\beta\alpha\delta} \right),\nonumber
\end{equation}
where we have $\bar{m}^{\alpha}\bar{m}^{\beta}m^{\gamma}m^{\delta}C_{\alpha\beta\gamma\delta} = 0$ and the above equation becomes,
\begin{equation}
 C_{\alpha\beta\gamma\delta}~l^{\alpha}k^{\gamma}\bar{m}^{\beta}m^{\delta} + C_{\gamma\beta\alpha\delta}~l^{\alpha}k^{\gamma}\bar{m}^{\beta}m^{\delta}  =  C_{\gamma\beta\alpha\delta}~ m^{\gamma}\bar{m}^{\beta}\bar{m}^{\alpha}m^{\delta},\nonumber
\end{equation}
where the right-handed side of the above equation can be write as $C_{\gamma\beta\alpha\delta}~ m^{\gamma}\bar{m}^{\beta}\bar{m}^{\alpha}m^{\delta} = -  C_{\alpha\beta\gamma\delta}~m^{\alpha}\bar{m}^{\beta}m^{\gamma}\bar{m}^{\delta}$ and we obtain
\begin{equation}
\label{identidade_C_3}
 C_{\alpha\beta\gamma\delta}~l^{\alpha}k^{\gamma}\bar{m}^{\beta}m^{\delta} + C_{\gamma\beta\alpha\delta}~l^{\alpha}k^{\gamma}\bar{m}^{\beta}m^{\delta}  = -  C_{\alpha\beta\gamma\delta}~m^{\alpha}\bar{m}^{\beta}m^{\gamma}\bar{m}^{\delta}.
\end{equation}
Now let us return to equation (\ref{identidade_C_2}) and we can multiply it by $k^{\beta}l^{\delta}$ where it results in
\begin{equation}
 l^{\alpha}k^{\beta}k^{\gamma}l^{\delta}\left(C_{\alpha\beta\gamma\delta} + C_{\gamma\beta\alpha\delta} \right) =  \bar{m}^{\alpha}m^{\gamma}k^{\beta}l^{\delta}\left(C_{\alpha\beta\gamma\delta} + C_{\gamma\beta\alpha\delta} \right),\nonumber
\end{equation}
where we have $k^{\gamma}k^{\beta}l^{\alpha}l^{\delta}C_{\gamma\beta\alpha\delta} = 0$ and the above equation becomes,
\begin{equation}
 C_{\alpha\beta\gamma\delta}~k^{\beta}l^{\delta}\bar{m}^{\alpha}m^{\gamma} + C_{\gamma\beta\alpha\delta}~k^{\beta}l^{\delta}\bar{m}^{\alpha}m^{\gamma}  =  - C_{\alpha\beta\gamma\delta}~ l^{\alpha}k^{\beta}l^{\gamma}k^{\delta},\nonumber
\end{equation}
\begin{equation}
 C_{\beta\alpha\delta\gamma}~k^{\beta}l^{\delta}\bar{m}^{\alpha}m^{\gamma} + C_{\delta\alpha\beta\gamma}~k^{\beta}l^{\delta}\bar{m}^{\alpha}m^{\gamma}  =  - C_{\alpha\beta\gamma\delta}~ l^{\alpha}k^{\beta}l^{\gamma}k^{\delta},\nonumber
\end{equation}
where it results
\begin{equation}
\label{identidade_C_4}
C_{\gamma\beta\alpha\delta}~l^{\alpha}k^{\gamma}\bar{m}^{\beta}m^{\delta} + C_{\alpha\beta\gamma\delta}~l^{\alpha}\bar{m}^{\beta}k^{\gamma}m^{\delta}  =  - C_{\alpha\beta\gamma\delta}~ l^{\alpha}k^{\beta}l^{\gamma}k^{\delta}.
\end{equation}
When we compare the equations (\ref{identidade_C_3}) and the abobe equation (\ref{identidade_C_4}), we obtain that,
\begin{equation}
\label{identidade_C_5}
  C_{\alpha\beta\gamma\delta}~ l^{\alpha}k^{\beta}l^{\gamma}k^{\delta} =  C_{\alpha\beta\gamma\delta}~m^{\alpha}\bar{m}^{\beta}m^{\gamma}\bar{m}^{\delta}.
\end{equation}
With the above equation (\ref{identidade_C_5}) we can rewrite the equation (\ref{Psi_2_1}) as
\begin{equation}
\label{Psi_2_2}
 \Psi_{2} = \frac{1}{3}\left( -C_{\alpha\beta\gamma\delta}~l^{\alpha}\bar{m}^{\beta}k^{\gamma}m^{\delta} + C_{\alpha\beta\gamma\delta}~l^{\alpha}k^{\beta}m^{\gamma}\bar{m}^{\delta} + C_{\alpha\beta\gamma\delta}~l^{\alpha}k^{\beta}l^{\gamma}k^{\delta}\right).
\end{equation}
or 
\begin{equation}
\label{Psi_2_3}
 \Psi_{2} = \frac{1}{3}\left(-C_{\alpha\beta\gamma\delta}~l^{\alpha}\bar{m}^{\beta}k^{\gamma}m^{\delta} + C_{\alpha\beta\gamma\delta}~l^{\alpha}k^{\beta}m^{\gamma}\bar{m}^{\delta} + C_{\alpha\beta\gamma\delta}~m^{\alpha}\bar{m}^{\beta}m^{\gamma}\bar{m}^{\delta}\right).
\end{equation}
We can get a way to rewrite the term  $C_{\gamma\beta\alpha\delta}~l^{\alpha}\bar{m}^{\beta}k^{\gamma}m^{\delta}$ from equations (\ref{identidade_C_3}) and  (\ref{identidade_C_4}) by permutations of indixes,
\begin{equation}
\label{identidade_C_6}
 C_{\gamma\beta\alpha\delta}~l^{\alpha}\bar{m}^{\beta}k^{\gamma}m^{\delta} = C_{\alpha\delta\gamma\beta}~l^{\alpha}\bar{m}^{\beta}k^{\gamma}m^{\delta} = - C_{\alpha\delta\beta\gamma}~l^{\alpha}\bar{m}^{\beta}k^{\gamma}m^{\delta}.
\end{equation}
If we use the equation of the first Biachi identity (\ref{first_Bianchi_equation}),
\begin{equation}
 C_{\alpha\beta\gamma\delta} + C_{\alpha\gamma\delta\beta} + C_{\alpha\delta\beta\gamma} = 0, \nonumber 
\end{equation}
and we multiply the above equation by $l^{\alpha}\bar{m}^{\beta}k^{\gamma}m^{\delta}$, then we obtain, 
\begin{equation}
 C_{\alpha\delta\beta\gamma}~l^{\alpha}\bar{m}^{\beta}k^{\gamma}m^{\delta} = - C_{\alpha\beta\gamma\delta}~l^{\alpha}\bar{m}^{\beta}k^{\gamma}m^{\delta} - C_{\alpha\gamma\delta\beta}~l^{\alpha}\bar{m}^{\beta}k^{\gamma}m^{\delta} = - C_{\alpha\beta\gamma\delta}~l^{\alpha}\bar{m}^{\beta}k^{\gamma}m^{\delta} -  C_{\alpha\beta\gamma\delta}~l^{\alpha}k^{\beta}m^{\gamma}\bar{m}^{\delta}.\nonumber
\end{equation}
With equation (\ref{identidade_C_6}) the above equation becomes,
\begin{equation}
  C_{\gamma\beta\alpha\delta}~l^{\alpha}\bar{m}^{\beta}k^{\gamma}m^{\delta} =  C_{\alpha\beta\gamma\delta}~l^{\alpha}\bar{m}^{\beta}k^{\gamma}m^{\delta} +  C_{\alpha\beta\gamma\delta}~l^{\alpha}k^{\beta}m^{\gamma}\bar{m}^{\delta}.
\end{equation}
Thus, if we replace the above equation into equation (\ref{identidade_C_4}) we have the below equation
\begin{equation}
 \label{identidade_C_7}
2 C_{\alpha\beta\gamma\delta}~l^{\alpha}\bar{m}^{\beta}k^{\gamma}m^{\delta} + C_{\alpha\beta\gamma\delta}~l^{\alpha}k^{\beta}m^{\gamma}\bar{m}^{\delta} =  - C_{\alpha\beta\gamma\delta}~ l^{\alpha}k^{\beta}l^{\gamma}k^{\delta},
\end{equation}
and we can put it into equation (\ref{Psi_2_2}) such we have to $\Psi_{2}$,
\begin{eqnarray}
 \Psi_{2} &=& \frac{1}{3}\left(\frac{1}{2}C_{\alpha\beta\gamma\delta}~l^{\alpha}k^{\beta}m^{\gamma}\bar{m}^{\delta} + \frac{1}{2}C_{\alpha\beta\gamma\delta}~ l^{\alpha}k^{\beta}l^{\gamma}k^{\delta} + C_{\alpha\beta\gamma\delta}~l^{\alpha}k^{\beta}m^{\gamma}\bar{m}^{\delta} + C_{\alpha\beta\gamma\delta}~l^{\alpha}k^{\beta}l^{\gamma}k^{\delta}\right)\cr
 &=&  \frac{1}{2}\left(C_{\alpha\beta\gamma\delta}~l^{\alpha}k^{\beta}m^{\gamma}\bar{m}^{\delta} + C_{\alpha\beta\gamma\delta}~ l^{\alpha}k^{\beta}l^{\gamma}k^{\delta}\right)\cr
 &=& -\frac{1}{2}\left(C_{\alpha\beta\gamma\delta}~k^{\alpha}l^{\beta}m^{\gamma}\bar{m}^{\delta} + C_{\alpha\beta\gamma\delta}~ k^{\alpha}l^{\beta}l^{\gamma}k^{\delta}\right), \nonumber
\end{eqnarray}
finally we have,
\begin{equation}
\label{Psi_2_4}
 \Psi_{2} = \frac{1}{2}C_{\alpha\beta\gamma\delta}~k^{\alpha}l^{\beta}  \left(k^{\gamma}l^{\delta} - m^{\gamma}\bar{m}^{\delta} \right).
\end{equation}
In Newman-Penrose complex null tetrads we from the above equation that,
\begin{equation}
\label{Psi_2_5}
 \Psi_{2} = \frac{1}{2} \left( C_{0101} - C_{0123}\right) \hspace*{1cm} \mbox{and} \hspace*{1cm} \bar{\Psi}_{2} = \frac{1}{2} \left( C_{0101} + C_{0123}\right). 
\end{equation}

If we recall the equation (\ref{identidade_C_7})  and we isolate the term $C_{\alpha\beta\gamma\delta}~l^{\alpha}k^{\beta}m^{\gamma}\bar{m}^{\delta} $ we have that,
\begin{equation}
 C_{\alpha\beta\gamma\delta}~l^{\alpha}k^{\beta}m^{\gamma}\bar{m}^{\delta} = -2 C_{\alpha\beta\gamma\delta}~l^{\alpha}\bar{m}^{\beta}k^{\gamma}m^{\delta} - C_{\alpha\beta\gamma\delta}~ l^{\alpha}k^{\beta}l^{\gamma}k^{\delta}, \nonumber
\end{equation}
we can replace it into equation (\ref{Psi_2_2}), then we have other identity for $\Psi_{2}$,
\begin{equation}
\label{Psi_2_7}
 \Psi_{2} = -C_{\alpha\beta\gamma\delta}~k^{\alpha}m^{\beta}  l^{\gamma}\bar{m}^{\delta} .
\end{equation}
In Newman-Penrose complex null tetrads we from the above equation that,
\begin{equation}
\label{Psi_2_8}
 \Psi_{2} =  -C_{0213} \hspace*{1cm} \mbox{and} \hspace*{1cm}\bar{\Psi}_{2} = -C_{0312}. 
\end{equation}

Also, we can recall the equation (\ref{Psi_2_4}) and replace the identity (\ref{identidade_C_5}) to obtain other equation for $\Psi_{2}$, such as below equation,
\begin{equation}
\label{Psi_2_9}
 \Psi_{2} = \frac{1}{2}C_{\alpha\beta\gamma\delta}~m^{\alpha}\bar{m}^{\beta}  \left(m^{\gamma}\bar{m}^{\delta} - k^{\gamma}l^{\delta} \right),
\end{equation}
where in Newman-Penrose complex null tetrads we from the above equation that,
\begin{equation}
\label{Psi_2_10}
 {\Psi}_{2} = \frac{1}{2} \left( C_{2323} - C_{0123}\right) \hspace*{1cm} \mbox{and} \hspace*{1cm} \bar{\Psi}_{2} = \frac{1}{2} \left( C_{2323} + C_{0123}\right). 
\end{equation}

Finally we can collect the results of the equations (\ref{Psi_0_3}), (\ref{Psi_4_2}), (\ref{Psi_1_4}), (\ref{Psi_1_5}), (\ref{Psi_1_6}), (\ref{Psi_3_4}), (\ref{Psi_3_5}), (\ref{Psi_3_6}), (\ref{Psi_2_4}), (\ref{Psi_2_7}) and (\ref{Psi_2_9}) to group the five complex components of the Weyl curvature tensor in Newman-Penrose formalism as follows,
\begin{equation}
 \begin{cases}
  \Psi_{0} =  C_{\alpha\beta\gamma\delta} k^{\alpha}m^{\beta}k^{\gamma}m^{\delta}, \cr
  \Psi_{1} =  C_{\alpha\beta\gamma\delta} k^{\alpha}m^{\beta}k^{\gamma}l^{\delta} = C_{\alpha\beta\gamma\delta} k^{\alpha}m^{\beta} \bar{m}^{\gamma}m^{\delta} = C_{\alpha\beta\gamma\delta} k^{\alpha}l^{\beta}k^{\gamma}m^{\delta}, \cr
  \Psi_{2} = \frac{1}{2}C_{\alpha\beta\gamma\delta}~k^{\alpha}l^{\beta}  \left(k^{\gamma}l^{\delta} - m^{\gamma}\bar{m}^{\delta} \right) = \frac{1}{2}C_{\alpha\beta\gamma\delta}~m^{\alpha}\bar{m}^{\beta}  \left(m^{\gamma}\bar{m}^{\delta} - k^{\gamma}l^{\delta} \right) = -C_{\alpha\beta\gamma\delta}~k^{\alpha}m^{\beta}  l^{\gamma}\bar{m}^{\delta},\cr
  \Psi_{3} =  -C_{\alpha\beta\gamma\delta} l^{\alpha}\bar{m}^{\beta} k^{\gamma}l^{\delta} =   -C_{\alpha\beta\gamma\delta} l^{\alpha}\bar{m}^{\beta}\bar{m}^{\gamma} m^{\delta} =  -C_{\alpha\beta\gamma\delta} k^{\alpha}l^{\beta}l^{\gamma}\bar{m}^{\delta}, \cr
  \Psi_{4} =  C_{\alpha\beta\gamma\delta} l^{\alpha}\bar{m}^{\beta}l^{\gamma}\bar{m}^{\delta}.
 \end{cases}
\end{equation}

%%%%%%%%%%%%%%%%%%%%%%%%%%%%%%%%%%%%%%%%%%%%%%%%%%%%%%%%%%%%%%
%%%%%%%%%%%%%%%%%%%%%%%%%%%%%%%%%%%%%%%%%%%%%%%%%%%%%%%%%%%%%%
%%%%%%%%%%%%%%%%%%%%%%%%%%%%%%%%%%%%%%%%%%%%%%%%%%%%%%%%%%%%%%
%%%%%%%%%%%%%%%%%%%%%%%%%%%%%%%%%%%%%%%%%%%%%%%%%%%%%%%%%%%%%%
%%%%%%%%%%%%%%%%%%%%%%%%%%%%%%%%%%%%%%%%%%%%%%%%%%%%%%%%%%%%%%
%%%%%%%%%%%%%%%%%%%%%%%%%%%%%%%%%%%%%%%%%%%%%%%%%%%%%%%%%%%%%%
%%%%%%%%%%%%%%%%%%%%%%%%%%%%%%%%%%%%%%%%%%%%%%%%%%%%%%%%%%%%%%
%%%%%%%%%%%%%%%%%%%%%%%%%%%%%%%%%%%%%%%%%%%%%%%%%%%%%%%%%%%%%%
%%%%%%%%%%%%%%%%%%%%%%%%%%%%%%%%%%%%%%%%%%%%%%%%%%%%%%%%%%%%%%
%%%%%%%%%%%%%%%%%%%%%%%%%%%%%%%%%%%%%%%%%%%%%%%%%%%%%%%%%%%%%%
%%%%%%%%%%%%%%%%%%%%%%%%%%%%%%%%%%%%%%%%%%%%%%%%%%%%%%%%%%%%%%
%%%%%%%%%%%%%%%%%%%%%%%%%%%%%%%%%%%%%%%%%%%%%%%%%%%%%%%%%%%%%%
%%%%%%%%%%%%%%%%%%%%%%%%%%%%%%%%%%%%%%%%%%%%%%%%%%%%%%%%%%%%%%
%%%%%%%%%%%%%%%%%%%%%%%%%%%%%%%%%%%%%%%%%%%%%%%%%%%%%%%%%%%%%%
%%%%%%%%%%%%%%%%%%%%%%%%%%%%%%%%%%%%%%%%%%%%%%%%%%%%%%%%%%%%%%
%%%%%%%%%%%%%%%%%%%%%%%%%%%%%%%%%%%%%%%%%%%%%%%%%%%%%%%%%%%%%%
%%%%%%%%%%%%%%%%%%%%%%%%%%%%%%%%%%%%%%%%%%%%%%%%%%%%%%%%%%%%%%
%%%%%%%%%%%%%%%%%%%%%%%%%%%%%%%%%%%%%%%%%%%%%%%%%%%%%%%%%%%%%%
%%%%%%%%%%%%%%%%%%%%%%%%%%%%%%%%%%%%%%%%%%%%%%%%%%%%%%%%%%%%%%
%%%%%%%%%%%%%%%%%%%%%%%%%%%%%%%%%%%%%%%%%%%%%%%%%%%%%%%%%%%%%%
%%%%%%%%%%%%%%%%%%%%%%%%%%%%%%%%%%%%%%%%%%%%%%%%%%%%%%%%%%%%%%
%%%%%%%%%%%%%%%%%%%%%%%%%%%%%%%%%%%%%%%%%%%%%%%%%%%%%%%%%%%%%%
%%%%%%%%%%%%%%%%%%%%%%%%%%%%%%%%%%%%%%%%%%%%%%%%%%%%%%%%%%%%%%
%%%%%%%%%%%%%%%%%%%%%%%%%%%%%%%%%%%%%%%%%%%%%%%%%%%%%%%%%%%%%%

\subsection{Example of Vaidya spacetime}
 
An example that we can verify about this mathematical formalism, it is the Vaidya spacetime. This spacetime can represent a spherical radiating star. It solution is a spacetime related to Schwarzschild, 
\begin{equation}
 \label{Schwarzschild_1}
 ds^2 = -\left(1-\frac{2m}{r} \right)dt^2 + \left(1-\frac{2m}{r} \right)^{-1}dr^2 + r^2\left(d\theta^2 + \sin^2\theta~d\phi^2 \right).
\end{equation}
In the similar way that we can adopt to obtain the Eddington–Finkelstein coordinates \cite{Hobson}, the lightcone structure of the the paths of radially incoming and outgoing photons in the Schwarzschild spacetime can be obtained by $ds^2=0$.
From the metric (\ref{Schwarzschild_1}), for a radially moving photon we have that 
\begin{equation}
 \label{dt}
 dt = \pm \left(1 - \frac{2m}{r} \right)^{-1} dr
\end{equation}
where the plus sign corresponds to a photon that is outgoing and the minus sign corresponds to a photon that is incoming. On integrating over plus sign, we obtain
\begin{equation}
 \label{outgoing_photon_1}
 u = t - r - 2m \ln \left(\frac{r}{2m} - 1\right),
\end{equation}
where the parameter $u$ is a integration constant, but that from now on it is assumed as null coordinate. With this, the differential $dt$ is given by,
\begin{equation}
 \label{outgoing_photon_2}
 dt = du + \left(1 - \frac{2m}{r} \right)^{-1} dr,
\end{equation}
where it is for outgoing photons in the Schwarzschild spacetime. With this we have that
\begin{equation}
 dt^2 = du^2 + \left(1 - \frac{2m}{r} \right)^{-2}~ dr^2 + 2\left(1 - \frac{2m}{r} \right)^{-1}~ du~dr, \nonumber
\end{equation}
and replacing the above equation in the equation (\ref{Schwarzschild_1}) we obtain the spherical Schwarzschild spacetime given by,
\begin{equation}
 \label{Schwarzschild_2}
 ds^2 = -\left(1-\frac{2m}{r}\right)~du^2 -2du~dr + r^2\left(d\theta^2 + \sin^2\theta~d\phi^2 \right).
\end{equation}
The above metric is especially convenient for calculating
the paths of radial null geodesics, for which 
$ds = d\theta = d\phi = 0$, and describe outgoing photons 
from spherical gravitational sources. The metric (\ref{Schwarzschild_2}) is called retarded Schwarzschild metric. 

If we integrate the differential equation (\ref{dt}) over minus sign, then we obtain
\begin{equation}
 \label{incoming_photon_1}
 v = t + r + 2m \ln \left(\frac{r}{2m} - 1\right),
\end{equation}
again, the parameter $v$ is a integration constant, but that from now on it is assumed as null coordinate. With this the differential $dt^2$ is given by,
\begin{equation}
 dt^2 = dv^2 + \left(1 - \frac{2m}{r} \right)^{-2}~ dr^2 - 2\left(1 - \frac{2m}{r} \right)^{-1}~ dv~dr, \nonumber
\end{equation}
where it is for incoming photons in the Schwarzschild spacetime. Thus, the spherical Schwarzschild spacetime is given by
\begin{equation}
 \label{Schwarzschild_3}
 ds^2 = -\left(1-\frac{2m}{r}\right)~dv^2 + 2dv~dr + r^2\left(d\theta^2 + \sin^2\theta~d\phi^2 \right),
\end{equation}
in such a way that the above metric is especially convenient for calculating the paths of radial null geodesics, for which 
$ds = d\theta = d\phi = 0$, and describe incoming photons 
from spherical gravitational sources. The metric (\ref{Schwarzschild_3}) is called advanced Schwarzschild metric.

These Schwarzschild spacetimes represented by equations (\ref{Schwarzschild_2}) and (\ref{Schwarzschild_3}) are util to describe the exact solutions  of the Einstein-Maxwell equations \cite{Griffiths}. Each metric (\ref{Schwarzschild_2}) and (\ref{Schwarzschild_3}) can express the outgoing and incoming radiation given by,
\begin{equation}
 \label{Vaidya_retarded}
 ds^2 = -\left(1-\frac{2m(u)}{r}\right)~du^2 - 2du~dr + r^2\left(d\theta^2 + \sin^2\theta~d\phi^2 \right)
\end{equation}
and 
\begin{equation}
 \label{Vaidya_advanced}
 ds^2 = -\left(1-\frac{2m(v)}{r}\right)~dv^2 + 2dv~dr + r^2\left(d\theta^2 + \sin^2\theta~d\phi^2 \right).
\end{equation}
The retarded solution of Vaidya metric (\ref{Vaidya_retarded}) corresponds to a spherical distribution of pure radiation, as we will see below, by emission of a central mass source. In this case the retarded solution of Vaidya metric can represent a radiation star or white hole. While the advanced solution of Vaidya metric (\ref{Vaidya_advanced}) is especially useful in black hole physics.

We can calculate the Weyl component and the Ricci component of the retarded Vaidya spacetime (\ref{Vaidya_retarded}) by Cartan's method. We can recall the equation (\ref{metric_N-P_01}), where we have $ \mbox{\bf g} = -2 {\bm \theta}^{0} \otimes {\bm \theta}^{1} + 2 {\bm \theta}^{2} \otimes {\bm \theta}^{3}$, where we have from equation (\ref{Vaidya_retarded}) that,
\begin{equation}
 2 {\bm \theta}^{0} \otimes {\bm \theta}^{1} = \left(1-\frac{2m(u)}{r}\right)~du\otimes du  + 2du\otimes dr.
\end{equation}
Thus we have that,
\begin{eqnarray}
 2 {\bm \theta}^{0} \otimes {\bm \theta}^{1} &=& 2\left({\omega^{0}}_{\mu} dx^{\mu}  \right)\otimes \left({\omega^{1}}_{\nu} dx^{\nu}  \right) \cr
 &=&  2\left({\omega^{0}}_{u} du + {\omega^{0}}_{r} dr  \right)\otimes \left({\omega^{1}}_{u} du + {\omega^{1}}_{r} dr  \right) \cr
 &=& 2\left({\omega^{0}}_{u} {\omega^{1}}_{u}  \right) du\otimes du + 2\left({\omega^{0}}_{r} {\omega^{1}}_{r}  \right) dr\otimes dr + 2\left({\omega^{0}}_{u} {\omega^{1}}_{r} + {\omega^{0}}_{r} {\omega^{1}}_{u} \right) du\otimes dr.
\end{eqnarray}
It results in three equations 
\begin{equation}
\label{Vaidya_tetrad_01}
 2\left({\omega^{0}}_{u} {\omega^{1}}_{u}  \right) = \left(1-\frac{2m(u)}{r}\right), 
\end{equation}
\begin{equation}
\label{Vaidya_tetrad_02}
 2\left({\omega^{0}}_{u} {\omega^{1}}_{r} + {\omega^{0}}_{r} {\omega^{1}}_{u} \right) = 2
\end{equation}
and
\begin{equation}
\label{Vaidya_tetrad_03}
{\omega^{0}}_{r} {\omega^{1}}_{r}   = 0.
\end{equation}
In the above equation (\ref{Vaidya_tetrad_03}) we choose ${\omega^{1}}_{r} = 0$, consequently in the equation (\ref{Vaidya_tetrad_02}) it results in ${\omega^{0}}_{r} {\omega^{1}}_{u} = 1$ and we write ${\omega^{0}}_{r} = 1$ and ${\omega^{1}}_{u} = 1$ and then we have from equation (\ref{Vaidya_tetrad_01}) that,
\begin{equation}
 {\omega^{0}}_{u} = \frac{1}{2} \left(1-\frac{2m(u)}{r}\right). \nonumber
\end{equation}
Thus we have that the matrix of tetrad follows as
\begin{equation}
 \left({\omega^{\alpha}}_{\mu} \right) = \begin{pmatrix}
    \dfrac{1}{2} \left(1-\dfrac{2m(u)}{r}\right) & 1 & 0 & 0 \cr
    1 & 0 & 0 & 0 \cr
    0 & 0 & \dfrac{r}{\sqrt{2}} & \dfrac{i}{\sqrt{2}} r\sin\theta \cr
    0 & 0 & \dfrac{r}{\sqrt{2}} & \dfrac{-i}{\sqrt{2}} r\sin\theta
                                         \end{pmatrix}
\end{equation}
and the coordinate dual basis $\tilde{\bm\theta}^{\alpha} = {\omega^{\alpha}}_{\mu} dx^{\mu}$ are seen as,
\begin{equation}
\label{Vaidya_tetrad_04}
 \begin{cases}
  \tilde{\bm\theta}^{0} = dr + \dfrac{1}{2} \left(1-\dfrac{2m(u)}{r}\right) du\cr
  \tilde{\bm\theta}^{1} = du \cr
  \tilde{\bm\theta}^{2} = \frac{1}{\sqrt{2}}\left(r~d\theta + ir\sin\theta~d\phi \right) \cr
  \tilde{\bm\theta}^{3} = \frac{1}{\sqrt{2}}\left(r~d\theta - ir\sin\theta~d\phi \right),
 \end{cases}
\end{equation}
where $\overline{\tilde{\bm\theta}^{2}} = \tilde{\bm\theta}^{3}$. 

Operating an exterior derivative on $\tilde{\bm\theta}^{0}$ results in,
\begin{equation}
\label{Vaidya_tetrad_05}
 d \tilde{\bm\theta}^{0} = \frac{m}{r^2} ~ dr \wedge du.
\end{equation}
and the exterior product between $\tilde{\bm\theta}^{0}$ and $\tilde{\bm\theta}^{1}$ follows as
\begin{equation}
 \tilde{\bm\theta}^{0} \wedge \tilde{\bm\theta}^{1} = \left[ dr + \dfrac{1}{2} \left(1-\dfrac{2m(u)}{r}\right) du \right] \wedge du = dr \wedge du, \nonumber
\end{equation}
so that with this equation,  we have that equation (\ref{Vaidya_tetrad_05}) becomes,
\begin{equation}
\label{Vaidya_tetrad_06}
 d \tilde{\bm\theta}^{0} = \frac{m}{r^2} ~  \tilde{\bm\theta}^{0} \wedge \tilde{\bm\theta}^{1}.
\end{equation}

The exterior derivative for the second one-form of dual basis (\ref{Vaidya_tetrad_04}), results in,
\begin{equation}
\label{Vaidya_tetrad_07}
 d \tilde{\bm\theta}^{1} = 0.
\end{equation}

In order to obtain $d\tilde{\bm\theta}^{2}$ and $d\tilde{\bm\theta}^{3}$
we need express $dr$, $d\theta$ and $d\phi$ in terms of one-forms of the dual basis (\ref{Vaidya_tetrad_04}). To isolate $d\theta$ in the system (\ref{Vaidya_tetrad_04}), it is straightforward to express $d\theta$ through of the below equation,
\begin{equation}
\label{Vaidya_tetrad_08}
 d \theta = \frac{1}{r\sqrt{2}}\left(\tilde{\bm\theta}^{2} + \tilde{\bm\theta}^{3} \right),
\end{equation}
For the one-form $d\phi$ we obtain that,
\begin{equation}
\label{Vaidya_tetrad_09}
 d \phi = \frac{-i}{r\sin\theta \sqrt{2}}\left(\tilde{\bm\theta}^{2} - \tilde{\bm\theta}^{3} \right).
\end{equation}
And for the one-form $dr$ we obtain that,
\begin{equation}
\label{Vaidya_tetrad_10}
 dr = \tilde{\bm\theta}^{0} - \frac{1}{2}\left(1-\frac{2m}{r} \right)\tilde{\bm\theta}^{1}.
\end{equation}
Thus, from system of equation (\ref{Vaidya_tetrad_04}) we have that de two-forms 
$d\tilde{\bm\theta}^{2}$ and $d\tilde{\bm\theta}^{3}$ become,
\begin{equation}
 d\tilde{\bm\theta}^{2} = \frac{1}{\sqrt{2}}\left(dr\wedge d\theta + i\sin\theta~dr\wedge d\phi + ir\cos\theta~d\theta\wedge d\phi \right) \nonumber
\end{equation}
and
\begin{equation}
 d\tilde{\bm\theta}^{3} = \frac{1}{\sqrt{2}}\left(dr\wedge d\theta - i\sin\theta~dr\wedge d\phi - ir\cos\theta~d\theta\wedge d\phi \right) = \overline{d\tilde{\bm\theta}^{2}}. \nonumber
\end{equation}
We can substitute equations (\ref{Vaidya_tetrad_08}), (\ref{Vaidya_tetrad_09}) and (\ref{Vaidya_tetrad_10}) in equation for 2-form $d\tilde{\bm\theta}^{2}$ and we obtain that,
\begin{eqnarray}
 d\tilde{\bm\theta}^{2} &=&  \frac{1}{\sqrt{2}}\left[\tilde{\bm\theta}^{0} - \frac{1}{2}\left(1-\frac{2m}{r} \right)\tilde{\bm\theta}^{1}\right] \wedge \left[ \frac{1}{r\sqrt{2}}\left(\tilde{\bm\theta}^{2} + \tilde{\bm\theta}^{3} \right)\right] + \frac{i\sin\theta}{\sqrt{2}} \left[\tilde{\bm\theta}^{0} - \frac{1}{2}\left(1-\frac{2m}{r} \right)\tilde{\bm\theta}^{1}\right]\wedge \left[\frac{-i}{r\sin\theta \sqrt{2}}\left(\tilde{\bm\theta}^{2} - \tilde{\bm\theta}^{3} \right)\right] \cr
 & & +\frac{i r \cos\theta}{\sqrt{2}}\left[\frac{1}{r\sqrt{2}}\left(\tilde{\bm\theta}^{2} + \tilde{\bm\theta}^{3} \right)\wedge \frac{(-i)}{r\sin\theta \sqrt{2}}\left(\tilde{\bm\theta}^{2} - \tilde{\bm\theta}^{3} \right) \right], \nonumber
\end{eqnarray}
such that it reduces to,
\begin{equation}
\label{Vaidya_tetrad_11}
 d\tilde{\bm\theta}^{2} = \frac{1}{r} ~\tilde{\bm\theta}^{0} \wedge \tilde{\bm\theta}^{2} -\frac{1}{2r}\left(1-\frac{2m}{r} \right)\tilde{\bm\theta}^{1} \wedge \tilde{\bm\theta}^{2} - \frac{1}{\sqrt{2}}\frac{\cot\theta}{r}~ \tilde{\bm\theta}^{2} \wedge \tilde{\bm\theta}^{3},
\end{equation}
and for $d\tilde{\bm\theta}^{3}$ we obtain,
\begin{equation}
\label{Vaidya_tetrad_12}
 d\tilde{\bm\theta}^{3} = \frac{1}{r} ~\tilde{\bm\theta}^{0} \wedge \tilde{\bm\theta}^{3} -\frac{1}{2r}\left(1-\frac{2m}{r} \right)\tilde{\bm\theta}^{1} \wedge \tilde{\bm\theta}^{3} + \frac{1}{\sqrt{2}}\frac{\cot\theta}{r}~ \tilde{\bm\theta}^{2} \wedge \tilde{\bm\theta}^{3}.
\end{equation}

With these two-forms we can use the first Cartan equation (\ref{1a_equacao_de_Cartan}) to calculate connection 1-form ${\bm \Gamma}_{\alpha\beta}$,
\begin{equation}
\label{1a_equacao_de_Cartan_2}
 \begin{cases}
  d\tilde{\bm\theta}^0 =  -{\bm \Gamma}_{01}\wedge \tilde{\bm\theta}^0 + {\bm \Gamma}_{12}\wedge\tilde{\bm\theta}^2
  + \overline{\bm \Gamma}_{12} \wedge\tilde{\bm\theta}^3 \cr
d\tilde{\bm\theta}^1  = {\bm \Gamma}_{01}\wedge \tilde{\bm\theta}^1 + {\bm \Gamma}_{02}\wedge \tilde{\bm\theta}^2
+ \overline{\bm \Gamma}_{02}\wedge \tilde{\bm\theta}^3 \cr
d\tilde{\bm\theta}^2  = \overline{\bm \Gamma}_{02}\wedge \tilde{\bm\theta}^0 + \overline{\bm \Gamma}_{12}\wedge \tilde{\bm\theta}^1
+ {\bm \Gamma}_{23}\wedge \tilde{\bm\theta}^2. 
 \end{cases} 
\end{equation}
For the first equation we have from (\ref{Vaidya_tetrad_06}) that,
\begin{equation}
 -{\bm \Gamma}_{01}\wedge \tilde{\bm\theta}^0 + {\bm \Gamma}_{12}\wedge\tilde{\bm\theta}^2
  + \overline{\bm \Gamma}_{12} \wedge\tilde{\bm\theta}^3 =  \frac{m}{r^2} ~  \tilde{\bm\theta}^{0} \wedge \tilde{\bm\theta}^{1} \nonumber
\end{equation}
and we can identify the 1-form ${\bm \Gamma}_{01}$ as
\begin{equation}
 \label{Vaidya_tetrad_13}
 {\bm \Gamma}_{01} = \frac{m}{r^2} ~  \tilde{\bm\theta}^{1}.
\end{equation}
From the first equation in (\ref{1a_equacao_de_Cartan_2}) results too
\begin{equation}
 {\bm \Gamma}_{12}\wedge\tilde{\bm\theta}^2
  = - \overline{\bm \Gamma}_{12} \wedge\tilde{\bm\theta}^3 \nonumber
\end{equation}
since $ \overline{\bm \Gamma}_{12} = {\bm \Gamma}_{13}$ this equation results in,
\begin{equation}
\label{Vaidya_tetrad_14}
 {\bm \Gamma}_{12}\wedge\tilde{\bm\theta}^2
  = - {\bm \Gamma}_{13} \wedge\tilde{\bm\theta}^3 .
\end{equation}
From the second equation in (\ref{1a_equacao_de_Cartan_2}), with $d\tilde{\bm\theta}^1  = 0$ from (\ref{Vaidya_tetrad_07}) and the equation (\ref{Vaidya_tetrad_13}) it results
\begin{equation}
\label{Vaidya_tetrad_15}
 {\bm \Gamma}_{02}\wedge\tilde{\bm\theta}^2
  = -{\bm \Gamma}_{03} \wedge\tilde{\bm\theta}^3.
\end{equation}
We can substitute the equation (\ref{Vaidya_tetrad_11}) ino the third equation of (\ref{1a_equacao_de_Cartan_2}), we obtain,
\begin{equation}
\label{Vaidya_tetrad_16}
 \frac{1}{r} ~\tilde{\bm\theta}^{0} \wedge \tilde{\bm\theta}^{2} -\frac{1}{2r}\left(1-\frac{2m}{r} \right)\tilde{\bm\theta}^{1} \wedge \tilde{\bm\theta}^{2} - \frac{1}{\sqrt{2}}\frac{\cot\theta}{r}~ \tilde{\bm\theta}^{2} \wedge \tilde{\bm\theta}^{3} = \overline{\bm \Gamma}_{02}\wedge \tilde{\bm\theta}^0 + \overline{\bm \Gamma}_{12}\wedge \tilde{\bm\theta}^1
+ {\bm \Gamma}_{23}\wedge \tilde{\bm\theta}^2 
\end{equation}
where we can identify,
\begin{equation}
\label{Vaidya_tetrad_17}
 \overline{\bm \Gamma}_{02} = -\frac{1}{r}~ \tilde{\bm\theta}^{2} = {\bm \Gamma}_{03}
\end{equation}
and 
\begin{equation}
\label{Vaidya_tetrad_18}
 \overline{\bm \Gamma}_{12} = \frac{1}{2r}\left(1-\frac{2m}{r} \right)\tilde{\bm\theta}^{2} = {\bm \Gamma}_{13}.
\end{equation}
In the equation (\ref{Vaidya_tetrad_17})  we observe that $\overline{\overline{\bm \Gamma}}_{02} = {\bm \Gamma}_{02} = -\dfrac{1}{r}~ \tilde{\bm\theta}^{3} $ and this together with the equation (\ref{Vaidya_tetrad_17}) satisfy the equation (\ref{Vaidya_tetrad_15}).
Performing the same calculation in the equation  (\ref{Vaidya_tetrad_18})  we observe that $\overline{\overline{\bm \Gamma}}_{12} = {\bm \Gamma}_{12} = \dfrac{1}{2r}\left(1-\dfrac{2m}{r} \right)\tilde{\bm\theta}^{3}$ and this together with the equation (\ref{Vaidya_tetrad_18}) satisfy the equation (\ref{Vaidya_tetrad_14}).

To calculate ${\bm \Gamma}_{23}$ it is necessary use $d\tilde{\bm\theta}^3= {\bm \Gamma}_{02}\wedge \tilde{\bm\theta}^0 + {\bm \Gamma}_{12}\wedge \tilde{\bm\theta}^1
- {\bm \Gamma}_{23}\wedge \tilde{\bm\theta}^3$ by applying a complex conjugate operation in third equation of (\ref{1a_equacao_de_Cartan_2}), that with aid of equation (\ref{Vaidya_tetrad_12}) we obtain
\begin{equation}
\label{Vaidya_tetrad_19}
 \frac{1}{r} ~\tilde{\bm\theta}^{0} \wedge \tilde{\bm\theta}^{3} -\frac{1}{2r}\left(1-\frac{2m}{r} \right)\tilde{\bm\theta}^{1} \wedge \tilde{\bm\theta}^{3} + \frac{1}{\sqrt{2}}\frac{\cot\theta}{r}~ \tilde{\bm\theta}^{2} \wedge \tilde{\bm\theta}^{3} = {\bm \Gamma}_{02}\wedge \tilde{\bm\theta}^0 + {\bm \Gamma}_{12}\wedge \tilde{\bm\theta}^1
- {\bm \Gamma}_{23}\wedge \tilde{\bm\theta}^3. 
\end{equation}
Thus, with this equation (\ref{Vaidya_tetrad_19}) and the equation (\ref{Vaidya_tetrad_16}) we have,
\begin{equation}
  \frac{1}{\sqrt{2}}\frac{\cot\theta}{r}~ \tilde{\bm\theta}^{2} \wedge \tilde{\bm\theta}^{3} + \frac{1}{\sqrt{2}}\frac{\cot\theta}{r}~ \tilde{\bm\theta}^{2} \wedge \tilde{\bm\theta}^{3} = - {\bm \Gamma}_{23}\wedge \tilde{\bm\theta}^2 - {\bm \Gamma}_{23}\wedge \tilde{\bm\theta}^3 \nonumber
\end{equation}
that results in
\begin{equation}
\label{Vaidya_tetrad_20}
    {\bm \Gamma}_{23} = \frac{1}{\sqrt{2}}\frac{\cot\theta}{r} \left( \tilde{\bm\theta}^3 - \tilde{\bm\theta}^2\right).
\end{equation}

We recall the system of 2-forms equations of the second equation of Cartan in equation (\ref{2a_equacao_de_Cartan})
\begin{equation}
\label{2a_equacao_de_Cartan_2}
 \begin{cases}
  \bm \Theta_{03} = d\bm\Gamma_{03} + {\bm \Gamma}_{03} \wedge ( \bm\Gamma_{01}+ \bm \Gamma_{23})\cr
   \bm \Theta_{12} = d\bm\Gamma_{12} - \bm\Gamma_{12} \wedge (\bm\Gamma_{01} +  \bm \Gamma_{23} )\cr
   \bm \Theta_{01} +  \bm \Theta_{23} = d( \bm\Gamma_{01}+ \bm\Gamma_{23}) - 2\bm \Gamma_{03}\wedge \bm\Gamma_{12},
 \end{cases}
\end{equation}
in order to obtain the components of the curvature tensor. The first calculation to be performed is a exterior derivative operation in $\bm\Gamma_{03} $ of equation (\ref{Vaidya_tetrad_17}),
\begin{equation}
 d\bm\Gamma_{03} = d\left[-\frac{1}{r}  \tilde{\bm\theta}^2 \right] = \frac{1}{r^2} dr\wedge  \tilde{\bm\theta}^2 - \frac{1}{r}~  d\tilde{\bm\theta}^2, \nonumber
\end{equation}
where we can use equation (\ref{Vaidya_tetrad_10}) and (\ref{Vaidya_tetrad_11}), and the above 2-form results in
\begin{equation}
    d\bm\Gamma_{03} = \frac{1}{\sqrt{2}}\frac{\cot\theta}{r^2}~ \tilde{\bm\theta}^2 \wedge \tilde{\bm\theta}^3. \nonumber
\end{equation}
The respective exterior product ${\bm \Gamma}_{03} \wedge ( \bm\Gamma_{01}+ \bm \Gamma_{23})$ results in,
\begin{equation}
 {\bm \Gamma}_{03} \wedge ( \bm\Gamma_{01}+ \bm \Gamma_{23}) = -\frac{1}{r}~\tilde{\bm\theta}^2 \wedge \left[\frac{m}{r^2}~\tilde{\bm\theta}^1 + \frac{1}{\sqrt{2}}\frac{\cot\theta}{r}\left(\tilde{\bm\theta}^3 - \tilde{\bm\theta}^2 \right) \right] = \frac{m}{r^3}~\tilde{\bm\theta}^1 \wedge \tilde{\bm\theta}^2 - \frac{1}{\sqrt{2}}\frac{\cot\theta}{r^2}~ \tilde{\bm\theta}^2 \wedge \tilde{\bm\theta}^3.\nonumber
\end{equation}
Putting the two above equations in the first equation of system (\ref{2a_equacao_de_Cartan_2}), the curvature 2-form $\bm \Theta_{03}$ follows as 
\begin{equation}
 \label{Vaidya_tetrad_21}
 \bm \Theta_{03} = \frac{m}{r^3}~\tilde{\bm\theta}^1 \wedge \tilde{\bm\theta}^2.
\end{equation}

The second calculation to be performed is a exterior derivative operation in $\bm\Gamma_{12} = \dfrac{1}{2r}\left(1-\dfrac{2m}{r} \right)\tilde{\bm\theta}^{3} $,
\begin{equation}
 d\bm\Gamma_{12} = d\left[\dfrac{1}{2r}\left(1-\dfrac{2m}{r} \right)\tilde{\bm\theta}^{3} \right] = \left[-\frac{1}{2r^2} + \frac{2m}{r^3}  \right]dr\wedge \tilde{\bm\theta}^{3} + \frac{1}{2r}\left(-\frac{2\dot{m}}{r} \right)du\wedge  \tilde{\bm\theta}^{3} + \frac{1}{2r}\left(1-\frac{2m}{r} \right)d \tilde{\bm\theta}^{3} , \nonumber
\end{equation}
where $\dot{m}=\dfrac{\partial m}{\partial u}$. We must use the equations  (\ref{Vaidya_tetrad_10}), (\ref{Vaidya_tetrad_12}) and $du = \tilde{\bm\theta}^{1}$ in order to get,
\begin{equation}
 d\bm\Gamma_{12} = \frac{m}{r^3}~\tilde{\bm\theta}^{0}\wedge \tilde{\bm\theta}^{3} - \frac{m}{2r^3}\left(1-\frac{2m}{r} \right)\tilde{\bm\theta}^{1}\wedge \tilde{\bm\theta}^{3} - \frac{\dot{m}}{r^2}~\tilde{\bm\theta}^{1}\wedge \tilde{\bm\theta}^{3} +\frac{\cot\theta}{2\sqrt{2}~r^2}\left(1-\frac{2m}{r} \right)\tilde{\bm\theta}^{2}\wedge \tilde{\bm\theta}^{3}. \nonumber
\end{equation}
The respective exterior product ${\bm \Gamma}_{12} \wedge ( \bm\Gamma_{01}+ \bm \Gamma_{23})$ results in,
\begin{equation}
 {\bm \Gamma}_{12} \wedge ( \bm\Gamma_{01}+ \bm \Gamma_{23}) =  \dfrac{1}{2r}\left(1-\dfrac{2m}{r} \right)\tilde{\bm\theta}^{3} \wedge \left[\frac{m}{r^2}~\tilde{\bm\theta}^1 + \frac{1}{\sqrt{2}}\frac{\cot\theta}{r}\left(\tilde{\bm\theta}^3 - \tilde{\bm\theta}^2 \right) \right] = -\frac{m}{2r^3}\left(1-\frac{2m}{r}\right)\tilde{\bm\theta}^1 \wedge \tilde{\bm\theta}^3 + \frac{\cot\theta}{2\sqrt{2}~r^2}\left(1-\frac{2m}{r}\right) \tilde{\bm\theta}^2 \wedge \tilde{\bm\theta}^3.\nonumber
\end{equation}
Putting the two above equations in the second equation of system (\ref{2a_equacao_de_Cartan_2}), the curvature 2-form $\bm \Theta_{12}$ follows as 
\begin{equation}
 \label{Vaidya_tetrad_22}
 \bm \Theta_{12} = \frac{m}{r^3}~\tilde{\bm\theta}^0 \wedge \tilde{\bm\theta}^3 - \frac{\dot{m}}{r^2}~\tilde{\bm\theta}^{1}\wedge \tilde{\bm\theta}^{3}.
\end{equation}

To the third equation of the system (\ref{2a_equacao_de_Cartan_2}), it becomes necessary to calculate the exterior derivatives in $\bm\Gamma_{01}$ of equation (\ref{Vaidya_tetrad_13}) and $\bm\Gamma_{23}$ of equation (\ref{Vaidya_tetrad_20}). Thus, the exterior derivative of $\bm\Gamma_{01}$ is
\begin{equation}
 d\bm\Gamma_{01} = d\left[\frac{m}{r^2}~\tilde{\bm\theta}^{1}\right] = \frac{\dot{m}}{r^2} du\wedge  \tilde{\bm\theta}^{1} - \frac{2m}{r^3} dr\wedge \tilde{\bm\theta}^{1} +\frac{m}{r^2} d \tilde{\bm\theta}^{1}  , \nonumber
\end{equation}
with $du=\tilde{\bm\theta}^{1}$, the equations (\ref{Vaidya_tetrad_10}) and 
(\ref{Vaidya_tetrad_07}) the above equation reduces to,
\begin{equation}
 d\bm\Gamma_{01} = - \frac{2m}{r^3} \tilde{\bm\theta}^{0}\wedge \tilde{\bm\theta}^{1}. \nonumber
\end{equation}
The exterior derivative of $\bm\Gamma_{23}$ is
\begin{equation}
 d\bm\Gamma_{23} = d\left[\frac{1}{\sqrt{2}}\frac{\cot\theta}{r} \left( \tilde{\bm\theta}^3 - \tilde{\bm\theta}^2\right)\right] = \frac{1}{\sqrt{2}}\left[-\frac{\cot\theta}{r^2}~dr\wedge \left( \tilde{\bm\theta}^3 - \tilde{\bm\theta}^2\right) \right] - \frac{\csc^2\theta}{r}~d\theta \wedge \left( \tilde{\bm\theta}^3 - \tilde{\bm\theta}^2\right) + \frac{\cot\theta}{r}\left( d\tilde{\bm\theta}^3 - d\tilde{\bm\theta}^2\right). \nonumber
\end{equation}
With aid of equations (\ref{Vaidya_tetrad_08}), (\ref{Vaidya_tetrad_10}), (\ref{Vaidya_tetrad_11}) and (\ref{Vaidya_tetrad_12}), the above 2-form reduces to,
\begin{equation}
 d\bm\Gamma_{23} = -\frac{1}{r^2}~\tilde{\bm\theta}^2 \wedge \tilde{\bm\theta}^3 . \nonumber
\end{equation}
The respective exterior product ${\bm \Gamma}_{03} \wedge  \bm\Gamma_{12}$ of the third equation of system (\ref{2a_equacao_de_Cartan_2}) is 
\begin{equation}
{\bm \Gamma}_{03} \wedge  \bm\Gamma_{12} = \left(-\frac{1}{r}~\tilde{\bm\theta}^2\right)\wedge\left[\frac{1}{2r}\left(1-\frac{2m}{r}\right) \tilde{\bm\theta}^3 \right] = -\frac{1}{2r^2}~\tilde{\bm\theta}^2\wedge \tilde{\bm\theta}^3 + \frac{m}{r^3}~\tilde{\bm\theta}^2\wedge \tilde{\bm\theta}^3.\nonumber
\end{equation}
Finally, we can substitute the above results in third equation of system (\ref{2a_equacao_de_Cartan_2}) where the curvature 2-form $\bm \Theta_{01} +  \bm \Theta_{23} $ is given by
\begin{equation}
 \label{Vaidya_tetrad_23}
 \bm \Theta_{01} +  \bm \Theta_{23} = -\frac{2m}{r^3}~\tilde{\bm\theta}^0\wedge \tilde{\bm\theta}^1 - \frac{2m}{r^3}~\tilde{\bm\theta}^2\wedge \tilde{\bm\theta}^3. 
\end{equation}

With the curvature 2-forms (\ref{Vaidya_tetrad_21}),  (\ref{Vaidya_tetrad_22}) and (\ref{Vaidya_tetrad_23}) we obtain the components of Riemann curvature by using,
\begin{equation}
\label{2-form-Riemann_Tensor}
 \bm \Theta_{\alpha\beta} = \frac{1}{2} R_{\alpha\beta\gamma\delta}~\tilde{\bm\theta}^{\gamma}\wedge ~\tilde{\bm\theta}^{\delta}. 
\end{equation}
For the equation  (\ref{Vaidya_tetrad_21}), where $\bm \Theta_{03} = \dfrac{m}{r^3}~\tilde{\bm\theta}^1 \wedge \tilde{\bm\theta}^2$ we have that
\begin{equation}
 \bm \Theta_{03} = \frac{1}{2} R_{0312}~\tilde{\bm\theta}^1 \wedge \tilde{\bm\theta}^2 + \frac{1}{2} R_{0321}~\tilde{\bm\theta}^2 \wedge \tilde{\bm\theta}^1 = R_{0312}~\tilde{\bm\theta}^1 \wedge \tilde{\bm\theta}^2. \nonumber 
\end{equation}
So we can identify the following component of the Riemann curvature tensor with the respective permutations of their indices,
\begin{equation}
\label{Vaidya_Riemann_Tensor_1}
 R_{0312} = \frac{m}{r^3}.
\end{equation}
For the equation  (\ref{Vaidya_tetrad_22}), where $\bm \Theta_{12} = \dfrac{m}{r^3}~\tilde{\bm\theta}^0 \wedge \tilde{\bm\theta}^3 - \dfrac{\dot{m}}{r^2}~\tilde{\bm\theta}^{1}\wedge \tilde{\bm\theta}^{3}$, we obtain,
\begin{equation}
 \bm \Theta_{12} = \frac{1}{2} R_{1203}~\tilde{\bm\theta}^0 \wedge \tilde{\bm\theta}^3 + \frac{1}{2} R_{1230}~\tilde{\bm\theta}^3 \wedge \tilde{\bm\theta}^0 + \frac{1}{2} R_{1213}~\tilde{\bm\theta}^1 \wedge \tilde{\bm\theta}^3 + \frac{1}{2} R_{1231}~\tilde{\bm\theta}^3 \wedge \tilde{\bm\theta}^1 = R_{1203}~\tilde{\bm\theta}^0 \wedge \tilde{\bm\theta}^3 + \frac{1}{2} R_{1213}~\tilde{\bm\theta}^1 \wedge \tilde{\bm\theta}^3. \nonumber 
\end{equation}
So we can identify the following component of the Riemann curvature tensor with the respective permutations of their indices,
\begin{equation}
\label{Vaidya_Riemann_Tensor_2}
 R_{1203} = \frac{m}{r^3},
\end{equation}
and
\begin{equation}
\label{Vaidya_Riemann_Tensor_3}
 R_{1213} = -\frac{\dot{m}}{r^2}.
\end{equation}
And finally for the equation  (\ref{Vaidya_tetrad_22}), where $\bm \Theta_{01} +  \bm \Theta_{23} = -\dfrac{2m}{r^3}~\tilde{\bm\theta}^0\wedge \tilde{\bm\theta}^1 - \dfrac{2m}{r^3}~\tilde{\bm\theta}^2\wedge \tilde{\bm\theta}^3$, which results in the following component of the Riemann curvature tensor with the respective permutations of their indices,
\begin{equation}
\label{Vaidya_Riemann_Tensor_4}
 R_{0101} =  -\frac{2m}{r^3} \hspace*{1cm}\mbox{and}\hspace*{1cm} R_{2323} =  -\frac{2m}{r^3}.
\end{equation}
We identify the components (\ref{Vaidya_Riemann_Tensor_1}), (\ref{Vaidya_Riemann_Tensor_2}) and (\ref{Vaidya_Riemann_Tensor_4}) as components of Weyl tensor,  because these terms do not contribute to the contraction $\gamma^{\epsilon\zeta}R_{\epsilon\alpha\zeta\beta} = R_{\alpha\beta}$. These terms, $R_{1203}$, $R_{0101}$ and $R_{2323}$ are components of Weyl tensor. For example,
\begin{equation}
 R_{01} = \gamma^{01}R_{1001} + \gamma^{10}R_{0011} + \gamma^{23}R_{3021} + \gamma^{32}R_{2031} = -\frac{2m}{r^3} + \frac{m}{r^3} + \frac{m}{r^3} -= 0, \nonumber
\end{equation}
and so on. 
For comparing $R_{1203}$, $R_{0101}$ and $R_{2323}$ with (\ref{Psi_2_5}), (\ref{Psi_2_8}) and (\ref{Psi_2_10}) we identify for example that,
$R_{0312}=C_{0312}$ and then we get from equation (\ref{Psi_2_8}) that 
$\bar{\Psi}_{2} = -C_{0312}$ or,
\begin{equation}
\label{Psi_2_11}
 \Psi_{2} = -\frac{m}{r^3}.
\end{equation}
The first  Bianchi identity (\ref{first_Bianchi_equation}), $R_{\alpha\beta\gamma\delta} +  R_{\alpha\gamma\delta\beta} + R_{\alpha\delta\beta\gamma} = 0$, yields,
\begin{equation}
 R_{0312}+R_{0123}+R_{0231} = \frac{m}{r^3} + R_{0123} - \frac{m}{r^3} = 0, \nonumber
\end{equation}
where we have $R_{0123} = C_{0123} = 0$. So we have from equation (\ref{Psi_2_10}) the same result $\Psi_2 = \dfrac{1}{2}C_{2323}= -\dfrac{m}{r^3}$.

There is a single nonzero component of Ricci tensor, $ R_{\alpha\beta}= \gamma^{\epsilon\zeta}R_{\epsilon\alpha\zeta\beta}$ that we can obtain from equation (\ref{Vaidya_Riemann_Tensor_3}) performing the calculation,
\begin{equation}
\label{Ricci_component_Vaidya}
 R_{11} = \gamma^{01}R_{1101} + \gamma^{10}R_{0111}+ \gamma^{23}R_{3121} + \gamma^{32}R_{2131} = -\frac{\dot{m}}{r^2} -\frac{\dot{m}}{r^2} = -\frac{2\dot{m}}{r^2}. 
\end{equation}
Thus, the Ricci tensor in Newman-Penrose coordinate basis is given by
\begin{equation}
 \bm {Ric} = R_{\alpha\beta}~\tilde{\bm\theta}^{\alpha}\otimes ~\tilde{\bm\theta}^{\beta}. \nonumber
\end{equation}
We can express the Ricci tensor in a coordinate basis by $\tilde{\bm\theta}^{\alpha} = {\omega^{\alpha}}_{\mu} dx^{\mu}$,
such that,
\begin{equation}
 \bm {Ric} = R_{\alpha\beta}({\omega^{\alpha}}_{\mu} dx^{\mu})\otimes ({\omega^{\beta}}_{\nu} dx^{\nu}) = R_{11} {\omega^{1}}_{\mu}{\omega^{1}}_{\nu}  dx^{\mu}\otimes dx^{\nu}, \nonumber
\end{equation}
we can recall the equation (\ref{dual_basis}) where ${\omega^{1}}_{\nu} = k_{\mu}$, it reduces to
\begin{equation}
 \bm {Ric} = -\frac{2\dot{m}}{r^2} k_{\mu}k_{\nu} dx^{\mu}\otimes dx^{\nu},
\end{equation}
where we can identify the components of the Ricci tensor in coordinate basis
\begin{equation}
 R_{\mu\nu} = -\frac{2\dot{m}}{r^2} ~k_{\mu}k_{\nu}.
\end{equation}
The scalar curvature vanishes with $k_{\mu}k^{\mu}=0$, then the Einstein' field equation of Vaidya spacetime results in a Einstein-Maxwell solution, where the energy-momentum tensor is
\begin{equation}
 T_{\mu\nu} = - \frac{\dot{m}}{4\pi G r^2} ~k_{\mu}k_{\nu}.
\end{equation}
The above  tensor is a energy-momentum tensor of null dust or pure radiation. In fact the Vaidya spacetime is  spherically symmetric, that in this example we have analyzed a retarded spacetime (\ref{Vaidya_retarded}), representing a emitting  null dust, the flux of massless particles, i.e. photons.

In Newman-Penrose formalis we have obtained for Einstein-Maxwell solution for pure electromagnetic radiation, the equation (\ref{Energy_momentum_tensor_7}), where we have
\begin{equation}
 T_{\alpha\beta} = 2~\Phi_{2}\overline\Phi_{2}~ k_{\alpha} k_{\beta},\nonumber
\end{equation}
and we can identify,
\begin{equation}
 \label{Vaidya_tetrad_24}
 |\Phi_{2}|^2 = \frac{|\dot{m}|}{8\pi G r^2}.
\end{equation}
As stated before, the null electromagnetic field has $\Phi_{0}=\Phi_{1} = 0$, and the electromagnetic tensor is ${\cal F}_{\alpha\beta} = 2 \Phi_{2} V_{\alpha\beta}$.

Also in the Newman-Penrose formalism we have obtained the equation (\ref{Einstein_Maxwell_NP_1})
\begin{equation}
  \Phi_{AB} = 8\pi G ~ \Phi_{A}\overline{\Phi}_{B}, \hspace*{1cm} \mbox{where}~~A,B=0,1,2. \nonumber 
\end{equation}
For the null electromagnetic field the only component nonzero is $\Phi_{2}$, such that we have the equation (\ref{Einstein_Maxwell_NP_Phi_22}), 
\begin{equation}
  \Phi_{22} = 8\pi G ~ \Phi_{2}\overline{\Phi}_{2}. \nonumber 
\end{equation}
In fact we have from equation (\ref{Ricci_Phi_22})
\begin{equation}
 \Phi_{22} = \frac{1}{2} R_{11}, \nonumber
\end{equation}
where we can use the result of equation (\ref{Ricci_component_Vaidya}) and therefore the component of Ricci tensor $\Phi_{22}$ in Newman-Penrose formalis becomes,
\begin{equation}
\label{Ricci_component_Vaidya_2}
 \Phi_{22} = -\frac{\dot{m}}{r^2}.
\end{equation}
If one substitute it in the Einstein-Maxwell equation $\Phi_{22} = 8\pi G ~ \Phi_{2}\overline{\Phi}_{2}$ we repeat the value of equation (\ref{Vaidya_tetrad_24}).

%%%%%%%%%%%%%%%%%%%%%%%%%%%%%%%%%%%%%%%%%%%%%%%%%%%%%%%%%%%%%%
%%%%%%%%%%%%%%%%%%%%%%%%%%%%%%%%%%%%%%%%%%%%%%%%%%%%%%%%%%%%%%
%%%%%%%%%%%%%%%%%%%%%%%%%%%%%%%%%%%%%%%%%%%%%%%%%%%%%%%%%%%%%%
%%%%%%%%%%%%%%%%%%%%%%%%%%%%%%%%%%%%%%%%%%%%%%%%%%%%%%%%%%%%%%
%%%%%%%%%%%%%%%%%%%%%%%%%%%%%%%%%%%%%%%%%%%%%%%%%%%%%%%%%%%%%%
%%%%%%%%%%%%%%%%%%%%%%%%%%%%%%%%%%%%%%%%%%%%%%%%%%%%%%%%%%%%%%

%%%%%%%%%%%%%%%%%%%%%%%%%%%%%%%%%%%%%%%%%%%%%%%%%%%%%%%%%%%%%%
%%%%%%%%%%%%%%%%%%%%%%%%%%%%%%%%%%%%%%%%%%%%%%%%%%%%%%%%%%%%%%
%%%%%%%%%%%%%%%%%%%%%%%%%%%%%%%%%%%%%%%%%%%%%%%%%%%%%%%%%%%%%%
%%%%%%%%%%%%%%%%%%%%%%%%%%%%%%%%%%%%%%%%%%%%%%%%%%%%%%%%%%%%%%
%%%%%%%%%%%%%%%%%%%%%%%%%%%%%%%%%%%%%%%%%%%%%%%%%%%%%%%%%%%%%%
%%%%%%%%%%%%%%%%%%%%%%%%%%%%%%%%%%%%%%%%%%%%%%%%%%%%%%%%%%%%%%

%%%%%%%%%%%%%%%%%%%%%%%%%%%%%%%%%%%%%%%%%%%%%%%%%%%%%%%%%%%%%%
%%%%%%%%%%%%%%%%%%%%%%%%%%%%%%%%%%%%%%%%%%%%%%%%%%%%%%%%%%%%%%
%%%%%%%%%%%%%%%%%%%%%%%%%%%%%%%%%%%%%%%%%%%%%%%%%%%%%%%%%%%%%%
%%%%%%%%%%%%%%%%%%%%%%%%%%%%%%%%%%%%%%%%%%%%%%%%%%%%%%%%%%%%%%
%%%%%%%%%%%%%%%%%%%%%%%%%%%%%%%%%%%%%%%%%%%%%%%%%%%%%%%%%%%%%%
%%%%%%%%%%%%%%%%%%%%%%%%%%%%%%%%%%%%%%%%%%%%%%%%%%%%%%%%%%%%%%

\section{Conclusion}

This review was intended to give mathematical details in a didactic way about the bivectors of General Relativity, starting from the properties of the electromagnetic bivector, from which it is possible to build a basis of bivectors. 
These concepts covered in this review in tensor language and differential forms are the first steps to understanding advanced General Relativity in an elegant framework of the spinor formalism \cite{Carmeli, Penrose_Rindler, Stewart}.
In addition to the possibility of studying the exact solutions of Einstein's field equations, it becomes possible in this context of the Newman-Penrose formalism, the algebraic classification of different types of spacetimes, through the algebraic classification of the Weyl tensor or Petrov classification \cite{Kramer, Carmeli, Hall, Stewart}.
In the example discussed in this review, the Vaidya spacetime, we calculate the complex component of the Weyl tensor in the Newman-Penrose formalism, $\Psi_{2} = -\dfrac{m}{r^3}$ in equation (\ref{Psi_2_11}), where the physical interpretation is that $\Psi_{2}$ is a Newton-like potential component of gravitational field \cite{Griffiths, Szekeres}.
Because the trace of Weyl curvature is zero, ${C^{\alpha}}_{\beta\alpha\delta} = C_{\beta\delta} = 0 $, the Weyl complex coefficients $\Psi_{0}$, $\Psi_{1}$, $\Psi_{2}$, $\Psi_{3}$ and  $\Psi_{5}$ are terms of gravitational field in vacuum. The physical meanings of   the Weyl complex coefficients $\Psi_{0}$, $\Psi_{1}$, $\Psi_{2}$, $\Psi_{3}$ and  $\Psi_{5}$ are properly treated in the reference \cite{Szekeres}.
Finally through the Newman-Penrose formalism we can interpret the Ricci tensor component $\Phi_{22}$ obtained for the Vaidya spacetime, as an exact solution of the Einstein-Maxwell equation: $\Phi_{22} = 8 \pi G \Phi_{2}\bar{\Phi}_{2}$, 
which should be interpreted as electromagnetic radiation, whose electromagnetic field component is $\Phi_{2}$, emitted by a radianting nonrotating star related to Vaidya spacetime.

%%%%%%%%%%%%%%%%%%%%%%%%%%%%%%%%%%%%%%%%%%%%%%%%%%%%%%%%%%%%%%
%%%%%%%%%%%%%%%%%%%%%%%%%%%%%%%%%%%%%%%%%%%%%%%%%%%%%%%%%%%%%%
%%%%%%%%%%%%%%%%%%%%%%%%%%%%%%%%%%%%%%%%%%%%%%%%%%%%%%%%%%%%%%
%%%%%%%%%%%%%%%%%%%%%%%%%%%%%%%%%%%%%%%%%%%%%%%%%%%%%%%%%%%%%%
%%%%%%%%%%%%%%%%%%%%%%%%%%%%%%%%%%%%%%%%%%%%%%%%%%%%%%%%%%%%%%
%%%%%%%%%%%%%%%%%%%%%%%%%%%%%%%%%%%%%%%%%%%%%%%%%%%%%%%%%%%%%%

%\bibliographystyle{elsarticle-num}

\end{document}